%% file: Adaptive_Caching_ToN.tex
\documentclass[journal]{IEEEtran}
\input{preambleToN}

\ifCLASSOPTIONcompsoc
  \usepackage[nocompress]{cite}
\else
  \usepackage{cite}
\fi

\ifCLASSINFOpdf
\else
\fi

\begin{document}

\bstctlcite{IEEEexample:BSTcontrol}

%
\title{Adaptive TTL-Based Caching for Content Delivery}


\author{Soumya~Basu,~\IEEEmembership{}
        Aditya~Sundarrajan,~\IEEEmembership{}
        Javad~Ghaderi,~\IEEEmembership{}
        Sanjay~Shakkottai,~\IEEEmembership{Fellow,~IEEE}
        and~Ramesh~Sitaraman~\IEEEmembership{}
\thanks{A short version of this work has appeared as a two-page extended abstract~\cite{soumya2017adaptive}
in the Proceedings of ACM Sigmetrics, Urbana, IL,  June 2017.}
\IEEEcompsocitemizethanks{\IEEEcompsocthanksitem S. Basu and S. Shakkottai are with the Department
of Electrical and Computer Engineering, The University of Texas at Austin, TX 78712. (E-mail: basusoumya@utexas.edu and shakkott@austin.utexas.edu).%
\IEEEcompsocthanksitem A. Sundarrajan and R. Sitaraman are with the College of Information and Computer Sciences, University of Massachusetts Amherst, MA 01003. (E-mail: asundar@cs.umass.edu and ramesh@cs.umass.edu)%
\IEEEcompsocthanksitem J. Ghaderi is with the Department of Electrical Engineering, Columbia University, New York, NY 10027. (E-mail: jghaderi@ee.columbia.edu)}%
}

\maketitle

\begin{abstract}
Content Delivery Networks (CDNs) cache and serve a majority of the user-requested content on the Internet. Designing caching algorithms that {\em automatically} adapt to the heterogeneity, burstiness, and non-stationary nature of real-world content requests is a major challenge and is the focus of our work. While there is much work on caching algorithms for stationary request traffic, the work on non-stationary request traffic is very limited. Consequently, most prior models are inaccurate for non-stationary production CDN traffic. We propose two TTL-based caching algorithms that provide provable performance guarantees for request traffic that is bursty and non-stationary. The first algorithm called d-TTL dynamically adapts a TTL parameter using stochastic approximation.  Given a feasible target hit rate, we show that d-TTL converges to its target value for a general class of bursty traffic that allows Markov dependence over time and non-stationary arrivals. The second algorithm called f-TTL uses two caches, each with its own TTL. The first-level cache adaptively filters out non-stationary traffic, while the second-level cache stores frequently-accessed stationary traffic. Given feasible targets for both the hit rate and the expected cache size, f-TTL asymptotically achieves both targets. We evaluate both d-TTL and f-TTL using an extensive trace containing more than 500 million requests from a production CDN server.  We show that both d-TTL and f-TTL converge to their hit rate targets with an error of about 1.3\%. But, f-TTL requires a significantly smaller cache size than d-TTL to achieve the same hit rate, since it effectively filters out non-stationary content.
\end{abstract}



%
\IEEEpeerreviewmaketitle

\input{Intro-v4-SB}
\input{System5-SB}

\input{algorithm-v5}

\input{MainResult4-JG}

\input{implementation}

\input{empirical}
\input{related}
\input{conclusions}


%

\section*{Acknowledgment}
This work is partially supported by the US Dept. of Transportation supported D-STOP Tier 1 University Transportation Center, NSF grants CNS-1652115 and CNS-1413998.

\ifCLASSOPTIONcaptionsoff
  \newpage
\fi



\bibliographystyle{IEEEtran}
\bibliography{mybib,cdnbib}
\clearpage
\renewcommand\thesubsectiondis{\thesectiondis.\arabic{subsection}}
\appendices
\input{proof}
\input{bhr}
\input{sensitivity}

\end{document}

%% file: preambleToN.tex
\usepackage[utf8]{inputenc}

\usepackage{amsmath, amsthm, amssymb}
\usepackage{bbm}
\usepackage{bm}
\usepackage{enumitem}
\usepackage{accents}
\usepackage{cuted}
\usepackage{float}
\usepackage{comment}
\usepackage{url}
\usepackage{multirow}
\usepackage{algorithm}
\usepackage{algorithmicx}
\usepackage[noend]{algpseudocode}
\makeatletter
\let\OldStatex\Statex
\renewcommand{\Statex}[1][3]{%
  \setlength\@tempdima{\algorithmicindent}%
  \OldStatex\hskip\dimexpr#1\@tempdima\relax}
\makeatother
\usepackage{graphicx}
\graphicspath{{./fig/}}
\usepackage[monochrome]{color}
\newcommand{\hi}[1]{\colorbox{yellow}{#1}}
\newcommand{\s}[1]{\textcolor{black}{#1}}
\usepackage{float}
\floatstyle{plain}
\newfloat{eqholder}{ht}{loe}
\floatname{eqholder}{Equation}

\newcommand{\noop}[1]{} 
\usepackage{array}
\makeatletter
\newtheorem*{rep@theorem}{\rep@title}
\newcommand{\newreptheorem}[2]{%
\newenvironment{rep#1}[1]{%
 \def\rep@title{#2 \ref{##1}}%
 \begin{rep@theorem}}%
 {\end{rep@theorem}}}
\makeatother
\newtheorem{theorem}{Theorem}
\newreptheorem{theorem}{Theorem}
\newtheorem{lemma}{Lemma}
\newreptheorem{lemma}{Lemma}
\newtheorem{proposition}{Proposition}
\newtheorem{claim}{Claim}
\newtheorem{corollary}{Corollary}
\theoremstyle{definition}

\newtheorem{definition}{Definition}
\theoremstyle{definition}

\theoremstyle{definition}

\newtheorem{assumptiongrp}{Assumption}
\theoremstyle{definition}
\usepackage{chngcntr}
\newcounter{preassumption}
\counterwithin{assumptiongrp}{preassumption}

\newcommand{\assumptiongroup}{\refstepcounter{preassumption}}

\newtheorem{condition}{Condition}
\theoremstyle{definition}
\newtheorem{algo}{Algorithm}
\theoremstyle{remark}

\theoremstyle{remark}
\newtheorem{remark}{Remark}

\newcommand{\mc}[1]{\mathcal{#1}}
\newcommand{\st}[1]{{\scriptstyle{#1}}}
\newcommand{\mcs}[1]{{\scriptstyle\mathcal{#1}}}
\newcommand{\ttl}{TTL}
\newcommand{\dm}[1]{{\displaystyle #1}}
\newcommand{\up}{\Gamma}
\newcommand{\vc}[3]{\overset{#2}{\underset{#3}{#1}}}

\newcommand*{\dt}[1]{%
  \accentset{\mbox{\large\bfseries .}}{#1}}
\usepackage{subcaption}

%% file: Intro-v4-SB.tex
\section{Introduction}
\label{sec:intro}
By caching and delivering content to millions of end users around the
world, content delivery networks (CDNs)~\cite{dilley2002globally} are an integral part
of the Internet infrastructure. A large CDN such as Akamai~\cite{nygren2010akamai} serves several trillion user requests a day from 170,000+ servers located in 1500+ networks in 100+ countries around the world.  The majority of today's Internet traffic  is delivered by CDNs. CDNs are expected to deliver nearly two-thirds of the Internet traffic by 2020 \cite{cisco-videogrowth}.

The main function of a CDN server is to cache and serve content requested by users. The effectiveness of a caching algorithm is measured by its achieved hit rate in relation to its cache size. There are two primary ways of measuring the hit rate. The {\em object hit rate (OHR)} is the fraction of the requested objects that are served from cache and the {\em byte hit rate (BHR)} is the fraction of the requested content bytes that are served from cache. We devise algorithms capable of operating with both notions of hit rate in our work.

The major technical challenge in designing caching algorithms for a modern CDN is
{\em adapting} to the sheer heterogeneity of the content that is accessed by users. The accessed content falls into multiple traffic classes that include web pages, videos, software downloads, interactive applications, and social networks. The classes differ widely in terms of the object size distributions and content access patterns. The popularity of the content also varies by several orders of magnitude with some objects accessed millions of times (e.g, an Apple iOS download), and other objects  accessed once or twice (e.g, a photo in a Facebook gallery). In fact, as shown in Figure~\ref{fig:object_pop_dist}, 70\% of the objects served by a CDN server are only requested once over a period of multiple days! Further, the requests served by a CDN server can change rapidly over time as different traffic mixes are routed to the server by the CDN's load balancer in response to Internet events.

Request statistics clearly play a key role in determining the hit rate of a CDN server. However,  when request patterns vary rapidly across servers and time, a one-size-fits-all approach provides inferior hit rate performance in a production CDN setting.  Further, manually tuning the caching algorithms for each individual server to account for the varying request statistics is prohibitively expensive. {Thus, our goal is to devise \underline{self-tuning} caching algorithms that can automatically learn and adapt to the request traffic and \underline{provably} achieve any feasible hit rate and cache size, even when the request traffic is bursty and non-stationary.

Our work fulfills a long-standing deficiency in the current state-of-art in the modeling and analysis of caching algorithms. Even though real-world CDN traffic is known to be heterogeneous, with bursty, non-stationary and transient request statistics, there are no known caching algorithms that provide theoretical performance guarantees for such traffic.\footnote{\color{black} We note that LRU cache has been previously studied under non-stationary models, e.g. box model~\cite{olmos2014catalog}, shot noise model~\cite{leonardi2015least}. However these works do not capture the transient requests that we study here.}  In fact, much of the known formal models and analyses assume that the traffic follows the \emph{Independent Reference Model} (IRM)\footnote{The inter arrival times are i.i.d. and the object request on each arrival are chosen independently from the same distribution. }.
{\color{black} However, when it comes to production traces such models lose their relevance. The following example highlights the stark inaccuracy of one popular model corroborating similar observations in \cite{olmos2014catalog, leonardi2015least, gast2016asymptotically}, among others.}

{\bf Deficiency of current models and analyses.}  Time-to-live (TTL)-based caching algorithms
\cite{jung2003modeling,  fofack2012analysis, berger2014exact, berger2015maximizing, garetto2016unified, dehghan2016utility}  use a  TTL parameter to determine how long an object may remain in cache. 
{\color{black} TTL caches have emerged as useful mathematical tools to analyze the performance of traditional capacity-based caching algorithms such as LRU, FIFO, etc. The cornerstone of such analyses is the work by Fagin~\cite{fagin1977asymptotic} that relates the cache hit rate with the expected cache size and characteristic time for IRM traffic, which is also popularly known as Che's approximation after the follow-up work~\cite{che2002hierarchical}. Under this approximation, a LRU cache has the same expected size and hit rate as a TTL-cache with the TTL value equal to its characteristic time. Che's approximation is known to be accurate in cache simulations that use synthetic IRM traffic and is commonly used in the design of caching algorithms for that reason \cite{fricker2012impact, fricker2012versatile, garetto2016unified,  bianchi2013check, guillemin2013experimental}.} 

However, we show that Che's approximation produces erroneous results for actual production CDN traffic that is neither stationary nor IRM across the requests. We used an extensive 9-day request trace from a production server in Akamai's CDN and derived TTL values for multiple hit rate targets using Che's approximation\footnote{\color{black} Under the assumption that traffic is IRM with memoryless arrival we compute the TTL/characteristic time that corresponds to the traget hit rate.}. We then simulated a cache with those TTL values on the production traces to derive the {\em actual} hit rate that was achieved. For a target hit rate of 60\%, we observed that a fixed-TTL algorithm that uses the TTL computed from Che's approximation achieved a hit rate of 68.23\% whereas the dynamic TTL algorithms proposed in this work achieve a hit rate of 59.36\% (see Section~\ref{sec:accuracy} for a complete discussion). This difference between the target hit rate and that achieved by fixed-TTL highlights the  inaccuracy of the current state-of-the-art theoretical modeling on production traffic.
\subsection{Main Contributions}
{\color{black} We propose two TTL-based algorithms: d-TTL (for ``dynamic TTL'') and f-TTL (for ``filtering TTL'') that provably achieve a target cache hit rate and cache size. Rather than statically deriving the required TTL values by inferring the request statistics, our algorithms {\em dynamically} adapt the TTLs  to the request patterns. To more accurately model real traffic, we allow the request traffic to be non-independent and have non-stationary components. Further, we allow content to be classified into {\em types}, where each type has a target hit rate (OHR or BHR) and an average target cache size. In practice, a type can consist of all objects of a specific kind from a specific provider, e.g. CNN webpages, Facebook images, CNN video clips, etc.} 
Our main contributions are as follows:\\
1) \textbf{d-TTL: A one-level TTL algorithm.}  Algorithm d-TTL maintains a single TTL value
for each type, and dynamically adapts this value upon each arrival
(new request) of an object of this type. Given a hit rate that is
``feasible'' (i.e. there exists a static
genie-settable TTL parameter that can achieve this hit rate), we show that
d-TTL almost surely converges to this target hit rate. Our result holds for a general class of bursty traffic
(allowing Markov dependence over time), and even in the presence of
non-stationary arrivals. To the best of our knowledge, this is the
first adaptive TTL algorithm that can provably achieve a target hit rate with such
stochastic traffic.

However, our empirical results show that
non-stationary and unpopular objects can contribute significantly to
the cache size, while they contribute very little to the cache hit rate
(heuristics  that use Bloom filters
to eliminate such traffic \cite{MaggsSitaraman2015} support this
observation).\\
2) \textbf{f-TTL: A two-level TTL algorithm.} The need to
achieve both a target hit rate and a target cache size motivates the f-TTL
algorithm. f-TTL comprises a pair of caches: a lower-level adaptive TTL cache
that filters rare objects based on arrival history, and a
higher-level adaptive TTL cache that stores filtered objects. We
design an adaptation mechanism for a pair of TTL values
(higher-level and lower-level) per type, and show that we can
asymptotically achieve the desired hit rate (almost surely), under
similar traffic conditions as with d-TTL.
If the {\em stationary} part of the traffic is Poisson,
we have the following stronger
property. Given any feasible (hit rate, expected cache size) pair\footnote{Feasibility here is with
  respect to any static two-level TTL algorithm that achieves a (target hit
  rate, target expected cache size) pair.}, the f-TTL algorithm
asymptotically achieves a corresponding pair that dominates the given
target\footnote{A pair dominates another pair if hit rate is at least
  equal to the latter and expected size is at most equal to the
  latter.}. Importantly, with non-stationary traffic, the two-level
adaptive TTL strictly outperforms the one-level TTL cache with respect
to the expected cache size.

Our proofs use  a two-level stochastic approximation
technique (along with a latent observer idea inspired from
actor-critic algorithms \cite{konda2003onactor}), and provide the
first theoretical justification for the deployment of two-level caches
such as ARC~\cite{megiddo2004outperforming}   in production systems with non-stationary traffic.\\
3) \textbf{Implementation and empirical evaluation:} We implement both d-TTL and f-TTL  and evaluate them using an extensive 9-day trace consisting of more than 500 million requests from a production Akamai CDN server. We observe that both d-TTL and f-TTL adapt well to the bursty and non-stationary nature of production CDN traffic. For a range of target object hit rate, both d-TTL and f-TTL converge to that target with an error of about 1.3\%. For a range of target byte hit rate, both d-TTL and f-TTL converge to that target with an error that ranges from 0.3\% to 2.3\%. While the hit rate performance of both d-TTL and f-TTL are similar, f-TTL shows a distinct advantage in cache size due to its ability to filter out non-stationary traffic. In particular, f-TTL requires a cache that is 49\% (resp., 39\%) smaller than d-TTL to achieve the same object (resp., byte) hit rate. This renders f-TTL useful to CDN settings where large amounts of non-stationary traffic can be filtered out to conserve cache space while also achieving target hit rates. \\

  Finally, from a practitioner's perspective, this work has the
  potential to enable new CDN pricing models. CDNs typically do not
  charge content providers on the basis of a guaranteed hit rate performance for
  their content, nor on the basis of the cache size that they use.
  Such pricing models have desirable properties, but do not commonly
  exist, in part, because current caching algorithms cannot provide
  such guarantees with low overhead. Our caching algorithms are the
  first to provide a theoretical guarantee on hit rate for each
  content provider, while controlling the cache space that they can
  use. Thus, our work removes a technical impediment to hit rate and  
  cache space based CDN pricing.

\subsection{Notations}
Some of the basic notations used in this paper are as follows. Bold font characters indicate vector variables and normal font characters indicate scalar variables.
We note $(x)^{+}= \max(0,x)$, $\mathbb{N}=\{1, 2,\dots\}$, and $[n]=\{1, 2,\dots, n\}$. The equality among two vectors means component-wise equality holds. Similarly, inequality among two vectors (denoted by $\preccurlyeq$) means the inequality holds  for each component separately. {\color{black} We use the term `w.p.' for `{\em with probability}', `w.h.p.' for  `{\em with high probability}', `a.s.' for `{\em almost surely}', and `a.a.s.' for `{\em asymptotically almost surely}'.}

%% file: System5-SB.tex
\section{System Model and Definitions}\label{sec:sys}
Every CDN server implements a cache that stores objects requested by users. When a user's request arrives at a CDN server, the requested object is served from its cache, if that object is present. Otherwise, the CDN server fetches the object from a remote origin server that has the original content and  then serves it to the user. In addition, the CDN server may place the newly-fetched object in its cache. In general, a caching algorithm decides which object to place in cache, how long objects need to be stored in cache, and which objects should be evicted from cache.

When the requested object is found in cache, it is a \textit{cache hit}, otherwise it is a \textit{cache miss}. A cache hit is desirable since the object can be retrieved locally from the proximal server and returned to the user with low latency. Additionally, it is often beneficial to maintain state (metadata such as the URL of the object or an object ID) about a recently evicted object for some period of time. Then, we experience a {\em cache virtual hit} if the requested object is not in cache but its metadata is in cache. Note that the metadata of an object takes much less cache space than the object itself.

Next, we formally describe the request arrival model, and the performance metrics: object (byte) hit rate and expected cache size, and formally state the objective of the paper. 

\input{request6-SB}
\subsection{Object (Byte) Hit Rate and Normalized Size}
There are two common measures of hit rate. {\color{blue} The \textbf{object hit rate (OHR)} is the fraction of requests that experience a cache hit. The \textbf{byte hit rate (BHR)} is the fraction of requested bytes that experience a cache hit. BHR measures the traffic reduction between the origin and the cache severs. Both measures can be computed for a single object or a group of objects. Here, we consider all the objects of one type as one separate group.}

\color{blue}
We formally define OHR and BHR as follows.
Given a caching algorithm, define $Y(l) = 1$ if the $l$-th arrival experiences a cache hit and $Y(l) = 0$ otherwise. Also, let $\mc{C}(\tau)$ be the set of objects in the cache at time $\tau$, for $\tau\geq 0$. 
\begin{definition}\label{def:ohr}
	The OHR for each type $t \in T$ is defined as
	\[h_t  = \liminf_{\tau \to \infty} \frac{\sum_{l: A(l) \leq \tau} \mathbbm{1}\left( c_{typ}(l) = t , Y(l) =1 \right) }{\sum_{l: A(l) \leq \tau}\mathbbm{1}\left( c_{typ}(l) = t\right)}\]
\end{definition}

\begin{definition}\label{def:bhr}
	The BHR for each type $t \in T$ is defined as
	\[h_t  = \liminf_{\tau \to \infty} \frac{\sum_{l: A(l) \leq \tau} w(l)\mathbbm{1}\left( c_{typ}(l) = t , Y(l) =1 \right) }{\sum_{l: A(l) \leq \tau}w(l)\mathbbm{1}\left( c_{typ}(l) = t\right)}\]
\end{definition}
The performance of a caching algorithm is often measured using its hit rate curve (HRC) that relates the hit rate that it achieves to the cache size (in bytes) that it requires. {\color{blue} 
In general, the hit rate depends on the request arrival rate which in turn affects the cache size requirement. We define a new metric called the \textit{\textbf{normalized size}} which is 
defined as the ratio of the time-average cache size (in bytes) utilized by the object(s) over the time-average arrival rate (in bytes/sec) of the object(s).}  The normalized size is formally defined below. 
\begin{definition}\label{def:normSize}
	For a caching algorithm, and each type $t\in T$, the \textbf{\textit{normalized size}} for type $t$ is defined as
		\begin{align*}
		s_t = \limsup_{\tau \to \infty} \frac{\int_{\tau' = 0}^{\tau} \sum_{\substack{c\in \mc{C}(\tau')}} w_c \mathbbm{1}(c_{typ} = t)  d\tau' }{\sum_{l: A(l) \leq \tau} w(l)\mathbbm{1}\left( c_{typ}(l) = t\right)} 
	\end{align*}
\end{definition}
}
\begin{remark}
Dividing both the numerator and the denominator by $\tau$ gives the interpretation of the normalized size as the average cache size utilized by the objects of type $t$ normalized by their aggregate arrival rate. For example, if a CDN operator wants to allocate an expected cache size of $100GB$ for type $t$ and its arrival rate is known to be $10GB/sec$, then the corresponding normalized size is $\frac{100GB}{10 GB/sec} = 10 sec$.
\end{remark}

\color{black}
\subsection{Design Objective} 
The fundamental challenge in cache design is striking a balance between two conflicting objectives: minimizing the cache size requirement and maximizing the cache hit rate. 
In addition, it is desirable to allow different Quality of Service (QoS) guarantees for different types of objects, i.e., different \emph{cache hit rates} and \emph{size targets} for different types of objects.  For example, a lower hit rate that results in a higher response time may be tolerable for a software download that happens in the background. But, a higher hit rate that results in a faster response time is desirable for a web page that is delivered in real-time to the user.  

In this work, \textbf{our objective} is to tune the {\ttl} parameters to asymptotically achieve a {\em target hit rate vector} $\mathbf{h}^{*}$ and a (feasible) {\em target normalized size vector} $\bm{s}^*$, without the prior knowledge of the content request process. The $t$-th components of $\mathbf{h}^{*}$ and $\bm{s}^*$, i.e., $h_t^{*}$ and ${s}^*_t$ respectively, denote the target hit rate and the target normalized size for objects of type $t\in [T]$. 

A CDN operator can group objects into types in an arbitrary way. If the objective is to achieve an overall hit rate and  cache size, all objects can be grouped into a single type. It should also be noted that the algorithms proposed in this work {\em do not} try to achieve the target hit rate with the smallest cache size; this is a non-convex optimization problem that is not the focus of this work. Instead, we only try to achieve a {\em given} target hit rate and target normalized size.

%% file: request6-SB.tex
\subsection{Content Request Model}
\label{ssec:Arr}
There are different types of content hosted on modern CDNs. A content type may represent a specific genre of content (videos, web pages, etc.) from a specific content provider (CNN, Facebook, etc.). A single server could be shared among dozens of content types. A salient feature of content hosted on CDNs is that the objects of one type can be very different from the objects of another type, in terms of their popularity characteristics, request patterns and object size distributions. Most content types exhibit a long tail of popularity where there is a smaller set of \emph{recurring objects} that demonstrate a stationary behavior in their popularity and are requested frequently by users, and a larger set of \emph{rare objects} that are unpopular and show a high degree of non-stationarity. Examples of rare objects include those that are requested infrequently or even just once, a.k.a. one-hit wonders~\cite{maggs2015algorithmic}. Another example is  an object that is rare in a temporal sense and is frequently accessed within a small time window, but is seldom accessed again. Such a bursty request pattern can occur during flash crowds~\cite{jung2002flash}. In this section, we present a content request model that captures these characteristics.

\noindent \textit{1) Content Description:}

We consider $T$ different types of content where each type consists of both recurring objects and rare objects. The set of recurring objects of type $t$ is denoted by $\mathcal{K}_t$ with $|\mathcal{K}_t| = K_t$ different objects, and the set of rare objects of type $t$ is denoted by $\mathcal{R}_t$. The entire universe of objects is represented as $\mathcal{U}\equiv \cup_{t\in T} \left(\mathcal{K}_t \cup \mathcal{R}_t\right)$, and the set of recurring objects is represented by a finite set $\mathcal{K}\equiv \cup_{t\in T}\mathcal{K}_t$. Let $ K \equiv  |\mathcal{K}|= \sum_{t\in T} K_t$.
In our model, The number of types $T$ is finite. For each type $t\in [T]$ there are finitely many recurring objects, i.e. $K_t$ is finite. However, the rare objects are allowed to be (potentially) countably infinite in number.

Each object $c\in \mathcal{U}$ is represented by a tuple, $c = (c_{i}, c_{typ}, c_{m})$, and its meta-data is represented as  $\tilde{c} = (c_i,c_{typ})$. Here, $c_{i}$ is the unique {\em label} for the object (e.g., its URL), $c_{typ}$ is the {\em type} that the object belongs to, and $c_{m}$ is the actual body of the object $c$. {\color{blue} If $c\in \mc{K}$, then w.l.o.g., we can index $c_i=k$ for some $k\in \{1, \dots K\}$.} The object meta-data, $\tilde{c} = (c_i,c_{typ})$, is assumed to have negligible size, and the size of object $c$ is denoted as $w_c=|c_m|$ (in bytes). Note that the object meta-data can be fully extracted from the incoming request. 
In our model, for all objects $c\in \mc{U}$, their sizes are uniformly bounded as $w_c\leq w_{\max}$. Moreover, we assume, for each type $t\in [T]$, all rare objects of type $t$ have equal size $\bar{w}_t$.\footnote{This could be relaxed to {\em average} size for type $t$ rare objects, as long as the average size over a large enough time window has $o(1)$ difference from the average, w.p. $1$.}

\noindent \textit{2) General Content Request Model:}  

We denote the object requested on $l$-th arrival as $$c(l)\equiv(\text{label}: c_i(l),\ \text{type}: c_{typ}(l),\ \text{size}:w(l)).$$  
Further, let $A(l)$ be the arrival time of the $l$-th request, and $X(l)$ be the $l$-th inter-arrival time, i.e., $X(l)=A(l)-A(l-1)$. {\color{blue}   We define a random variable $Z(l)$ which specifies the label of the $l$-th request if the request is for a recurrent object, and specifies its type if the request is for a rare object (i.e. $Z(l) = c_i(l)$ if $c(l)\in \mc{K}$, and $Z(l) = c_{typ}(l)$ otherwise). We also require the following two definitions:
\begin{eqnarray*}
X_{pre}(l) &=& \min \{A(l) - A(l'): l'<l, c(l')= c(l)\}\\
X_{suc}(l) &=& \min \{A(l') - A(l): l'>l, c(l')= c(l)\},
\end{eqnarray*}
hence $X_{pre}(l)$ and $X_{suc}(l)$ represent the preceding and succeeding inter-arrival time for the object requested on $l$-th arrival, respectively. By convention, $\min\{\emptyset\} = \infty$.

For any constant $R>0$, and $l\geq1$, define the set of objects that arrived within $R$ units of time from the $l$-th arrival, as
$$\mc{A}(l;R) = \{c(l'): l'\in \mathbb{N}, A(l')\leq A(l) - R\}.$$ We also define, for all $R>0$ and type $t \in [T]$, the \textit{bursty arrival indicator} $\beta_{t}(l;R)$ as the indicator function of the event: (1) the $l$-th request is for some rare object $c$ of type $t$, \textit{and} (2) the previous request of the same rare object $c$ happened (strictly) less than $R$ units of time earlier. Specifically, $\beta_{t}(l;R) = \mathbbm{1}( c(l)\in \mathcal{R}_t, X_{pre}(l)<R)$. Note that $\beta_{t}(l;R)$ does not depend on a specific $c \in \mathcal{R}_t$, but accumulates over all rare objects of type $t$.}

The general content request model is built on a Markov renewal process $\left(A(l), Z(l)\right)_{l\in \mathbb{N}}$~\cite{fox1968semi} (to model the stationary components and potential Markovian dependence on the object requests), followed by rare object labeling to model non-stationary components. Formally, our general content request model, parameterized by constant $R>0$, is as follows.

\assumptiongroup
\begin{assumptiongrp}
\label{a:Arr}

General Content Request Model ($R$):

\begin{itemize}[leftmargin=*]
%
\item\textbf{Markov renewal process $\left(A(l), Z(l)\right)_{l\in \mathbb{N}}$}
\begin{itemize}
\item[(i)] The inter-arrival times $X(l)=A(l)-A(l-1)$, $l\in \mathbb{N}$, are identically distributed, independently of each other and $Z(l)$.  The inter-arrival time distribution follows a probability density function (p.d.f.), $f(x)$ which is \emph{absolutely continuous} w.r.t a Lebesgue measure on $(\mathbb{R},+ )$ and has \emph{simply connected support}, {\color{blue} i.e. if $f(x)>0, f(y)>0$ then $f(z)>0$ for all $z\in (x,y)$.} The inter-arrival time has a \emph{nonzero finite mean} denoted by $1/\lambda$.

\item[(ii)]
\color{black}
The process $Z(l)$ is a Markov chain over $(K+T)$ states indexed by $1,\cdots,K+T$.
The first $K$ states represent the $K$ recurring objects. The rare objects (possibly infinite in number) are grouped according to their types, thus producing the remaining $T$ states, i.e, the states $K+1,\cdots, K+T$ represent rare objects of types $1, \cdots, T$, respectively.
The transition probability matrix of the Markov chain $Z(l)$ is given by $P$, where
$$P(c,c'):=P(Z(l)=c'|Z(l-1)=c),\ \forall c, c'\in[K+T].$$
We assume that the diagonal entries $P(c,c)>0$, hence the Markov chain is aperiodic. Also the Markov chain is assumed to be irreducible, thus it possesses a stationary distribution denoted by $\bm{\pi}$.
\color{black}
\end{itemize}
\item \textbf{Object labeling process $c(l)$}
\begin{itemize}
\item[(i)]\textit{Recurrent objects:} On the $l$-th arrival, if the Markov chain $Z(l)$ moves to a state $k\in [K]$, the arrival is labeled by the recurrent object $k$, i.e. $c_i(l) = k$.

\item[(ii)]\textit{Rare objects:} On the $l$-th arrival, if the Markov chain $Z(l)$ moves to a state $K+t$, $t\in[T]$, the arrival is labeled by a rare object of type $t$, chosen from $\mc{R}_t$ such that the label assignment has no stationary behavior in the time-scale of $O(1)$ arrivals and it shows a {\em rarity} behavior in large time-scales. Formally, on $l$-th arrival, given $Z(l) = K+t$,
\begin{itemize}[leftmargin=*]
\item[-] if $\sum_{l' =1}^{l}\beta_t(l';R) = O(\sqrt{l})$: select any rare object of type $t$ (arbitrarily), i.e., $c_i(l) \in \mc{R}_t$
\item[-] else: select any rare object of type $t$ that was not requested within $R$ time units, i.e., $c_i(l) \in \mc{R}_t\setminus \mc{A}(l;R)$.
\end{itemize}   
\end{itemize}
\end{itemize}
\end{assumptiongrp}
The above labeling of rare objects respects a more general \textit{$R$-rarity condition} defined below (which is sufficient for our theoretical  results):
\begin{definition}[$R$-rarity condition]
	\label{def:rarity}
	For any type $t\in T$, and a finite $R>0$,
	\begin{align} \label{eq:rarity}
		\lim_{m\to \infty}\tfrac{1}{N_m^{t}} \sum_{l=m}^{m+N_m^{t}} \beta_{t}(l;R) = 0, ~\text{w.p. } 1.
	\end{align}
	for any $N_m^{t} = \omega(\sqrt{m})$.
\end{definition}

For any type $t$, let $\alpha_t$ be the aggregate fraction of total request arrivals for rare objects of type $t$ in the long run. Note that by the Markov renewal construction, $\alpha_t=\pi_{(K+t)}$, where $\bm{\pi}$ is the stationary distribution of the process $Z(l)$. If $\alpha_t>0$, then for the the $R$-rarity condition to hold, it is sufficient to have infinitely many rare objects of the same type $t$ (over an infinite time horizon).

\begin{remark}[Comment on the $R$-rarity condition]
{\color{blue} The ``$R$-rarity condition'' states that asymptotically (i.e, after $m$-th arrival, for large enough $m$) for each type $t$, requests for rare objects of that type can still arrive as bursty arrivals (i.e., request for any particular \emph{rare object} is separated by less than $R$ time units), as long as over large time windows (windows of size $N_m^t = \omega(\sqrt{m})$) the number of such bursty arrivals becomes infrequent (i.e. $o(N_m^t)$ w.p. $1$).} 	
Note that the definition of bursty arrival and the associated ``$R$-rarity condition'' is not specified for a particular constant $R$, but is {\em parameterized by $R$} and we shall specify the specific value later. If $R$-rarity condition holds then $R'$-rarity condition also holds for any $R'\in [0, R)$, which easily follows from the definition.
\end{remark}

\begin{remark}[Relevance of the $R$-rarity condition]
The condition (\ref{def:rarity}) is fairly general, as at any point in time, no matter how large, it allows the existence of rare objects which may exhibit burstiness for small time windows. {\color{blue} Trivially, if the inter-arrival time of each rare object is greater than $R$, then  $R$-rarity condition is satisfied.} More interestingly, the following real-world scenarios satisfy the (rare) object labeling process in Assumption~\ref{a:Arr}.
\begin{itemize}[leftmargin=*]
\item {\em One-hit wonders~\cite{maggs2015algorithmic}.} For each type $t$, a constant fraction $\alpha_t$ of total arrivals consists of rare objects that are requested \emph{only once}. As the indicator $\beta_{t}(l;R)$ is zero for the first and only time that an object is requested,  $\sum_{l' = 1}^{l} \beta_{t}(l';R) = 0$, for all $l\geq 1$ and type $t\in [T]$.
\item \emph{Flash crowds~\cite{jung2002flash}.} Constant size bursts {\color{blue}(i.e. a collection of $O(1)$ number of bursty arrivals)} of requests for rare objects  may occur over time, with $O(\sqrt{\tau})$ number of such bursts up to time $\tau$. This allows for infinitely many such bursts. In this scenario, almost surely,   {\color{blue} for any type $t$, $\sum_{l' = 1}^{l} \beta_{t}(l';R) = O(\sqrt{l})$.  Therefore, it is a special case of our model.}
\end{itemize}
\end{remark}

\begin{remark}[Generalization of rare object labeling]
	In our proofs we only require that the $R$-rarity condition holds, for a certain value of $R$.  Therefore, we can generalize our result to any rare object labeling process that satisfies the $R$-rarity condition (Definition~\ref{def:rarity}), for that specific value of $R$. Further, it is possible to weaken the rarity condition by requiring the condition to hold with high probability instead of w.p. $1$.
\end{remark}

\begin{remark}[Relevance of the content request model]
	Most of the popular inter-arrival time distributions, e.g., Exponential, Phase-type, Weibull, satisfy the inter-arrival model in Assumption~\ref{a:Arr}. Moreover, it is easy to see that any i.i.d. distribution for content popularity, including Zipfian distribution, is a special case of our object labeling process. In fact, the labeling process is much more general in the sense that it can capture the influence of different objects on other objects, which may span across various types.
\end{remark}

\vspace{5 em}
\noindent \textit{3) Special case: Poisson Arrival with Independent Labeling:}

We next consider a specific model for the arrival process which is a well-studied special case of  Assumption~\ref{a:Arr}. We will later show that under this arrival process we can achieve stronger guarantees on the system performance.


\begin{assumptiongrp}
\label{a:ArrPoisson}
Poisson Arrival with Independent Labeling:

\begin{itemize}[leftmargin=*]
\item The inter arrival times are i.i.d. and {\emph {exponentially}} distributed with rate $\lambda > 0$.
\item The labels for the recurring objects are determined independently. At each request arrival, the request is labeled a recurring object $c$ with probability $\pi_c$, and is labeled a rare object of type $t$ with probability $\alpha_t$, {\color{blue} following the same rare object labeling process for rare objects of type $t$, as in Assumption~\ref{a:Arr}.}
\item For each recurrent object $c$, its size is given by $w_c$, which is non-decreasing w.r.t. probability $\pi_c$ and at most $w_{\mathrm{max}}$. For each type $t\in [T]$, all rare objects of type $t$ have size $\bar{w}_t$.
\end{itemize}
\end{assumptiongrp}
\assumptiongroup

%% file: algorithm-v5.tex
\section{Adaptive TTL-based Algorithms}\label{sec:cachingalgorithms}
A  TTL-based caching algorithm works as follows. When a new object is requested, the object is placed in cache and is associated with a time-to-live ({\ttl}) value.  If no new requests are received for that object, the {\ttl} value is decremented in real-time and the object is evicted when the {\ttl} becomes zero.
If a cached object is requested, the {\ttl} is reset to its original value.
In a TTL cache, the {\ttl} helps balance the cache size and hit rate objectives. When the {\ttl} increases, the object stays in cache for a longer period of time, increasing the cache hit rate, at the expense of a larger cache size. The opposite happens when {\ttl} decreases.


 
 We propose two adaptive TTL algorithms. First, we present a dynamic TTL algorithm (d-TTL) that adapts its TTL to achieve a target hit rate $\mathbf{h}^{*}$. While d-TTL does a good job of achieving the target hit rate, it does this at the expense of caching rare and unpopular recurring content for an extended period of time, thus causing an increase in cache size without any significant contribution towards the cache hit rate. We present a second adaptive TTL algorithm called filtering TTL (f-TTL) that filters out rare content to achieve the target hit rate with a smaller cache size. To the best of our knowledge, both d-TTL and f-TTL are the first adaptive TTL-based caching algorithms that are able to achieve a target hit rate $\mathbf{h}^*$ and a feasible target normalized size $\mathbf{s}^*$ for non-stationary traffic. 

\input{blueprint}

%% file: blueprint.tex

	
\subsection{Dynamic TTL (d-TTL) Algorithm}
We propose a dynamic TTL  algorithm, d-TTL, that adapts a TTL parameter on each arrival  to achieve a target hit rate $\mathbf{h}^{*}$. 
\subsubsection{Structure} The d-TTL algorithm consists of a single TTL cache $\mc{C}$. It also maintains a {\em TTL vector} $\bm{\theta}(l) \in \mathbb{R}_+^{T}$, at the time of $l$-th arrival, where $\theta_t(\cdot)$ represents the {\ttl} value for type $t$. Every object $c$ present in the cache $\mc{C}$, has a timer $\psi^0_c$ that encodes its remaining {\ttl} and is decremented in real time. On the $l$-th arrival, if the requested object $c$ of type $t$ is present in cache, $\theta_t(l)$ is decremented, and if the requested object $c$ to type $t$ is not present in cache, object $c$ is fetched from the origin, cached in the server and $\theta_t(l)$ is incremented. In both cases, $\psi^0_c$ is set to the updated timer $\theta_t(l + 1)$ until the object is re-requested or evicted. As previously discussed, object $c$ is evicted from the cache when $\psi^0_{c}=0$.  

{\color{blue} 
	\subsubsection{Key Insights} To better understand the dynamic TTL updates, we consider a simple scenario where we have unit sized objects of a single type and a target hit rate $h^*$. 
	
	{\bf Adaptation based on stochastic approximation.} 
	Consider a TTL parameter $\theta$. Upon a cache miss, $\theta$ is incremented by $\eta h^*$ and upon a cache hit, $\theta$ is decremented by $\eta (1-h^*)$, where $\eta > 0$ is some positive step size. More concisely, $\theta$ is changed by $\eta(h^* - Y(l)))$, where $Y(l) = 1$ upon a cache hit and $Y(l)=0$ upon a cache miss. 
If the expected hit rate under a fixed TTL value  $\theta$ is $h$, then the expected change in the value of $\theta$ is given by $\eta((1-h)h^*-h(1-h^*))$. It is easy to see that this expected change approaches $0$, as $h$ approaches $h^*$. 
In a dynamic setting, $Y(l)$ provides a noisy estimate of $h$.
However, by choosing {\em decaying step size}, i.e. on $l$-th arrival $\eta = \eta(l)= \frac{1}{l^{\alpha}}$, for $\alpha \in (0.5,1]$,  we can still ensure convergence, by using results from stochastic approximation theory~\cite{kushner2003stochastic}. 
	
	{\bf Truncation in presence of rare objects.} In some scenarios, the target hit rate $h^*$ may be unattainable due to the presence of rare objects. Indeed, in the 9-day trace used in our paper, around $4\%$ of the requests are for one-hit wonders. Clearly, in this scenario, a hit rate of over $96\%$ is unachievable.
Whenever, $h^*$ is unattainable $\theta$ diverges with the above adaptation. Therefore, under {\em unknown amount of rare traffic} it becomes necessary to \textbf{truncate} $\theta$ with a large but finite value $L$ to make the algorithm robust.
}

\subsubsection{Adapting $\bm{\theta}(l)$} Following the above discussion, we restrict the {\ttl} value $\bm{\theta}(l)$ to $\bm{\theta}(l)\preceq \bm{L}$\footnote{ This gives an upper bound, typically a large one, over the size of the cache. Further it can control the staleness of objects.}. Here $\mathbf{L}$ is the truncation parameter of the algorithm and an increase in $\mathbf{L}$ increases the achievable hit rate (see Section \ref{sec:theorem} for details). For notational similarity with f-TTL, we introduce a latent variable $\bm{\vartheta}(l)\in \mathbb{R}^T$ where $\bm{\vartheta}(l) \in [0,1]$. Without loss of generality, instead of adapting $\bm{\theta}(l)$, we dynamically adapt each component of $\bm{\vartheta}(l)$ and set $\theta_t(l) = L_t \vartheta_t(l)$, where $\vartheta_t(\cdot)$ is the latent variable for objects of type $t$. The d-TTL algorithm is presented in Algorithm~\ref{alg:d-TTL}, where the value of  $\bm{\theta}(l)$ dynamically changes according to Equation (\ref{eq:ttlAdapt1}). 

\input{dTTL}

\subsection{Filtering TTL (f-TTL) Algorithm}
Although the d-TTL algorithm achieves the target hit rate, it might provide cache sizes which are excessively large. This is due to the observation that d-TTL still caches rare and unpopular content which might contribute to non-negligible portion of the cache size (for example one-hit wonders still enter the cache while not providing any cache hit). We propose a two-level filtering TTL algorithm (f-TTL) that efficiently filters non-stationary content to achieve the target hit rate along with the target normalized size.

\subsubsection{Structure} The two-level f-TTL algorithm maintains two caches:  a higher-level (or deep) cache $\mc{C}$ and a lower-level cache $\mc{C}_s$. {\color{blue} The higher-level cache (deep) cache $\mc{C}$ behaves similar to the single-level cache in d-TTL (Algorithm~\ref{alg:d-TTL}), whereas the lower-level cache $\mc{C}_s$ ensures that cache $\mc{C}$ stores mostly stationary content. Cache $\mc{C}_s$ does so by filtering out rare and unpopular objects, while suitably retaining bursty objects. To facilitate such filtering, it uses additional sub-level caches: {\em shadow cache} and {\em shallow cache}, each with their own dynamically adapted TTL value. The TTL value associated with the {\em shadow cache} is equal to the TTL value of deep cache $\mc{C}$, whereas the TTL associated with the {\em shallow cache} is smaller. 
	
{\bf TTL timers for f-TTL.} {\color{blue} The complete algorithm for f-TTL is given in Algorithm~\ref{alg:f-TTL}. 
	f-TTL maintains a time varying TTL-value {\em $\bm{\theta}^s(l)$ for shallow cache}, along TTL value {\em $\bm{\theta}(l)$ for both deep and shadow caches}. Every object $c$ present in f-TTL has an {\em exclusive} {\ttl} tuple $(\psi^{0}_c, \psi^{1}_c, \psi^{2}_c)$ indicating remaining TTL for that specific object: $\psi^{0}_c$ for deep cache $\mc{C}$, $\psi^1_{c}$ for the shallow cache of $\mc{C}_s$, and  $\psi^2_{c}$ for the shadow cache of $\mc{C}_s$. Object $c$ is evicted from $\mc{C}$ (resp., $\mc{C}_s$) when $\psi^0_c$ (resp., $\psi^1_c$) becomes $0$. Further, the metadata $\tilde{c}$ is evicted from $\mc{C}_s$ when $\psi^2_c$ equals $0$.}

{\color{blue} Suppose on the $l$-th arrival, the request is for object $c(l)$ (of type $t(l)$ and size $w(l)$). Let $c(l) = c$ and $t(l) = t$. The algorithm first updates the two TTL values to $\bm{\theta}^s(l+1)$ and  $\bm{\theta}^s(l+1)$, according to the update rules which will be described shortly. Then, it performs one of the operations below.
	
	{\bf Cache hit:} If a cache hit occurs, i.e., $c$ is either in the deep cache $\mc{C}$ or in the shallow cache of $\mc{C}_s$, then we cache object $c$ in the deep cache $\mc{C}$ with {\ttl} $\theta_t(l+1)$, thus setting the TTL tuple to $(\theta_t(l+1), 0, 0)$. Further, if $c$ was in shallow cache at the time of hit, the object $c$ and its metadata $\tilde{c}$ is removed from shallow cache and shadow cache of $\mc{C}_s$, resp. [lines 12-15 in Algorithm~\ref{alg:f-TTL}].

	{\bf Cache miss:} If both object $c$ and its meta-data $\tilde{c}$ is absent from $\mc{C}$ and $\mc{C}_s$, we have a cache miss. In this event, we cache object $c$ in shallow cache of $\mc{C}_s$ with {\ttl} $\theta^s_t(l+1)$ and its meta data $\tilde{c}$ in shadow cache of $\mc{C}_s$ with TTL $\theta_t(l+1)$; i.e. the TTL tuple is set to $(0, \theta^s_t(l+1), \theta_t(l+1))$ [lines 19-20 in Algorithm~\ref{alg:f-TTL}].

	{\bf Cache virtual hit:} Finally, if $\tilde{c}$ belongs to the shadow cache but object $c$ is absent from the shallow cache, a cache virtual hit occurs. Then we cache  $c$ in the deep cache $\mc{C}$  with {\ttl} tuple $(\theta_t(l+1), 0, 0)$,  and evict $\tilde{c}$ from $\mc{C}_s$ [lines 16-18 in Algorithm~\ref{alg:f-TTL}]. 
}

\input{fTTL}


\subsubsection{Key Insights} We  pause here to provide the essential insights behind the structure and adaptation rules in f-TTL. 

{\bf Normalized size of f-TTL algorithm.} We begin with characterization of the normalized size of the different types under the f-TTL  algorithm.
For the $l$-th request arrival, define $\hat{s}(l)$ to be the time that the requested object will spend in the cache until either it is evicted or the same object is requested again, whichever happens first. We call $\hat{s}(l)$ the normalized size of the $l$-th arrival. Therefore, the contribution of the $l$-th request toward the cache size is $w(l)\hat{s}(l)$, where $\hat{s}(l)=\min\{X_{suc}(l), \theta_{t(l)}^s(l+1)\}$ for cache miss and $\hat{s}(l)=\min\{X_{suc}(l), \theta_{t(l)}(l+1)\}$ for cache hit/virtual hit. Then the normalized size, defined in Def.~\ref{def:normSize}, can be equivalently characterized as 
\begin{equation}\label{eq:fTTLnormSize}
s_t = \limsup_{\tau \to \infty}\frac{\sum_{l: A(l)< \tau} w(l)\hat{s}(l)\mathbbm{1}(c_{typ}(l)= t)}{\sum_{l: A(l)< \tau} w(l)\mathbbm{1}(c_{typ}(l)= t)}, \forall t\in [T].
\end{equation} 

To explain the key insights, we consider a simple scenario: single type, unit sized objects, hit rate target $h^*$ and normalized size target $s^*$.

{\bf Shadow Cache for filtering rare objects.} The shadow cache and shallow cache in $\mc{C}_s$ play complementary roles in efficiently filtering out rare and unpopular objects. By storing the meta-data (with negligible size) with TTL $\theta$ upon a new arrival, the shadow cache simulates the deep cache but with negligible storage size. Specifically, on the second arrival of the same object, the presence of its meta-data implies that it is likely to result in cache hits if stored in $\mc{C}$ with TTL $\theta$. This approach is akin to ideas in Bloom filter~\cite{maggs2015algorithmic} and 2Q~\cite{johnson1994x3}. 

{\bf Shallow Cache for recurring bursty objects.} While using shadow cache filters rare objects (e.g. one-hit wonders) as desired, it has an undesirable impact as the first two arrivals of any object always result in cache miss, thus affecting the hit rate. In the absence of shallow cache, this can lead to higher TTL $\theta$ (for the deep cache), for a given target hit rate, compared to d-TTL. This problem is even more pronounced when one considers correlated requests (e.g. Markovian labeling in our model), where requests for an object typically follow an on-off pattern---a few requests come in a short time-period followed by a long time-period with no request.\footnote{Under our model, a lazy labelling Markov chain with $K$ states where the transitions are $i\rightarrow i$ w.p. 0.5 and $i \rightarrow(i+1)\mod K$ w.p. 0.5., for all $i\in [K]$, is such an example.} Inspired from multi-level caches such as LRU-K~\cite{o1993lru}, we use shallow cache to counter this problem. By caching new arrivals with a smaller TTL $\theta^s$ in shallow cache, f-TTL ensures that, on one hand, rare and unpopular objects are quickly evicted; while on the other, for correlated requests cache miss on the second arrival is avoided.   

\textbf{Two-level Adaptation.} In f-TTL, the TTL $\theta$ is dedicated to attain target hit rate $h^*$ and is adapted in the same way as in d-TTL. The TTL of shallow cache, $\theta^s$, is however adapted to attain a normalized size target $s^*$. 
Therefore, adaption must depend on the normalized size $\hat{s}(l)$. Consider the adaptation strategy: first create an online unbiased estimate for the normalized size, denoted by $s(l)$ for the $l$-th arrival, and then change $\theta^s$ as  $\theta^s\leftarrow \min \{(\theta^s+\eta_s( s^* - s(l)))^+, \theta\}$ for some decaying step size $\eta_s$. Clearly, as the expected normalized size  $s = \mathbb{E}[s(l)]$ approaches $s^*$ and the expected hit rate $h = \mathbb{E}[Y(l)]$ approaches $h^*$, the expected change in TTL pair $(\theta, \theta^s)$ approaches $(0,0)$.\footnote{It is not the only mode of convergence for $\theta^s$. Detailed discussion on the convergence of our algorithm will follow shortly.} 


\textbf{Two time-scale approach for convergence.} Due to the noisy estimates of the expected hit rate and the expected normalized size, $Y(l)$ and $s(l)$ resp., we use decaying step sizes $\eta(l)$ and $\eta_s(l)$. However, if $\eta(l)$ and $\eta_s(l)$ are of the same order, convergence is no longer guaranteed as adaptation noise for $\theta$ and $\theta^s$ are of the same order. For example, if for multiple $(\theta_i, \theta^s_i)$, the same target hit rate and normalized size can be attained, then the TTL pair may oscillate between these points. We avoid this by using $\eta(l)$ and $\eta_s(l)$ of different orders: on $l$-th arrival we update $\theta \leftarrow \min \{(\theta + (h^* - Y(l))/l^{\alpha})^{+}, L\}$ for $\alpha \in (0.5, 1)$ and $\theta^s \leftarrow \min\{(\theta^s + (s^* - s(l))/l)^{+} , \theta\}$.  By varying $\theta^s$ much slower than $\theta$, the adaptation behaves as if $\theta^s$ is fixed and it changes $\theta$ to attain the hit rate $h^*$. On the other hand, $\theta^s$ varies slowly to attain the normalized size while $h^*$ is maintained trough faster dynamics.

\textbf{Mode collapse in f-TTL with truncation.} Recall, in presence of rare objects TTL $\theta$ is truncated by a large but finite $L$.  Consider a scenario where f-TTL attains hit rate target $h^*$ if and only if  both $\tilde{\theta}>0$ and $\tilde{\theta}^s > 0$. Now let $s^*$ be set in such a way that it is too small to attain $h^*$. Under this scenario the TTL value $\theta^s$ constantly decreases and collapses to $0$, and the TTL value $\theta$ constantly increases and collapses to $L$. Mode collapse $(\theta, \theta^s) = (L,0)$ occurs while failing to achieve the achievable hit rate $h^*$. In order to avoid such mode collapse, it is necessary to intervene in the natural adaptation of $\theta^s$ and increase it whenever $\theta$ is close to $L$. But due to this intervention, the value of $\theta^s$ may change even if the expected normalized size estimate equals the target $s^*$, which presents a paradox!

\textbf{Two time-scale actor-critic adaptation.} To solve the mode collapse problem, we rely on the principle of separating {\em critics} (the parameters that evaluate performance of the algorithm and serve as memory of the system), and {\em actors} (the parameters that are functions of the critics and govern the algorithm). This is a key idea introduced in the Actor-critic algorithms~\cite{konda2003onactor}. Specifically, we maintain two critic parameters $\vartheta$ and $\vartheta^s$, whereas the parameters $\theta$ and $\theta^s$ play the role of actors.\footnote{It is possible to work with $\theta$ alone, without introducing $\vartheta$. However, having $\vartheta$ is convenient for defining the threshold function in \eqref{eq:thres}.} The critics are updated as discussed above but constrained in $[0,1]$, i.e on $l$-th arrival $\vartheta \leftarrow \min\{(\vartheta + (h^* - Y(l))/ l^{\alpha})^{+}, 1\}$, for $\alpha \in (0.5,1)$ and $\vartheta^s \leftarrow \min\{(\vartheta^s + (s^* - s(l))/ l)^{+}, 1\}$. The actors are updated as, 
 $\theta = L \vartheta$, and for some small $\epsilon>0$, (i) $\theta^s = L \vartheta^s$ if $\vartheta < 1 - 1.5\epsilon$, (ii) $\theta^s = L \vartheta$ if $\vartheta > 1 - 0.5\epsilon$, and (iii) smooth interpolation in between.  With this dynamics $\vartheta^s$ stops changing if the expected normalized size estimate equals $s^*$, which in turn fixes $\theta^s$ despite the external intervention.  

\subsubsection{Estimating the normalized size}
The update rule for $\theta^s$ depends on the normalized size $\hat{s}(l)$ which is not known upon the arrival of $l$-th request. Therefore, we need to estimate $\hat{s}(l)$. However, as $\hat{s}(l)$ depends on updated TTL values, and future arrivals, its online estimation is non-trivial. The term $s(l)$, defined in lines 5,7, and 9 in Algorithm~\ref{alg:f-TTL}, serves as an online estimate of $\hat{s}(l)$.\footnote{With slight abuse of notation, we use `$s$' in $s(l)$ and $\hat{s}(l)$ to denote `normalized size'; whereas in $\mc{C}_s$, $\theta^s(l)$, $\vartheta^s(l)$, and $\eta_s(l)$  `$s$' denotes `secondary cache'. } First, we construct an approximate upper bound for $\hat{s}(l)$ as $\theta_{t(l)}^s(l)$ for cache miss and $\theta_{t(l)}(l)$ otherwise. Additionally, if it is a deep (resp. shallow) cache  hit with remaining timer value $\psi_{c(l)}^0$ (resp. $\psi_{c(l)}^1$), we update the estimate to $(\theta_{t(l)}(l) - \psi_{c(l)}^0)$ (resp. $(\theta_{t(l)}(l) - \psi_{c(l)}^1)$), to correct for the past overestimation. Due to decaying step sizes, and bounded TTLs and object sizes, replacing $\hat{s}(l)$ by $s(l)$ in Eq.~\eqref{eq:fTTLnormSize} keeps $s_t$ unchanged $\forall t\in [T]$. We postpone the details to Appendix.
}

\subsubsection{Adapting $\bm{\theta}^s(l)$ and $\bm{\theta}(l)$}
The adaptation of the parameters $\bm{\theta}(l)$ and $\bm{\theta}^s(l)$ is done following the above actor-critic mechanism, where $\bm{\vartheta}(l)$ and $\bm{\vartheta}^s(l)$ are the two critic parameters lying in  $[0,1]^T$. 
Similar to d-TTL, the f-TTL algorithm adaptively decreases $\bm{\vartheta}(l)$ during cache hits and increases $\bm{\vartheta}(l)$ during cache misses. Additionally, f-TTL also increases $\bm{\vartheta}(l)$ during cache virtual hits. Finally, for each type $t$ and on each arrival $l$, the TTL $\theta_t(l) = L_t \vartheta_{t}(l)$ [line 10 in Algorithm~\ref{alg:f-TTL}]

The external intervention is implemented through a \emph{threshold function}, $\up(x,y;\epsilon) : [0,1]^2 \to [0, 1]$. Specifically, the parameter $\bm{\theta}^s(l)$ is defined in Equation~\ref{eq:ttldifAdapt1} as
$$\theta_t^s(l)= L_t\vartheta_t(l)\up\left(\vartheta_t(l), \vartheta^s_t(l);\epsilon\right)~\forall t \in [T].$$
Here, the threshold function $\up(x,y;\epsilon)$  takes value $1$ for $x\geq 1-\epsilon/2$ and value $y$ for $x\leq 1-3\epsilon/2$, and {\color{blue} the partial derivative w.r.t. $x$} is bounded by $4/\epsilon$. Additionally, it is twice differentiable and non-decreasing w.r.t. both $x$ and $y$. 

This definition maintains the invariant $\theta^s_t(l) \leq \theta_t(l)$  $ \forall t, l$. Note that, in the extreme case when $\bm{\vartheta}^s(l) = 0$, we only cache the metadata of the requested object on first access, but not the object itself. We call this the \emph{full filtering TTL}.

One such threshold function can be given as follows with the convention $0/0=1$, 
\begin{equation}\label{eq:thres}
\up(x,y;\epsilon)=\left(y+\frac{(1-y)((x-1+\tfrac{3\epsilon}{2})^+)^4}{((x-1+\tfrac{3\epsilon}{2})^+)^4 +  ((1-\tfrac{\epsilon}{2}-x)^+)^4} \right).
\end{equation} 
If the estimate $s(l) > s^*_{t(l)}$, following our intuition, we filter out more aggressively by decreasing $\vartheta_{t(l)}^s(l)$ and consequently  $\theta^s_{t(l)}(l)$. The opposite occurs when $s(l) < s^*_{t(l)}$ [line 11 in Algorithm~\ref{alg:f-TTL}].

%% file: dTTL.tex
\renewcommand{\algorithmicrequire}{\textbf{Input:}}
\renewcommand{\algorithmicensure}{\textbf{Output:}}
\def\NoNumber#1{{\def\alglinenumber##1{}\State #1}\addtocounter{ALG@line}{-1}}

{\color{black}
\begin{algorithm}[h]
\small
\begin{algorithmic}[1]
\Require
\Statex[0] Target hit rate $\mathbf{h}^{*}$, TTL upper bound $\mathbf{L}$.
\Statex[0] For $l$-th request, $l\in\mathbb{N}$, object $c(l)$, size $w(l)$ \& type $t(l)$.
\Ensure Cache requested object using dynamic TTL, $\bm{\theta}$.
\State {\bf Initialize:} Latent variable $\bm{\vartheta}(0)=\mathbf{0}$.
\ForAll {$l \in \mathbb{N}$}
	\If {Cache hit, $c(l)\in \mc{C}$}
		\State $Y(l) = 1$
	\Else { Cache miss}
		\State $Y(l) = 0$	
	\EndIf
	\State {\bf Update TTL $\theta_{t(l)}(l)$}:
	\Statex[1]
	\vspace{-10pt}
	\begin{equation}\label{eq:ttlAdapt1}
	\begin{aligned}
	\hspace{1.5ex}\vartheta_{t(l)}(l+1) &= \mc{P}_{[0,1]}\left(\vartheta_{t(l)}(l)+\eta(l)
	{\widehat{w}(l)}\left(h^*_{t(l)} - Y(l)\right)\right)\\
	\theta_{t(l)}(l+1) &= L_{t(l)}\vartheta_{t(l)}(l+1)
	\end{aligned}
	\end{equation}
	\vspace{-10pt}
	\Statex[1] {where,}
	\Statex[2] {$\eta(l)= \frac{\eta_0}{l^{\alpha}}$ is a decaying step size for $\alpha\in (1/2,1)$,}
	\Statex[2] {$\mc{P}_{[0,1]}(x)= \min\{1,\max\{0,x\}\}$,}
 	\Statex[2] {$\widehat{w}(l) = 1$ for OHR and $w(l)$ for BHR.}
	\NoNumber {}
	\State Cache $c$ with {\ttl} $\psi^0_{c(l)} = \theta_{t(l)}(l+1)$ in $\mc{C}$.
\EndFor
\end{algorithmic}
\caption{Dynamic TTL (d-TTL)}
\label{alg:d-TTL}
\end{algorithm}
}

%% file: fTTL.tex
\renewcommand{\algorithmicrequire}{\textbf{Input:}}
\renewcommand{\algorithmicensure}{\textbf{Output:}}
\def\NoNumber#1{{\def\alglinenumber##1{}\State #1}\addtocounter{ALG@line}{-1}}

\floatname{algorithm}{\color{black}Algorithm}
\makeatletter
\renewcommand\fs@ruled{\def\@fs@cfont{\bfseries}\let\@fs@capt\floatc@ruled
  \def\@fs@pre{{\color{black}\hrule height.8pt depth0pt \kern2pt}}%
  \def\@fs@post{{\color{black}\kern2pt\hrule\relax}}%
  \def\@fs@mid{{\kern2pt\color{black}\hrule\kern2pt}}%
  \let\@fs@iftopcapt\iftrue}
\makeatother

\begin{algorithm}[h]
\small
\begin{algorithmic}[1]
\color{black}
\Require
\Statex[0] Target hit rate $\mathbf{h}^{*}$, target normalized size $\mathbf{s}^{*}$, TTL  bound $\mathbf{L}$.
\Statex[0] For $l$-th request, $l\in\mathbb{N}$, object $c(l)$, size $w(l)$ \& type $t(l)$.
\Ensure Cache requested object using dynamic TTLs, $\bm{\theta}$ and $\bm{\theta}^s$.
\State {\bf Intialize:} Latent variables, $\bm{\vartheta}(0)=\bm{\vartheta}^s(0)=\mathbf{0}$.
\ForAll {$l \in \mathbb{N}$}
	\If {Cache hit, $c(l) \in \mc{C} \cup \mc{C}_s$}
		\State $Y(l) = 1$,	
		\State  $s(l) = \begin{cases} \theta_{t(l)}(l) - \psi^0_{c(l)}, & \text{ if } c\in \mc{C}\\
				\theta_{t(l)}(l) - \psi^1_{c(l)}, & \text{ if } c\in \mc{C}_s.  \end{cases}$
	\ElsIf {Virtual hit, $c(l) \notin \mc{C} \cup \mc{C}_s$ and $\tilde{c}(l) \in \mc{C}_s$}
		\State $Y(l) = 0$, $s(l) = \theta_{t(l)}(l)$.
	\Else { Cache miss}
		\State $Y(l) = 0$, $s(l) = \theta^s_{t(l)}(l)$.
	\EndIf
	\State {\bf Update TTL} $\theta_{t(l)}(l)$:
	\Statex[1]
	\vspace{-10pt}
	\begin{equation*}
	\begin{aligned}
	\hspace{2ex}\vartheta_{t(l)}(l+1) &= \mc{P}_{[0,1]}\left(\vartheta_{t(l)}(l)+\eta(l)
	{\widehat{w}(l)}\left(h^*_{t(l)} - Y(l)\right)\right)\\
	\theta_{t(l)}(l+1) &= L_{t(l)}\vartheta_{t(l)}(l+1),
	\end{aligned}
	\end{equation*}
	\vspace{-10pt}
	\Statex[1] {where,}
	\Statex[2] {$\eta(l)= \frac{\eta_0}{l^{\alpha}}$ is a decaying step size for $\alpha\in (1/2,1)$,}
	\Statex[2] {$\mc{P}_{[0,1]}(x)= \min\{1,\max\{0,x\}\}$,}
	\Statex[2] {$\widehat{w}(l) = 1$ for OHR and $w(l)$ for BHR.}
	\NoNumber {}
	\State {\bf Update TTL } $\theta_{t(l)}^s(l)$:
	\Statex[1] {\vspace{-10pt}
	\begin{equation}\label{eq:ttldifAdapt1}
	\begin{aligned}
	\hspace{3.5ex}\vartheta_{t(l)}^s(l+1)&= \mc{P}_{[0,1]}\left(\vartheta_{t(l)}^s(l)
	+\eta_s(l)w(l)(s^*_{t(l)}-s(l))\right)\\
	\theta_{t(l)}^s(l+1)&= L_{t(l)}\vartheta_{t(l)}(l+1)\up
	\left(\vartheta_{t(l)}(l+1), \vartheta^s_{t(l)}(l+1);\epsilon\right)
	\end{aligned}
	\end{equation}
	\vspace{-10pt}}
	\Statex[1] {where,}
	\Statex[2] {$\eta_s(l)=\frac{\eta_0}{l}$ and $\epsilon$ is a parameter of the algorithm,}
	\Statex[2] {$\up(\cdot,\cdot;\epsilon)$ is a \emph{threshold function}.}
	\NoNumber {}
	\If {Cache hit, $c(l) \in \mc{C} \cup \mc{C}_s$}
		\If {$c(l) \in \mc{C}_s$}
			\State Evict $\tilde{c}(l)$ from $\mc{C}_s$ and move $c(l)$ from $\mc{C}_s$ to $\mc{C}$.
		\EndIf
		\State Set {\ttl} tuple to $(\theta_{t(l)}(l+1),0,0)$.
	\ElsIf {Virtual hit, $c(l) \notin \mc{C} \cup \mc{C}_s$ and $\tilde{c}(l) \in \mc{C}_s$}
		\State Evict $\tilde{c}(l)$ from $\mc{C}_s$,
		\State Cache $c(l)$ in $\mc{C}$ and set {\ttl} tuple to $(\theta_{t(l)}(l+1),0,0)$.
	\Else { Cache Miss}
		\State Cache $c(l)$ and $\tilde{c}(l)$ in $\mc{C}_s$ and
		\Statex[2] set {\ttl} tuple to $(0,\theta^s_{t(l)}(l+1), \theta_{t(l)}(l+1))$.
	\EndIf
\EndFor
\end{algorithmic}
\caption{\color{black}Filtering TTL (f-TTL)}
\label{alg:f-TTL}
\end{algorithm}

%% file: MainResult4-JG.tex
\section{Analysis of Adaptive TTL-Based Algorithms}\label{sec:theorem}
In this section we present our main theoretical results.
%
%
We consider a setting where the {\ttl} parameters live in a compact
space. Indeed, if the {\ttl} values become unbounded, then objects
never leave the cache after entering it. This setting is captured
through the definition of $\mathbf{L}$ feasibility, presented below.

\begin{definition}
\label{def:feasible1}
\s{For an arrival process $\mc{A}$ and d-TTL
algorithm, object (byte) hit rate $\mathbf{h}$ is
`$\mathbf{L}$-feasible' if there exists a
$\bm{\theta} \preccurlyeq \mathbf{L}$ such that d-TTL algorithm with fixed {\ttl} $\bm{\theta}$
achieves $\mathbf{h}$ asymptotically almost surely under $\mc{A}$.}
\end{definition}

\begin{definition}
\label{def:feasible2}
\s{For an arrival process $\mc{A}$ and f-TTL caching algorithm,
object (byte) hit rate, normalized size tuple $(\bm{h},\bm{s})$ is
`$\mathbf{L}$-feasible'  if there
exist $\bm{\theta} \preccurlyeq \mathbf{L}$ and
$\bm{\theta}^s \preccurlyeq \bm{\theta}$, such that f-TTL algorithm
with fixed {\ttl} pair $(\bm{\theta},\bm{\theta}^s)$
achieves $(\bm{h},\bm{s})$ asymptotically  almost surely under
$\mc{A}$.}
\end{definition}


To avoid trivial cases (hit rate being 0 or 1), we
have the following definition.

\begin{definition}
A hit rate $\mathbf{h}$ is `typical' if $h_t\in (0, 1)$ for all types $t\in [T]$.
\end{definition}
\subsection{Main Results}
We now show that both d-TTL and f-TLL  asymptotically almost surely (a.a.s.) achieve any `feasible'
\textit{object (byte) hit rate}, $\mathbf{h}^{*}$ for the arrival process in Assumption \ref{a:Arr}, using stochastic approximation techniques.
Further, we prove a.a.s that f-TTL converges to a specified $(\bm{h}^*,\bm{s}^*)$ tuple for \textit{object (byte) hit rate} and \textit{normalized size}.
\begin{theorem}
\label{thm:SAConv}
\s{Under Assumption~\ref{a:Arr} with $\|\bm{L}\|_{\infty}$-rarity condition (i.e. $R=\|\bm{L}\|_{\infty}$)}:

\textbf{d-TTL:} if the hit rate target
$\bm{h}^*$ is both $\bm{L}$-feasible and `typical', then the d-TTL
algorithm with parameter $\bm{L}$ converges to a TTL value of
$\bm{\theta}^*$ a.a.s. Further, the average hit rate converges to $\bm{h}^*$ a.a.s.

\textbf{f-TTL:} if the target tuple of hit rate and
normalized size, $(\bm{h}^*, \bm{s}^*)$, is $(1-2\epsilon)\bm{L}$-feasible,
with $\epsilon>0$, and $\bm{h}^*$ is `typical', then the f-TTL algorithm with parameter $\bm{L}$
and $\epsilon$ converges to a TTL pair
$(\bm{\theta}^*,\bm{\theta}^{s^*})$
a.a.s. Further the average hit rate converges to $\bm{h^*}$
a.a.s., while the average normalized size converges to some
$\hat{\bm{s}}$ a.a.s.. Additionally, $\hat{\bm{s}}$ for each type $t$,
satisfies one of the following three conditions:
\begin{enumerate}
\item The average normalized size converges to $\hat{s}_t = s_t^*$ a.a.s.
\item The average normalized size converges to $\hat{s}_t > s^*_t$ a.a.s.
  and $\theta^{s^*}_t = 0$ a.a.s.
\item The average normalized size converges to $\hat{s}_t < s^*_t$ a.a.s.
and $\theta^{s^*}_t = \theta_t^*$ a.a.s.
\end{enumerate}
\end{theorem}

	As stated in Theorem~\ref{thm:SAConv}, the f-TTL algorithm converges
	to one of three scenarios. We refer to the second scenario as {\em collapse to full-filtering TTL}, because in this case, the lower-level cache contains only labels of objects instead of caching the objects themselves. We refer to the third scenario as {\em collapse to d-TTL}, because in this case, cached objects have equal TTL values in the deep, shadow and shallow
	caches.

{\color{blue}The f-TTL algorithm ensures that under Assumption~\ref{a:Arr}, with  $\|\bm{L}\|_{\infty}$-rarity condition, the rate at which the rare objects enter the deep cache $\mc{C}$ is a.a.s. zero (details deferred to Appendix), thus limiting the normalized size contribution of the rare objects to those residing in the shallow cache of $\mc{C}_s$.  Theorem~\ref{thm:SAConv} states that f-TTL converges to a filtration level which is within two extremes: \emph{full-filtering} f-TTL where rare objects are completely filtered (scenario 2) and d-TTL where no filtration occurs (scenario 3).}

We note that in f-TTL, scenario $1$ and scenario $3$
have `good' properties. Specifically, in each of these two scenario, the
f-TTL algorithm converges to an average normalized size which is smaller than or
equal to the target normalized size. However, in scenario $2,$ the average
normalized size converges to a normalized size larger than the given target under
general arrivals in Assumption~\ref{a:Arr}. However, under
Assumption~\ref{a:ArrPoisson}, we show that the scenario $2$ cannot
occur, as formalized in Corollary below.
\begin{corollary}
\label{corr:Poisson}
Assume the target tuple of hit rate and normalized size,
$(\bm{h}^*, \bm{s}^*)$, is $(1-2\epsilon)\bm{L}$-feasible with
$\epsilon>0$ and additionally, $\bm{h}^*$ is `typical'. Under
Assumption~\ref{a:ArrPoisson} with $\|\bm{L}\|_{\infty}$-rarity condition, a f-TTL algorithm with parameters
$\bm{L}$, $\epsilon$, achieves asymptotically almost surely a tuple
$(\bm{h}^*,\bm{s})$ with normalized size $\bm{s}\preccurlyeq \bm{s}^*$.
\end{corollary}


\subsection{Proof Sketch of Main Results}
Here we present a proof sketch of Theorem~\ref{thm:SAConv}, and Corollary~\ref{corr:Poisson}.
The complete proof can be found in Appendix\footnote{Due to lack of space we present the appendices as supplementary material to the main article.}. 

The proof of Theorem~\ref{thm:SAConv} consists
of two parts. The first part deals with the `static analysis' of the
caching process, where parameters $\bm{\vartheta}$ and
$\bm{\vartheta}^s$ both take fixed values in $[0,1]$ (i.e., no
adaptation of parameters). In the second part (the `dynamic
analysis'), employing techniques from the theory of stochastic
approximation~\cite{kushner2003stochastic}, we show that the TTL $\bm{\theta}$ for d-TTL and
the TTL pair $(\bm{\theta}, \bm{\theta}^s)$ for f-TTL converge almost surely.
Further, the average hit rate (and average normalized size for f-TTL)
satisfies Theorem~\ref{thm:SAConv}.

The evolution of the caching process is represented as a discrete time
stochastic process uniformized over the arrivals into the system. At
each arrival, the system state is completely described by the
following: {\em (1)} the timers of recurrent objects (i.e. $(\psi_c^0,\psi_c^1,\psi_c^2)$ for $c\in \mc{K}$), {\em (2)} the current value of the pair $(\bm{\vartheta}, \bm{\vartheta}^s),$ and
{\em (3)} the object requested on the last arrival. However, due to
the presence of a constant fraction of non-stationary arrivals in
Assumption~\ref{a:Arr}, we maintain a state with incomplete
information. Specifically, our (incomplete) state representation does
not contain the timer values of the rare objects present in the
system. This introduces a bias (which is treated as noise) between the
actual process, and the evolution of the system under incomplete
state information.

In the static analysis, we prove that the system with fixed
$\bm{\vartheta}$ and $\bm{\vartheta}^s$ exhibits uniform convergence
to a unique stationary distribution. Further, using techniques from
regeneration process and the `rarity condition' in Equation~\eqref{eq:rarity}, we
calculate the asymptotic average hit rates and the asymptotic
average normalized sizes of each type for the original caching process.
We then argue that asymptotic averages for both the hit rate and
normalized size of the incomplete state system is same as the original
system. This is important for the dynamic analysis because this characterizes the time averages of the
  adaptation of $\bm{\vartheta}$ and $\bm{\vartheta}^s$.

In the dynamic analysis, we analyze the system under variable
$\bm{\vartheta}$ and $\bm{\vartheta}^s$, using results of almost sure
convergence of (actor-critic) stochastic approximations with a two
timescale separation~\cite{bhatnagar2009natural}. The proof follows
the ODE method; the following are the key steps in the proof of
dynamic analysis:
\begin{enumerate}[leftmargin=*]
\item We show that the effects of the bias introduced by the
  non-stationary process
  satisfies Kushner-Clark condition~\cite{kushner2003stochastic}.

\item The expectation (w.r.t. the history up to step $l$) of the $l$-th
  update as a function of $(\bm{\vartheta}, \bm{\vartheta}^s)$ is
  Lipschitz continuous.

\item The incomplete information system is uniformly ergodic.

\item The ODE (for a fixed $\bm{\vartheta}^s$) representing the mean
  evolution of $\bm{\vartheta}$ has a unique limit point. Therefore, the limit point of this ODE is a unique function of $\bm{\vartheta}^s$.

\item (f-TTL analysis with two timescales) Let the ODE at the slower
  time scale, representing the mean evolution of $\bm{\vartheta}^s$,
  have stationary points $\{(\bm{\vartheta}, \bm{\vartheta}^s)_i\}$.  We characterize each stationary point, and show that it corresponds to one of the three cases stated in Theorem ~\ref{thm:SAConv}. {\color{blue} Finally, we prove all the limit points of the evolution are given by the stationary points of the ODE.}
\end{enumerate}


As stated in Theorem~\ref{thm:SAConv}, the f-TTL algorithm converges
to one of three scenarios, under general arrivals in Assumption~\ref{a:Arr}.
However, under Assumption~\ref{a:ArrPoisson}, we show that the scenario $2$ cannot
occur, as formalized in Corollary~\ref{corr:Poisson}. The proof of Corollary~\ref{corr:Poisson} follows from Theorem~\ref{thm:SAConv} and the following Lemma~\ref{lemm:filtering}.
\begin{lemma}
\s{Under Assumption~\ref{a:ArrPoisson} with $\|\mathbf{L}\|_{\infty}$-rarity condition and for any type $t$, suppose f-TTL algorithm achieves an average hit rate $h_t$ with two
  different TTL pairs, 1) $(\theta_t, \theta_t^s)$ with $\theta_t^s=0$
  (full filtering), and 2) $(\hat{\theta}_t, \hat{\theta}_t^s)$, with
  $\hat{\theta}_t^s>0$, where $\max\{\theta_t,\hat{\theta}_t\} \leq L_t$.}
  Then the normalized size achieved with the
  first pair is less or equal to the normalized size achieved with
  the second pair. Moreover, in the presence of rare objects of type
  $t$, i.e. $\alpha_t>0$, this inequality in achieved normalized size is
  strict.
\label{lemm:filtering}
\end{lemma}
The proof of this lemma is presented in Appendix~A 
and the technique, in its current form, is specific to
Assumption~\ref{a:ArrPoisson}.

%% file: implementation.tex
\section{Implementation of d- and f-TTL}\label{sec:T2P}
One of the main practical challenges in implementing d-TTL and f-TTL is adapting $\theta$ and $\theta^s$ to achieve the desired hit rate in the presence of unpredictable non-stationary traffic. We observe the following major differences between the theoretical and practical settings. First, the arrival process in practice changes over time (e.g. day-night variations) whereas our model assumes the stationary part is fixed. Second, the hit rate performance in finite time horizons is often of practical interest. While our content request model accounts for non-stationary behavior in finite time windows, the algorithms are shown to converge to the target hit rate {\em asymptotically}. But, this may not be true in finite time windows. 
We now discuss some modifications we make to translate theory to practice and evaluate these modification in Section \ref{sec:empirical}.


\textbf{Fixing the maximum TTL.} The truncation parameter (maximum TTL value) $\bm{L}$ defined in Section \ref{sec:cachingalgorithms} is crucial in the analysis of the system. However, in practice, we can choose an arbitrarily large value such that we let $\theta$ explore a larger space to achieve the desired hit rate in both d-TTL and f-TTL.

\textbf{Constant step sizes for $\theta$ and $\theta^s$ updates.} Algorithms~\ref{alg:d-TTL} and \ref{alg:f-TTL} use decaying step sizes $\eta(l)$ and $\eta_s(l)$ while adapting $\theta$ and $\theta^s$. This is not ideal in practical settings where the traffic composition is constantly changing, and we need $\theta$ and $\theta^s$ to capture those variations. Therefore, we choose carefully hand-tuned constant step sizes that capture the variability in traffic well. \s{We discuss the sensitivity of the d-TTL and f-TTL algorithms to changes in the step size in Section~\ref{sec:sensitivity}}. 

\textbf{Tuning normalized size targets.} In practice, f-TTL may not be able to achieve small normalized size targets in the presence of time varying and non-negligible non stationary traffic. In such cases, f-TTL uses the target normalized size to induce filtering.  For instance, when there is a sudden surge of non-stationary content, $\theta^s$ can be aggressively reduced by setting a small target normalized size. This in turn filters out a lot of non-stationary objects while an appropriate increase in $\theta$ maintains the target hit rate. Hence, the target normalized size can be used as a tunable knob in CDNs to adaptively filter out unpredictable non-stationary content. In our experiments in Section \ref{sec:empirical}, \s{we use a target normalized size that is 50\% of the normalized size of d-TTL. This forces f-TTL to achieve the same hit rate as d-TTL but at half the cache space, if feasible.} In practice, the normalized size targets are chosen based on performance requirements of different content types.  It should be noted that a target normalized size of 0 while most aggressive, is not necessarily the best target. This is because, a target normalized size of 0, sets $\theta^s$ to 0 and essentially increases the $\theta$ to attain the hit rate target. This may lead to an increase in the average cache size when compared to an f-TTL implementation with a non-zero target normalized size. Specifically, in our simulations at a target OHR of 40\%, setting a non-zero target normalized size leads to nearly 15\% decrease in the average cache size as compared to a target normalized size of 0.

%% file: empirical.tex
\section{Empirical Evaluation}\label{sec:empirical}

\begin{figure*}[!ht]
	\minipage{0.32\linewidth}
	\includegraphics[width=0.9\linewidth]{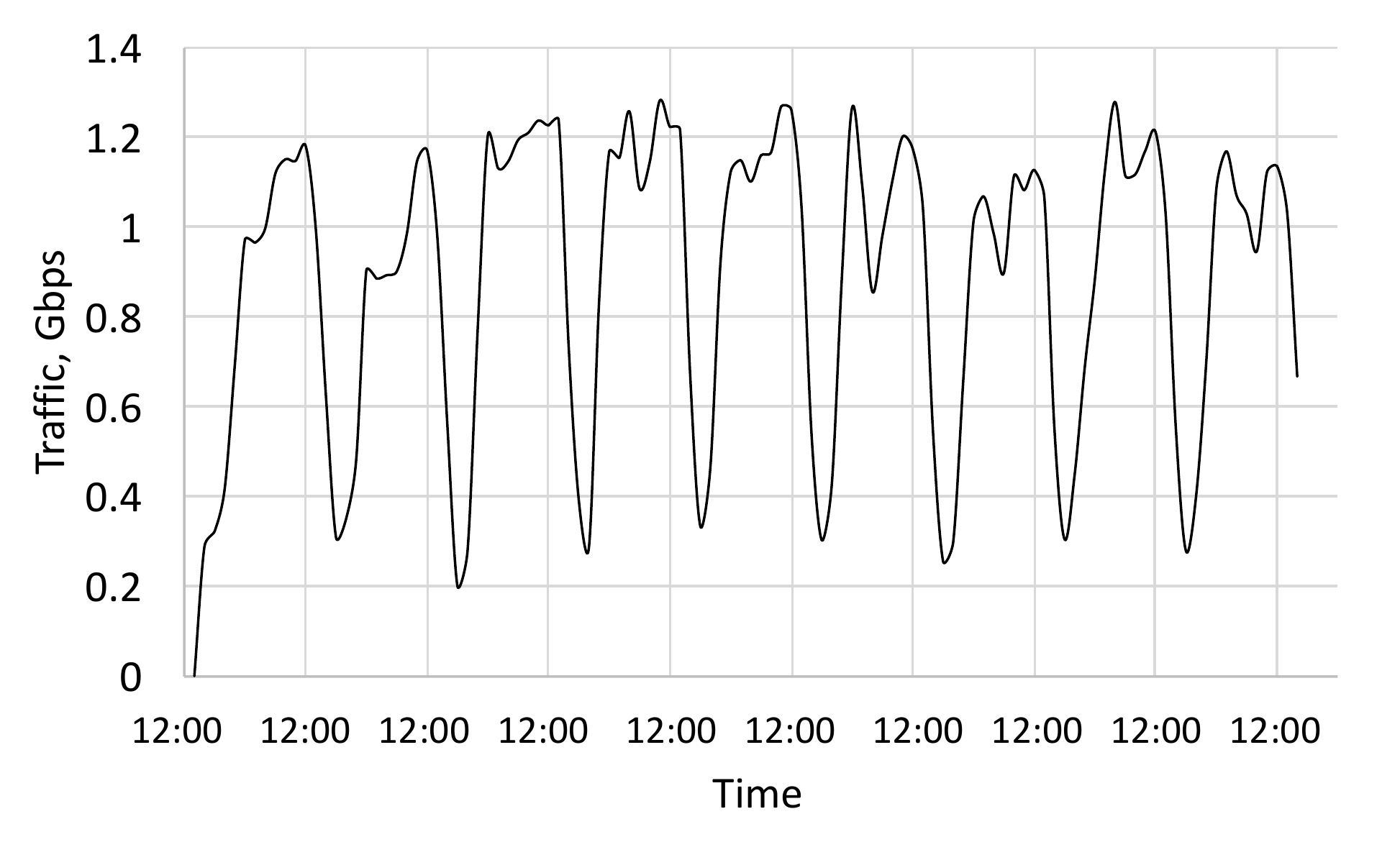}
	\small
	\caption{Content traffic served to users from the CDN server,  averaged every 2 hours. The traces were collected from $29^{th}$ January  to $6^{th}$ February 2015.}
	\label{fig:server-traffic-gbps}
	\endminipage\hfill
	\minipage{0.32\linewidth}
	\includegraphics[width=0.9\linewidth]{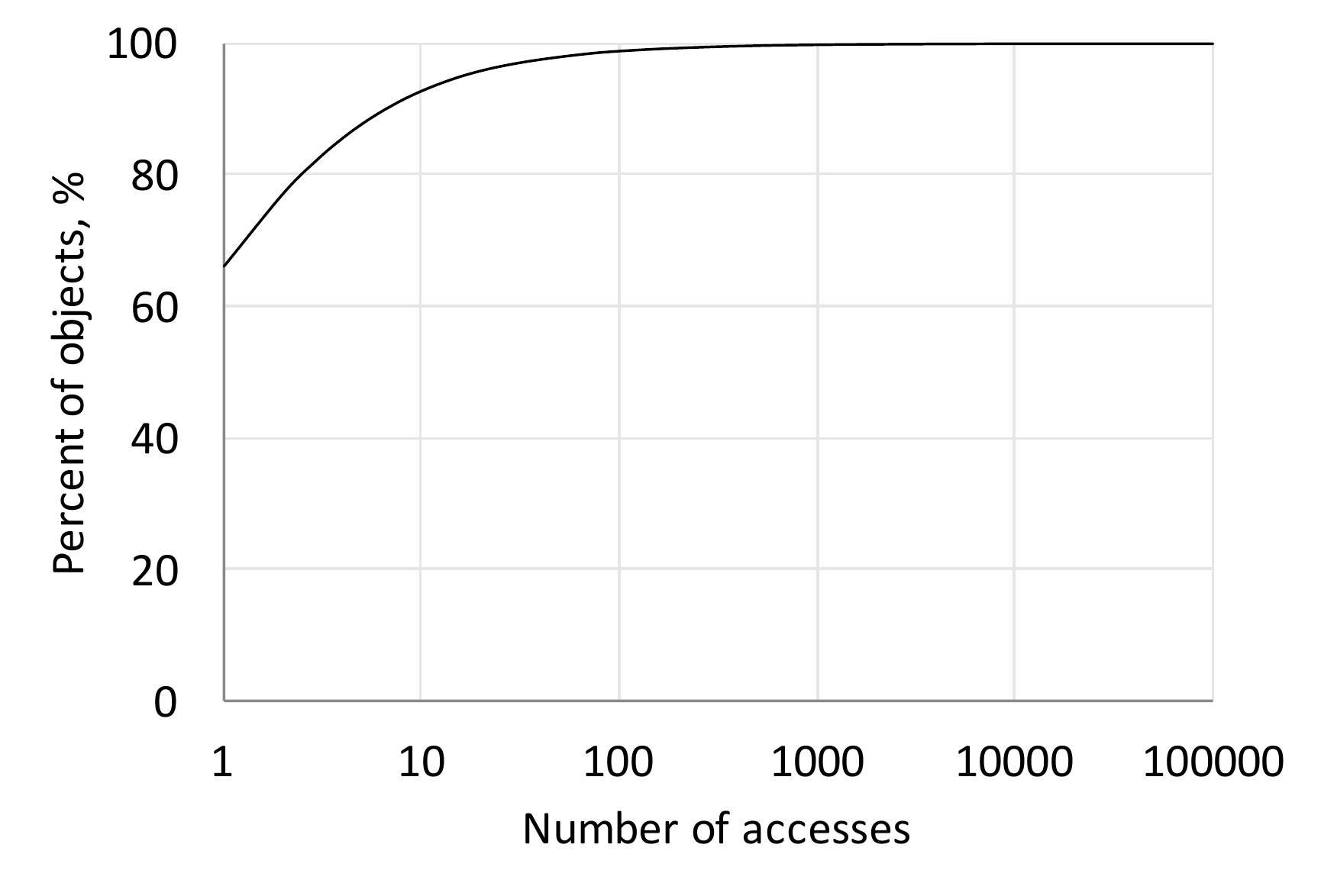}
	\small
	\caption{Popularity of content accessed by users in the 9-day period.}
	\label{fig:object_pop_dist}
	\endminipage\hfill
	\minipage{0.32\linewidth}
	\includegraphics[width=0.9\linewidth]{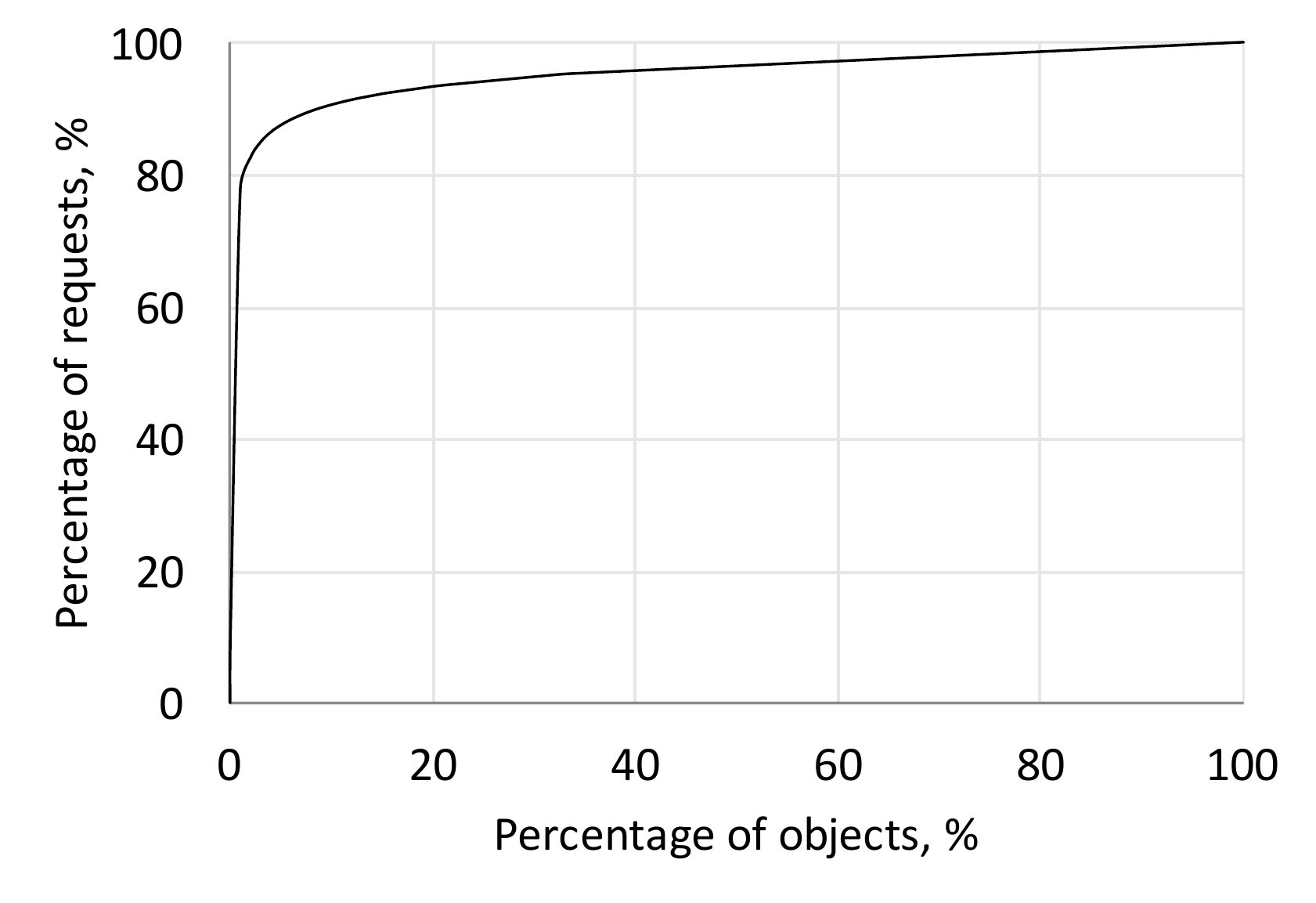}
	\small
	\caption{A large fraction of the requests are for a small fraction of the objects.}
	\label{fig:obj_req_frac}
	\endminipage
\end{figure*}

We evaluate the performance of d-TTL and f-TTL, both in terms of the hit rate achieved and the cache size requirements, using actual production traces from a major CDN.

\subsection{Experimental setup}\label{sec:exp-setup}
{\bf Content Request Traces.}
We use an extensive data set containing access logs for content requested by users that we collected from a typical production server in Akamai's commercially-deployed CDN \cite{nygren2010akamai}. The logs contain requests for predominantly web content (hence, we only compute TTLs for a single content type). Each log line corresponds to a single request and contains a timestamp,  the requested URL (anonymized), object size, and bytes served for the request.  The access logs were collected over a period of 9 days. The traffic served in Gbps captured in our data set is shown in Figure \ref{fig:server-traffic-gbps}. We see that there is a diurnal traffic pattern with the first peak generally around 12PM (probably due to increased traffic during the afternoon) and the second peak occurring around 10-11PM. There is a small dip in traffic during the day between 4-6PM. This could be during evening commute when there is less internet traffic. The lowest traffic is observed at the early hours of the morning between 4AM and 9AM. 
 
\s{The content requests traces used in this work contain 504 million requests (resp., 165TB) for 25 million distinct objects (resp., 15TB)}. 
{\color{blue}From Figure~\ref{fig:object_pop_dist}, we see that about 70\% of the objects in the trace are one-hit wonders. This indicates that a large fraction of objects need to be cached with no contribution towards the cache hit rate. Moreover, from Figure~\ref{fig:obj_req_frac}, we see that about 90\% of the requests are for only 10\% of the most popular objects indicating the the remaining 90\% of the objects contribute very little to the cache hit rate. Hence, both these figures indicate that the request trace has a large fraction of unpopular content. The presence of a significant amount of non-stationary traffic in the form of ``one-hit-wonders'' in production traffic is consistent with similar observations made in earlier work \cite{MaggsSitaraman2015}.}


{\bf Trace-based Cache Simulator.} We built a custom event-driven simulator to simulate the different TTL caching algorithms. The simulator takes as input the content traces and computes a number of statistics such as the hit rate obtained over time, the variation in $\theta$, $\theta^s$ and the cache size over time. We implement and simulate both d-TTL and f-TTL  using the parameters listed in Table~\ref{tab:sim-param}. {\color{blue} We consider a single type for our the empirical study.} 

We use constant step sizes, $\eta$=1e-2 and $\eta_s$=1e-9, while adapting the values of $\theta$ and $\theta^s$. The values chosen were found to capture the variability in our input trace well. \s{We evaluate the sensitivity of d-TTL and f-TTL to changes in $\eta$ and $\eta_s$ in Section~\ref{sec:sensitivity}}.

\begin{table}[h]
\small
\caption{Simulation parameters.  In this table $s^*$ is the target normalized size and $w_{\mathrm{avg}}$ is the average object size.}
\label{tab:sim-param}
\centering
\begin{tabular}{|p{1.0in}|p{0.35in}|p{1.0in}|p{0.4in}|}
\hline
Simulation length &  9 days & Number of requests & 504 m\\
\hline 
Min TTL value & 0 sec & Max TTL value & $10^7$ sec \\
\hline
Step size for $\theta$  & $\eta$ & Step size for $\theta^s$  & $\frac{\eta_s}{s^* w_{\mathrm{avg}}}$  \\
\hline
\end{tabular}
\end{table}

\subsection{How TTLs adapt over time}
\begin{figure} [!h]
\centering
\includegraphics[width= 0.95\linewidth, ]{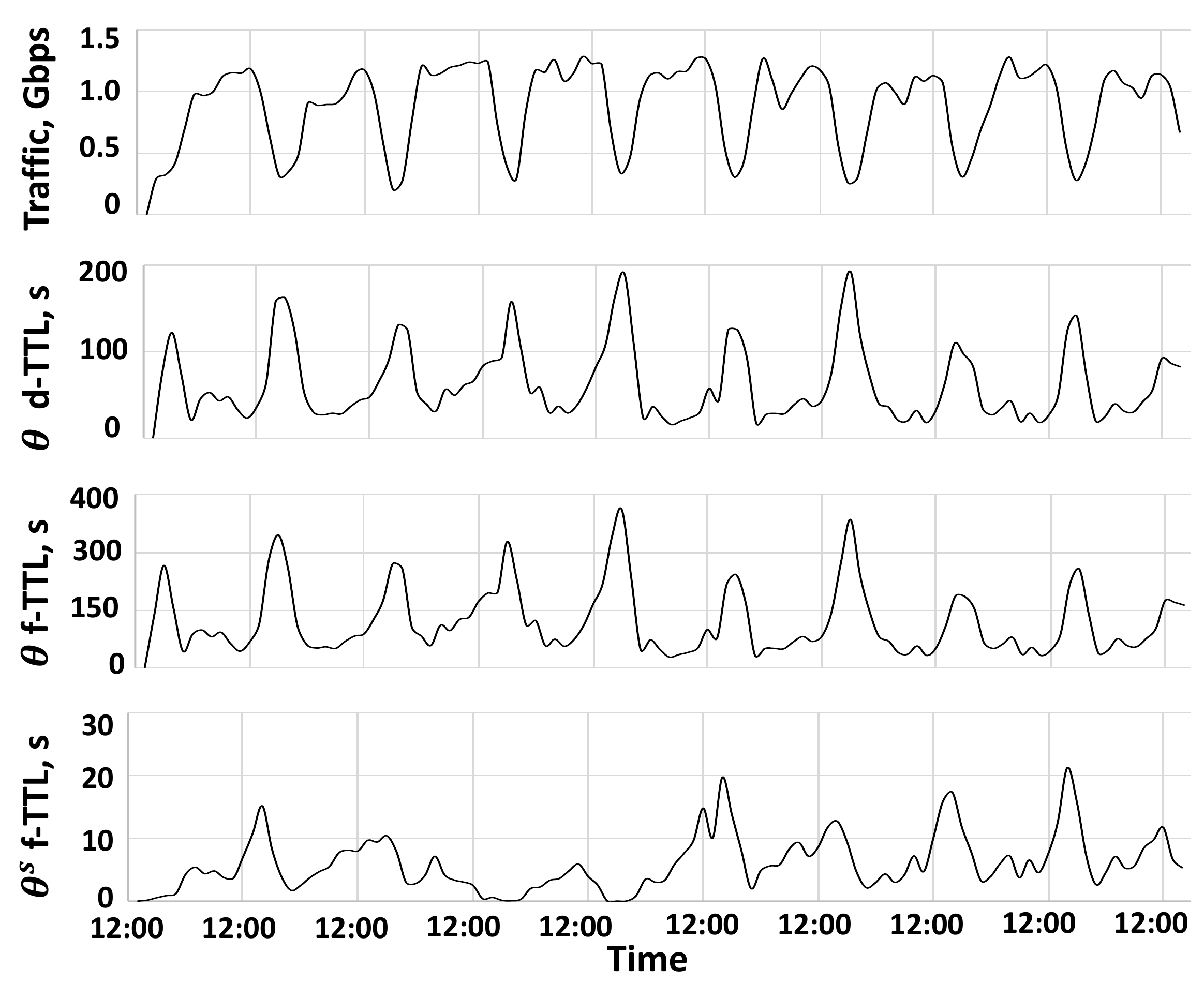}
\small
\caption{Variation in $\theta$ for d-TTL,  $\theta$ for f-TTL and $\theta^s$ over time with target object hit rate=60\%.}
\label{fig:t-ts-df-ohr-vos}
\end{figure}

To understand how d-TLL and f-TTL adapt their TTLs over time in response to request dynamics, we simulated these algorithms with a target object hit rate of 60\% and \s{a target normalized size that is 50\% of the normalized size achieved by d-TTL}.
In Figure \ref{fig:t-ts-df-ohr-vos} we plot the traffic in Gbps, the variation in $\theta$ for d-TTL, $\theta$ for f-TTL and $\theta^s$ over time, all averaged over 2 hour windows. We consider only the object hit rate scenario to explain the dynamics. We observe similar performance when we consider byte hit rates. 

\begin{figure*}[!ht]
	\minipage{0.29\linewidth}
	\centering
	\includegraphics[width=1\linewidth]{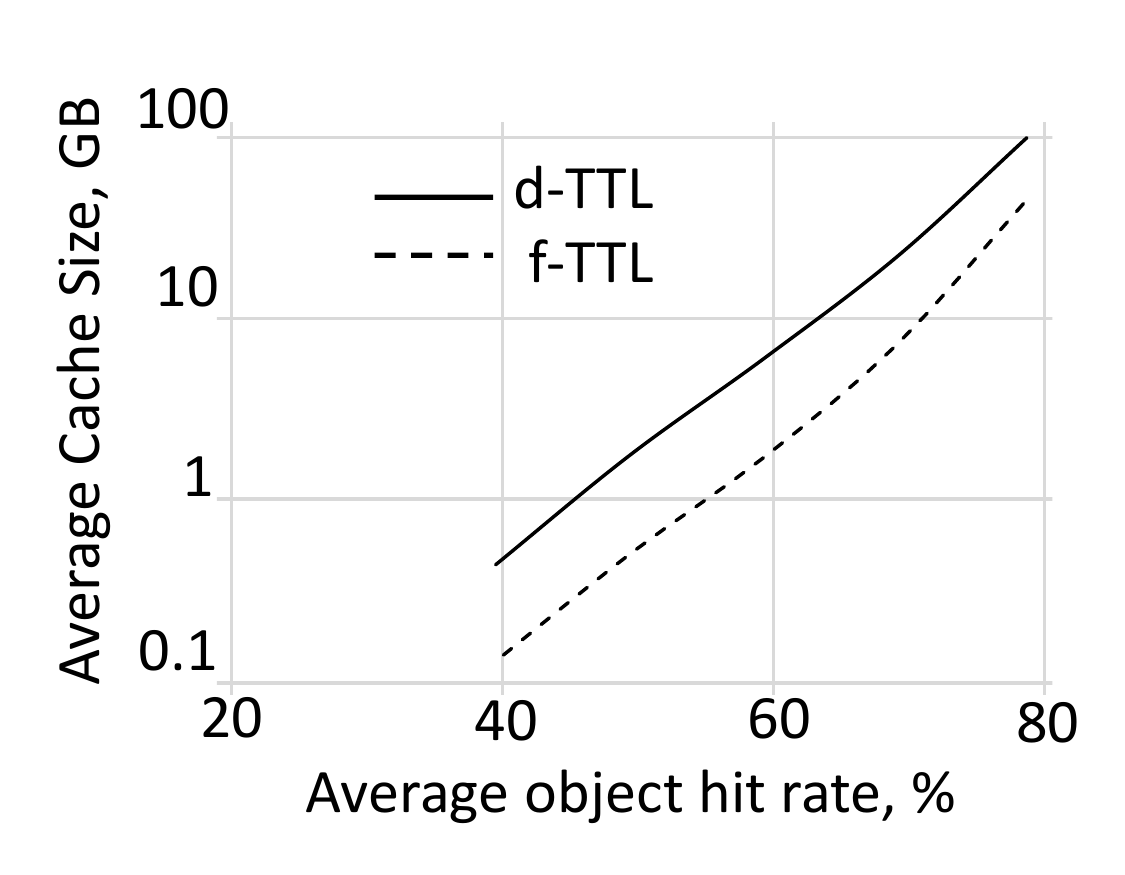}
	\small
	\caption{Hit rate curve for object hit rates.}
	\label{fig:hrc-ohr-vos}
	\endminipage\hfill
	\minipage{0.34\linewidth}
	\includegraphics[width=1\linewidth]{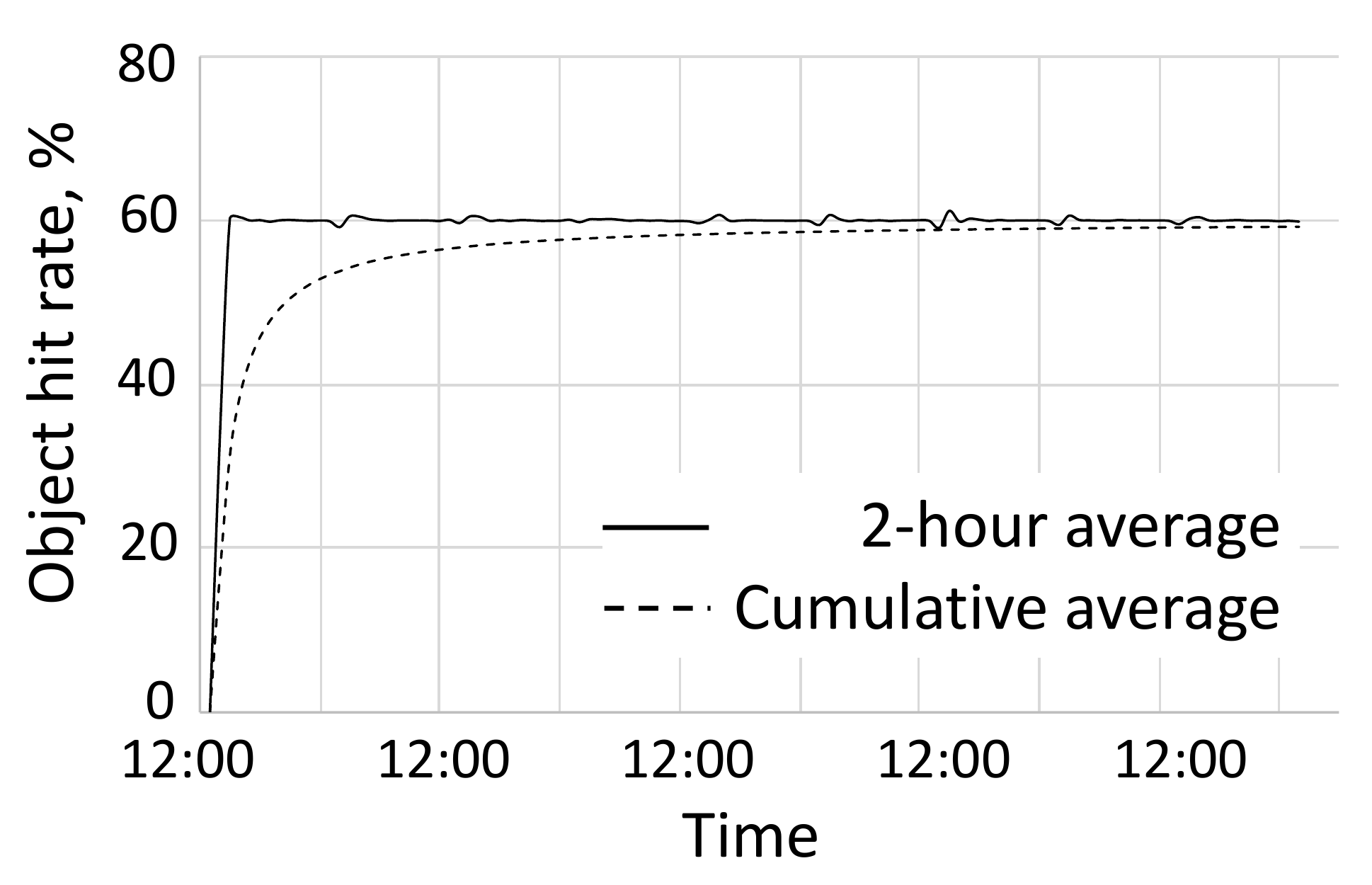}
	\small
	\caption{Object hit rate convergence over time for d-TTL; target object hit rate=60\%.}
	\label{fig:hc-ohr-vos-d-ttl}
	\endminipage\hfill
	\minipage{0.34\linewidth}
	\includegraphics[width=1\linewidth]{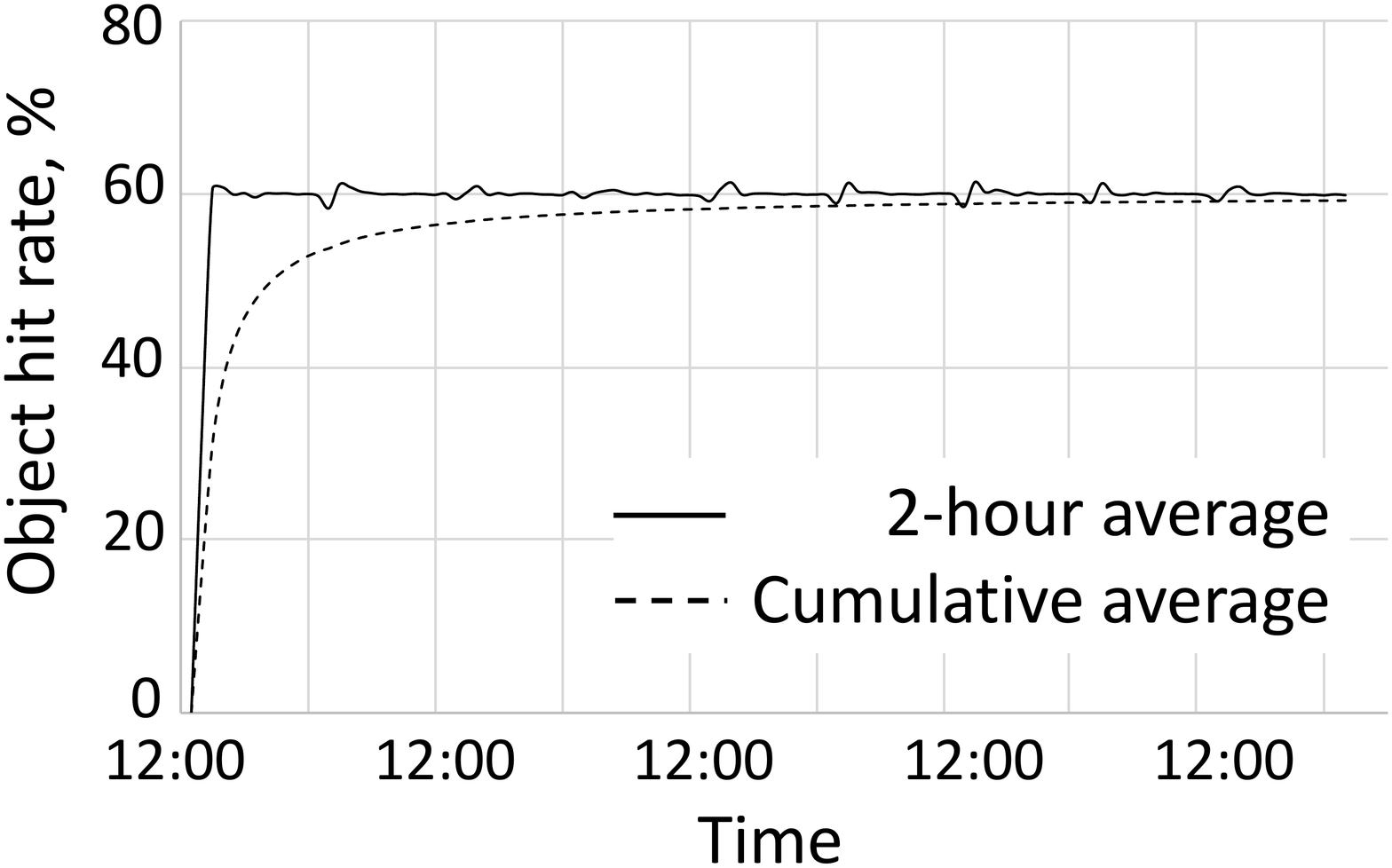}
	\small
	\caption{Object hit rate convergence over time for f-TTL; target object hit rate=60\%.}
	\label{fig:hc-ohr-vos-f-ttl}
	\endminipage
\end{figure*}

From Figure \ref{fig:t-ts-df-ohr-vos}, we see that the value of $\theta$ for d-TTL is smaller than that of f-TTL. This happens due to the fact that f-TTL filters out rare objects to meet the target normalized size, which can in turn reduce the hit rate, resulting in an increase in $\theta$ to achieve the target hit rate.
We also observe that $\theta$ for both d- and f-TTL is generally smaller during peak hours when compared to off-peak hours. This is  because the inter-arrival time of popular content is smaller during peak hours. Hence, a smaller $\theta$ is sufficient to achieve the desired hit rate. However, during off-peak hours, traffic drops by almost 70\%. With fewer content arrivals per second, $\theta$ increases to provide the same hit rate. In the case of f-TTL, the increase in $\theta$, increases the normalized size of the system, which in turn leads to a decrease in $\theta^s$. This matches with the theoretical intuition that d-TTL adapts $\theta$ only to achieve the target hit rate while f-TTL adapts both $\theta$ and $\theta^s$ to reduce the cache size while also achieving the target hit rate.

\subsection{Hit rate performance of d-TTL and f-TTL}
The performance of a caching algorithm is often measured by its hit rate curve (HRC) that relates its cache size with the (object or byte) hit rate that it achieves. HRCs are useful for CDNs as they help provision the right amount of cache space to obtain a certain hit rate. We compare the HRCs of d-TTL and f-TTL for object hit rates and show that f-TTL significantly outperforms d-TTL by filtering out the rarely-accessed non-stationary objects. The HRCs for byte hit rates are shown in  Appendix~B.

To obtain the HRC for d-TTL, we fix the target hit rate at 80\%, 70\%, 60\%, 50\% and 40\% and measure the hit rate and cache size achieved by the algorithm. Similarly, for f-TTL, we fix the target hit rates at 80\%, 70\%, 60\%, 50\% and 40\%. Further, we set {\emph the target normalized size of f-TTL to 50\% of the normalized size of d-TTL}. The HRCs for object hit rates are shown in Figures \ref{fig:hrc-ohr-vos}. The hit rate performance for byte hit rates is discussed in Appendix~B.
Note that the y-axis is presented in log scale for clarity.

From Figure ~\ref{fig:hrc-ohr-vos} we see that f-TTL always performs better than d-TTL i.e. for a given  hit rate, f-TTL requires lesser cache size on average than d-TTL. In particular, on average, f-TTL {\color{blue} with a target normalized size  equal to 50\% of d-TTL} requires a cache that is 49\% smaller than d-TTL to achieve the same object hit rate. {\color{blue}  In Appendix~C, we discuss the performance of f-TTL for other normalized size targets.}

\input{table2}
\input{table4}
\subsection{Convergence of d-TTL and  f-TTL}
For the dynamic TTL algorithms to be useful in practice, they need to converge to the target hit rate with low error. In this section we measure the object hit rate convergence over time, averaged over the entire time window and averaged over 2 hour windows for both d-TTL and f-TTL. We set the target object hit rate to 60\% and a target normalized size that is 50\% of the normalized size of d-TTL. The byte hit rate convergence is discussed in Appendix~B.



From Figures \ref{fig:hc-ohr-vos-d-ttl} and \ref{fig:hc-ohr-vos-f-ttl}, we see that the 2 hour averaged object hit rates achieved by both d-TTL and f-TTL have a cumulative error of less than 1.3\% while achieving the target object hit rate on average. We see that both d-TTL and f-TTL tend to converge to the target hit rate, which illustrates that both d-TTL and f-TTL are able to adapt well to the dynamics of the input traffic.

In general, we also see that d-TTL has lower variability for object hit rate compared to f-TTL due to the fact that d-TTL does not have any bound on the normalized size while achieving the target hit rate, while f-TTL is constantly filtering out non-stationary objects to meet the target normalized size while also achieving the target hit rate.

\subsection{Accuracy of d-TTL and f-TTL} \label{sec:accuracy}

\s{A key goal of the dynamic TTL algorithms (d-TTL and f-TTL) is to achieve a target hit rate, even in the presence of bursty and non-stationary requests. We evaluate the performance of both these algorithms by fixing the target hit rate and comparing the hit rates achieved by d-TTL and f-TTL with caching algorithms such as Fixed TTL (TTL-based caching algorithm that uses a constant TTL value) and LRU (constant cache size), provisioned using Che's approximation \cite{che2002hierarchical}. We only present the results for object hit rates (OHR) in Table~\ref{tab:ttl-che}. Similar behavior is observed for byte hit rates.}

\s{For this evaluation, we fix the target hit rates (column 1) and {\color{blue} analytically compute} the TTL (characteristic time) and cache size using Che's approximation (columns 2 and 6) on the request traces {\color{blue} assuming Poisson traffic}. We then {\color{blue} measure} the hit rate and cache size of Fixed TTL (columns 3 and 4) using the TTL computed in column 2, and the hit rate of LRU (column 5) using the cache size computed in column 6. Finally, we compute the hit rate and cache size achieved by d-TTL and f-TTL (columns 7-10) to achieve the target hit rates in column 1 and a target normalized size that is 50\% of that of d-TTL.}

\s{We make the following conclusions from Table ~\ref{tab:ttl-che}.}\\
1) \s{The d-TTL and f-TTL algorithms meet the target hit rates with a small error of 1.2\% on average. This is in contrast to the Fixed TTL algorithm which has a high error of 14.4\% on average and LRU which has even higher error of 20.2\% on average. This shows that existing algorithms such as Fixed TTL and LRU are unable to meet the target hit rates while using heuristics such as Che's approximation, which cannot account for non-stationary content.}\\
2) \s{The cache size required by d-TTL and f-TTL is 23.5\% and 12\% respectively, of the cache size estimated by Che's approximation and 35.8\% and 18.3\% respectively, of the cache size achieved by the Fixed TTL algorithm, on average. This indicates that both LRU and the Fixed TTL algorithm, provisioned using Che's approximation, grossly overestimate the cache size requirements.}

{\color{blue}We note that using complex heuristics, such as the shot noise model~\cite{leonardi2015least} or the advanced popularity estimation~\cite{olmos2016inverse}, may improve  accuracy over Che's approximation with Poisson traffic.}

\subsection{Robustness and sensitivity of d-TTL and f-TTL } \label{sec:sensitivity}
\s{We use constant step sizes while adapting the values of $\theta$ and $\theta_s$ in practical settings for reasons discussed in Section ~\ref{sec:T2P}. In this section, we evaluate the robustness and sensitivity of d-TTL and f-TTL to the chosen step sizes. The robustness captures the change in performance due to large changes in step size, whereas the sensitivity captures the change due to small perturbations around a specific step size. For ease of explanation, we only focus on two target object hit rates, 60\% and 80\% corresponding to medium and high hit rates. The observations are similar for other target hit rates and for byte hit rates.}

Table~\ref{tab:d-ttl-rob} illustrates the robustness of d-TTL to exponential changes in the step size $\eta$. For each target hit rate, we measure the average hit rate achieved by d-TTL, the average cache size and the 5\% outage fraction, for each value of step size. The 5\% outage fraction is defined as the fraction of time the hit rate achieved by d-TTL differs from the target hit rate by more than 5\%. 

From this table, we see that a step size of 0.01 offers the best trade-off among the three parameters, namely average hit rate, average cache size and 5\% outage fraction. Table~\ref{tab:d-ttl-sen} illustrates the sensitivity of d-TTL to small changes in the step size. We evaluate d-TTL at step sizes $\eta=0.01\times(1\pm0.05)$. We see that d-TTL is  insensitive to small changes in  step size.


\s{To evaluate the robustness and sensitivity of f-TTL, we fix the step size $\eta=0.01$ to update $\theta$ and evaluate the performance of f-TTL at different step sizes, $\eta_s$, to update $\theta_s$. The results for robustness and sensitivity are shown in Tables~\ref{tab:f-ttl-rob} and~\ref{tab:f-ttl-sen} respectively. For f-TTL, we see that a step size of $\eta_s$=1e-9 offers the best tradeoff among the different parameters namely average hit rate, average cache size and 5\% outage fraction. Like, d-TTL, f-TTL is insensitive to the changes in step size parameter $\eta_s$.}

{\color{blue}In Table~\ref{tab:d-ttl-rob} and Table~\ref{tab:f-ttl-rob},  a large step size makes the d-TTL and f-TTL algorithms more adaptive to the changes in traffic statistics. This results in reduced error in the average OHR and reduced 5\% outage fraction. However, during periods of high burstiness, a large step size can lead to a rapid increase in the cache size required to maintain the target hit rate. The opposite happens for small step sizes. }

%% file: table2.tex
\begin{table*}[!ht]
\small
\caption{\s{Comparison of target hit rate and average cache size achieved by d-TTL and f-TTL with Fixed-TTL and LRU.}
\label{tab:ttl-che}}
\centering
\begin{tabular}{|p{0.55in}|p{0.55in}|p{0.55in}|p{0.55in}|p{0.55in}|p{0.55in}|p{0.55in}|p{0.55in}|p{0.55in}|p{0.55in}|}
\hline
Target  & \multicolumn{3}{|c|}{Fixed TTL (Che's approx.)} & \multicolumn{2}{|c|}{LRU (Che's approx.)} & \multicolumn{2}{|c|}{d-TTL} & 
\multicolumn{2}{|c|}{f-TTL}\\
\cline{2-10}
OHR (\%) & TTL\,(s)  & OHR (\%) & Size (GB) & OHR (\%) & Size\,(GB)  & OHR (\%) & Size (GB) &
OHR (\%) & Size (GB)\\
\hline 
\hline
80 & 2784 & 83.29 & 217.11 & 84.65 & 316.81 & 78.72 & 97.67 & 78.55 & 55.08 \\ 
\hline
70 & 554 & 75.81 & 51.88 & 78.37 & 77.78 & 69.21 & 21.89 & 69.14 & 11.07 \\
\hline
60 & 161 & 68.23 & 16.79 & 71.64 & 25.79 & 59.36 & 6.00 & 59.36 & 2.96 \\
\hline
50 & 51 & 60.23 & 5.82 & 64.18 & 9.2 & 49.46 & 1.76 & 49.47 & 0.86 \\ 
\hline
40 & 12 & 50.28 & 1.68 & 54.29 & 2.68 & 39.56 & 0.44 & 39.66 & 0.20 \\ 
\hline
\end{tabular}
\end{table*}

%% file: table4.tex
\begin{table*}[!ht]
\small
\caption{\s{Impact of exponential changes in constant step size $\eta$ on the performance of d-TTL (robustness analysis).}}
\label{tab:d-ttl-rob}
\centering
\begin{tabular}{|p{0.55in}|p{0.55in}|p{0.55in}|p{0.55in}|p{0.55in}|p{0.55in}|p{0.55in}|p{0.55in}|p{0.55in}|p{0.55in}|}
\hline
Target  & \multicolumn{3}{|c|}{Average OHR (\%)} & \multicolumn{3}{|c|}{Average cache size (GB)} & \multicolumn{3}{|c|}{5\% outage fraction}  \\
\cline{2-10}
 OHR (\%) & $\eta$ = 0.1 & $\eta$ = 0.01 & $\eta$ = 0.001 & $\eta$ = 0.1 & $\eta$ = 0.01 & $\eta$ = 0.001 & $\eta$ = 0.1 & $\eta$ = 0.01 & $\eta$ = 0.001 \\
\hline 
\hline
60 & 59.35 & 59.36 & 59.17 & 9.03 & 6.00 & 5.41 & 0.01 & 0.01 & 0.05 \\
\hline
80 & 79.13 & 78.72 & 77.69 & 150.56 & 97.67 & 75.27 & 0.07 & 0.11 & 0.23 \\
\hline
\end{tabular}
\end{table*}

\begin{table*}[!ht]
\small
\caption{\s{Impact of linear changes in constant step size $\eta=0.01$ on the performance of d-TTL (sensitivity analysis).}}
\label{tab:d-ttl-sen}
\centering
\begin{tabular}{|p{0.55in}|p{0.55in}|p{0.55in}|p{0.55in}|p{0.55in}|p{0.55in}|p{0.55in}|p{0.55in}|p{0.55in}|p{0.55in}|}
\hline
Target  & \multicolumn{3}{|c|}{Average OHR (\%)} & \multicolumn{3}{|c|}{Average cache size (GB)} & \multicolumn{3}{|c|}{5\% outage fraction}  \\
\cline{2-10}
 OHR (\%) & $\eta$(1+0.05) & $\eta$ & $\eta$(1-0.05) & $\eta$(1+0.05) & $\eta$ & $\eta$(1-0.05) & $\eta$(1+0.05) & $\eta$ & $\eta$(1-0.05) \\
\hline 
\hline
60 & 59.36 & 59.36 & 59.36 & 5.98 & 6.00 & 6.02 & 0.01 & 0.01 & 0.01 \\
\hline
80 & 78.73 & 78.72 & 78.71 & 98.21 & 97.67 & 97.1 & 0.11 & 0.11 & 0.11 \\
\hline
\end{tabular}
\end{table*}

\begin{table*}[!ht]
\small
\caption{\s{Impact of exponential changes in constant step size $\eta_s$ on the performance of f-TTL (robustness analysis).}}
\label{tab:f-ttl-rob}
\centering
\begin{tabular}{|p{0.55in}|p{0.55in}|p{0.55in}|p{0.55in}|p{0.55in}|p{0.55in}|p{0.55in}|p{0.55in}|p{0.55in}|p{0.55in}|}
\hline
Target  & \multicolumn{3}{|c|}{Average OHR (\%)} & \multicolumn{3}{|c|}{Average cache size (GB)} & \multicolumn{3}{|c|}{5\% outage fraction}  \\
\cline{2-10}
 OHR (\%) & $\eta_s$ = 1e-8 & $\eta_s$ = 1e-9 & $\eta_s$ = 1e-10 & $\eta_s$ = 1e-8 & $\eta_s$ = 1e-9 & $\eta_s$ = 1e-10  & $\eta_s$ = 1e-8 & $\eta_s$ = 1e-9 & $\eta_s$ = 1e-10  \\
\hline 
\hline
60 & 59.36 & 59.36 & 59.36 & 5.46 & 2.96 & 1.88 & 0.01 & 0.01 & 0.02 \\
\hline
80 & 78.65 & 78.55 & 78.47 & 89.52 & 55.08 & 43.34 & 0.12 & 0.14 & 0.17 \\
\hline
\end{tabular}
\end{table*}

\begin{table*}[!ht]
\small
\caption{\s{Impact of linear changes in constant step size $\eta_s=\text{\normalfont 1e-9}$ on the performance of f-TTL (sensitivity analysis).}}
\label{tab:f-ttl-sen}
\centering
\begin{tabular}{|p{0.55in}|p{0.55in}|p{0.55in}|p{0.55in}|p{0.55in}|p{0.55in}|p{0.55in}|p{0.55in}|p{0.55in}|p{0.55in}|}
\hline
Target  & \multicolumn{3}{|c|}{Average OHR (\%)} & \multicolumn{3}{|c|}{Average cache size (GB)} & \multicolumn{3}{|c|}{5\% outage fraction}  \\
\cline{2-10}
 OHR (\%) & $\eta_s$(1+0.05) & $\eta_s$ & $\eta_s$(1-0.05) & $\eta_s$(1+0.05) & $\eta_s$ & $\eta_s$(1-0.05) & $\eta_s$(1+0.05) & $\eta_s$ & $\eta_s$(1-0.05) \\
 \hline 
\hline
60 & 59.36 & 59.36 & 59.36 & 3.01 & 2.96 & 2.91 & 0.01 & 0.01 & 0.01 \\
\hline
80 & 78.55 & 78.55 & 78.54 & 55.65 & 55.08 & 54.27 & 0.14 & 0.14 & 0.14 \\
\hline
\end{tabular}
\end{table*}

%% file: related.tex
\section{Related Work}\label{sec:related}
Caching algorithms have been studied for decades in different contexts such as CPU caches, memory caches, CDN caches and so on. We briefly review some relevant prior work. 

\textbf{TTL-based caching.} TTL caches have found their place in theory as a tool for analyzing capacity based caches~\cite{dehghan2016utility, garetto2016unified, fofack2012analysis}, starting from {\color{blue} characteristic time approximation of LRU caches~\cite{fagin1977asymptotic, che2002hierarchical}.} Recently, its generalizations~\cite{bianchi2013check, olmos2014catalog} have commented on its wider applicability. However, the generalizations hint towards the need for more tailored approximations and show that the vanilla {\color{blue} characteristic time approximation} can be inaccurate~\cite{olmos2014catalog}. On the applications side, recent works have demonstrated the use of TTL caches in utility maximization~\cite{dehghan2016utility} and hit ratio maximization~\cite{berger2015maximizing}. Specifically, in~\cite{dehghan2016utility} the authors provide an online TTL adaptation highlighting the need for adaptive algorithms. However,  unlike prior work, we propose the first adaptive TTL-based caching algorithms that provides provable hit rate and normalized size performance in the presence of non-stationary traffic such as one-hit wonders and traffic bursts.

We also review some capacity-based caching algorithms. 

\textbf{Capacity-based caching.} Capacity-based caching algorithms have been in existence for over 4 decades and have been studied both theoretically (\s{e.g. exact expressions~\cite{king1971, gelenbe1973unified, aven1987stochastic} and mean field approximations~\cite{hirade2010analysis, tsukada2012fluid, gast2015transient} for hit rates, {\color{blue} mixing time of caching algorithms~\cite{li2017accurate}} }) and empirically (in the context of web caching:~\cite{podlipnig2003survey}).  Various cache replacement algorithms have been proposed based on the frequency of object requests (e.g. LFU), recency of object requests (e.g. LRU) or a combination of the two parameters (e.g., LRU-K, 2Q, LRFU~\cite{o1993lru, johnson1994x3, lee1999existence}).
Given that CDNs see a lot of non-stationary traffic, cache admission policies such as those using bloom filters~\cite{maggs2015algorithmic} have also been proposed to maximize the hit rate under space constraints. Further, non-uniform distribution of object sizes have led to more work that admit objects based on the size (e.g., LRU-S, AdaptSize~\cite{starobinski2001probabilistic, berger2017adaptsize}). While several capacity-based algorithms have been proposed, most don't provide theoretical guarantees in achieving target hit rates.

\textbf{Cache Tuning and Adaptation.} Most existing adaptive caching algorithms require careful parameter tuning to work in practice. There have been two main cache tuning methods: (1) global search over parameters based on prediction model, e.g.~\cite{cidon2015dynacache, saemundsson2014dynamic}, and (2) simulation and parameter optimization based on shadow cache, e.g.~\cite{cidon2016cliffhanger}. The first method often fails in the presence of cache admission policies; whereas, the second method typically assumes stationary arrival processes to work well. However, with real traffic, static parameters are not desirable~\cite{megiddo2004outperforming} and an adaptive/self-tuning cache is necessary. The self-tuning heuristics include, e.g., ARC~\cite{megiddo2004outperforming},  CAR~\cite{bansal2004car}, PB-LRU~\cite{zhu2004pb}, which try to adapt cache partitions based on system dynamics. While these tuning methods are meant to deal with non-stationary traffic, they lack theoretical guarantees unlike our work, where we provably achieve a target hit rate and a feasible normalized size by dynamically changing the TTLs of cached content.

Finally, we discuss work related to cache hierarchies, highlighting  differences between those and the f-TTL algorithm.
       
\textbf{Cache hierarchies.} Cache hierarchies, made popular for web caches in~\cite{chankhunthod1996hierarchical, che2002hierarchical,SitaramanKasbekarLichtensteinEtAl2014}, consist of separate caches, mostly LRU~\cite{rosensweig2010approximate,SitaramanKasbekarLichtensteinEtAl2014} or TTL-based~\cite{fofack2012analysis}, arranged in multiple levels; with users at the lowest level and the server at the highest. A requested object is fetched from the lowest possible cache and, typically, replicated in all the caches on the request path. {\color{blue} Analysis for network of TTL-caches were presented in~\cite{fofack2012analysis, berger2014exact}.  In a related direction, the performance of complex networks of size based caches were approximated in~\cite{rosensweig2010approximate}.}                 

While similar in spirit, the f-TTL algorithm differs in its structure and operation from hierarchical caches. Besides the adaptive nature of the TTLs, the higher and lower-level caches are assumed to be co-located and no object is replicated between them---a major structural and operational difference. Further, the use of shadow cache and shallow cache in lower-level cache $\mc{C}_s$ distinguishes f-TTL from the above.

%% file: conclusions.tex
\section{Conclusions}\label{sec:conclusions}
In this paper we designed adaptive TTL caching algorithms that can automatically
learn and adapt to the request traffic and provably
achieve any feasible hit rate and cache size. Our work fulfills a long-standing deficiency in the modeling and analysis of caching algorithms in presence of bursty and non-stationary request traffic. In particular, we presented a theoretical justification for the use of two-level caches in CDN settings where large amounts of non-stationary traffic can be filtered out to conserve cache space while also achieving target hit rates. On the practical side, we evaluated our TTL caching algorithms using traffic traces from a production Akamai CDN server. The evaluation results show that our adaptive TTL algorithms can achieve the target hit rate with high accuracy; further, the two-level TTL algorithm can achieve the same target hit rate at much smaller cache size.

%% file: proof.tex
\section{Proof of Main Results}\label{sec:proof}
In this section we provide complete proofs of the results in the main paper. Firstly we prove our main result, Theorem~\ref{thm:SAConv}. Then we provide proofs for the special cases involving Poisson arrivals with independent labeling and sub-classes of that.  

The proof of Theorem~\ref{thm:SAConv} is separated in three main parts. In the first part we formally introduce the dynamic system as a discrete time Markov process on continuous state space, a.k.a. Harris chains. In the second part we analyze the static caching system---caching system with the parameters $\bm{\theta}$ and $\bm{\theta}^s$ fixed. In the last part, we prove that  the sufficient conditions for the convergence of the two timescale stochastic approximation algorithm is satisfied under the Assumption~\ref{a:Arr}. Our proof uses projected ODE based convergence methods~\cite{bhatnagar2001two}. The static analysis in the previous step plays a key role in characterizing the ODEs representing the mean evolution of the adaptive TTLs.  

Some notations used through out the proofs are given next. A scalar function when presented in boldface means we apply the same function to each coordinate of its vector argument, i.e. $\bm{f}(\bm{x})=(f(x_1),\cdots, f(x_n))$. Further, $\mathbf{1}_n$ denotes the all $1$ vector of size $n$ and $\mathbf{e}_{i}^{n}$ denotes the $i$-th standard basis for $\mathbb{R}^n$ ($i$-th coordinate is equal to $1$ and other coordinates are equal to $0$).

\input{process}

\input{static}

\input{dynamic}

\input{poisson} 

%% file: process.tex
\subsection{Caching Process}
The evolution of the caching system is presented as a discrete time Markov chain on a continuous state space, commonly known as Harris chains. We first observe that, by fixing $\bm{\vartheta}^s = \bm{1}$ the f-TTL quickly identifies with the d-TTL system (when we treat $\bm{\theta}^s = \bm{\theta}$). We call it f-TTL cache collapsing to d-TTL cache.\footnote{Technically, the f-TTL with $\bm{\vartheta}^s = \bm{1}$ gets coupled to d-TTL from time $\tau(l) = \|\bm{L}\|_{\infty}=L_{max}$ onwards with the d-TTL system. Further, any sample path (with non zero probability) for the d-TTL system is contained in at least one sample path (with non zero probability) of collapsed f-TTL. Therefore, any a.s. convergence in the collapsed f-TTL will imply an a.s. convergence in the d-TTL.}
This leads us to unify the two systems, d-TTL and  f-TTL system, where for d-TTL $\bm{\vartheta}^s(l) = \bm{1}$ for all $l$. 

We next develop the necessary notations for the rest of the proofs before formally describing the system evolution. 


\paragraph{Arrival Process Notation} The sequence of arrivals in the system is given by the sequence of increasing r.v. $\bm{A}=\{A(l): l \in \mathbb{N}\}$ where each inter arrival time $X(l) = A(l)-A(l-1)$ has identical distribution satisfying the conditions in Assumption~\ref{a:Arr}. By convention we define $A(0)=0$. 

Under Assumption~\ref{a:Arr}, the request labels have Markovian dependence across subsequent arrivals.
Specifically, the object labeling process $Z$ is a DTMC  given as $\{Z(l): l \in \mathbb{N}\}$, where $Z(l)$ denotes the $(l-1)$-th request label for {\em recurrent objects} and request type for {\em rare objects}, for $l\geq 1$. Let $Z(0)$ be the initial state. We define $L_c = \{l: c(l) = c, l \in \mathbb{N} \} $ as the sequence of instances where object $c$ is requested, for $c\in \mc{U}$---both {\em recurrent} and {\em rare} objects. The request arrival times of a object $c$ is given as $\bm{A}_c = \{A(l):l \in L_c\}$ and $A_c(j)$ denotes the $j$-th arrival of object $c$. We denote the $j$-th inter arrival time for object $c$ as $X_c(j)= A_c(j)- A_c(j-1)$. Further by $N_c(u) = |\{j : A_c(j) \leq u\}|$ we denote the number of arrival of object $c$ up to time $u$, for $u\geq 0$. 

For a {\em recurrent object} $c$, due to the irreducibility and aperiodicity of the DTMC $Z$, the inter arrival distribution has a stationary distribution with c.d.f. $p_c(\cdot)=\mathbb{P}\left( X_c(1)\leq \cdot\right)$ and mean $1/\lambda_c\equiv 1/(\pi_c\lambda)$. However, for the {\em rare objects} the inter arrival time distribution is not necessarily defined. But the rare objects of type $t$ have an aggregate rate of $\alpha_t \lambda$.

We now present the evolution of the system states and the TTL parameters.
\paragraph{System State Notation} The evolution of the system is a coupled stochastic process given as the sequence, 
${\{\left( \bm{\vartheta}(l), \bm{\vartheta}^s(l),\mc{S}(l)\equiv \left(\bm{\Psi}(l), Z(l)\right)\right): l \in \mathbb{N}\}}$. At the $l$-th arrival,  $\bm{\Psi}(l)$ represents the {\ttl} information of the recurrent objects and $Z(l)$ denotes which object (object-type for `rare' objects) is requested in the $(l-1)$-th arrival. The vector of timer tuples for  \emph{recurring objects} is $$\bm{\Psi}(l)=\left\{ \psi_c(l)\equiv(\psi^{0}_c(l), \psi^{1}_c(l), \psi^{2}_c(l)) : c\in \mc{K}\right\}.$$ 

We reemphasize that the labeling of the \emph{rare objects} is not included in the state space, only the information about $Z(l)$ is included while maintaining system state. The system under complete information is not necessarily stationary.

\paragraph{System State Evolution.} 
There exists an appropriate probability space $\left(\Omega, \mathcal{F}, \mathbb{P}_{\mathrm{sys}}\right)$ on which the complete system evolution, including the information of the {\em rare object} labels, is defined in the standard way. Let $\mc{F}(l)\subseteq \mathcal{F}$ be a filtration---a sequence of non decreasing $\sigma$-algebras---such that the history of the system adaptation upto the $l$-th arrival, $\mc{H}(l)=\{\bm{\vartheta}(i), \bm{\vartheta}^s(i), \mc{S}(i), X(i-1): i \leq l\}$ is $\mc{F}(l)$-measurable. By $\mathbb{E}_l$, we denote the conditional expectation conditioned on $\mc{F}(l)$, i.e.  $\mathbb{E}\left(\cdot\lvert\mc{F}(l)\right)$.

\begin{figure}
\centering
\includegraphics[width=0.8\linewidth]{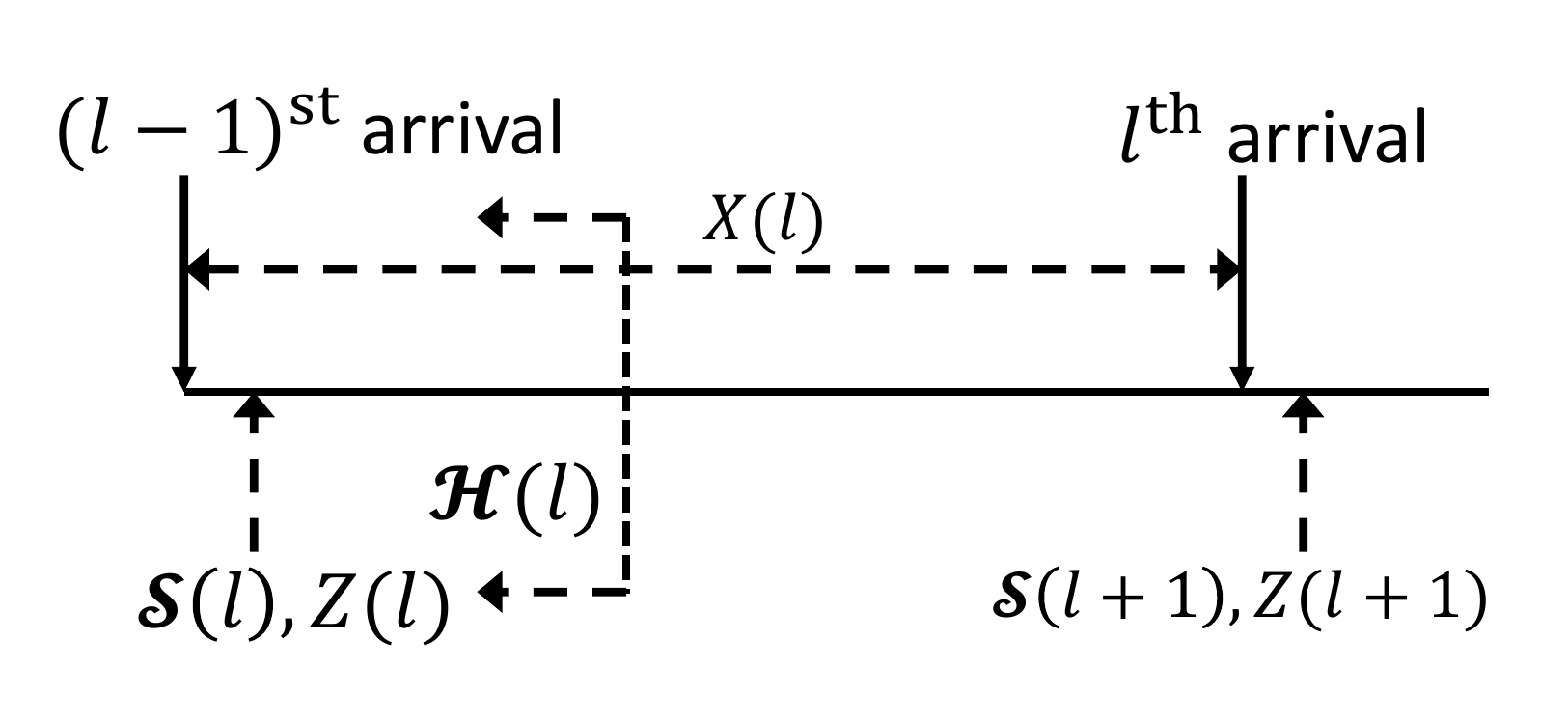}
\caption{System Evolution.}
\label{fig:evolution}
\end{figure}

Let the system be in state $\left( \bm{\vartheta}(l), \bm{\vartheta}^s(l),\mc{S}(l)\equiv \left(\bm{\Psi}(l), Z(l)\right)\right)$ after the $(l-1)$-th arrival. On $l$-th arrival, the following events can occur,
\begin{itemize}
\item Recurrent object $c\in \mc{K}$ is requested and a cache hit or cache virtual hit occurs. We have $\psi_{c}(l+1) = (\theta_{c_{typ}}(l),0,0)$, $Z(l+1)= c$, and for all $c' \neq c$, $c'\in \mc{K}$, $\psi_{c'}(l+1) = (\psi_{c'}(l)- X(l)(1,1,1))^+$. 
\item Recurrent object $c\in \mc{K}$ is requested and a cache miss happens. We have $\psi_{c}(l+1)= ( 0, \theta^s_{c_{typ}}(l), \theta_{c_{typ}}(l))$, $Z(l+1)= c$ and for all $c'\neq c$, $c'\in \mc{K}$, $\psi_{c'}(l+1) = (\psi_{c'}(l)- X(l)(1,1,1))^+$. 
\item Rare object $c\notin \mc{K}$ is requested and $c$ is of type $t$. We have $Z(l+1)= K+t$ (state representing rare objects of type $t$) and for all  $c\in \mc{K}$, $\psi_{c}(l+1) = (\psi_{c}(l)- X(l)(1,1,1))^+$. 
\end{itemize}

We next formalize the above. For notational convenience we define the cache miss indicator, $q^m_c(l)$, cache virtual hit indicator, $q^v_c(l)$ and, cache hit indicator for two caches $q^{h}_c(l)$, $q^{hs}_c(l)$,  for all $c\in \mc{K}$ and for all $l\in \mathbb{N}$ as given in the following equations. The cache hit event for $c\in \mc{K}$ is given by $(q^{h}_c(l)+q^{hs}_c(l))$.
\begin{gather*}
q^m_c(l) = \mathbbm{1}\left(\max\{\psi^{0}_c(l), \psi^{1}_c(l), \psi^{2}_c(l)\} < X(l)\right),\\
q^v_c(l) = \mathbbm{1}\left(\psi^{1}_c(l) < X(l) \land \psi^{2}_c(l) \geq X(l)\right),\\
q^{h}_c(l) = \mathbbm{1}\left(\psi^{0}_c(l) \geq X(l) \right),\
q^{hs}_c(l) = \mathbbm{1}\left(\psi^{1}_c(l) \geq X(l) \right).
\end{gather*}
As $X(l)\sim X$, it is convenient to express the average of $q^{\{m,v,h,hs\}}_c(l)$ conditional to $\mc{F}(l)$ as,
\begin{gather*}
\mathbb{E}_l q^m_c(l) = \mathbb{P}\left(X > \max\{\psi_c(l)\}\right),\\
\mathbb{E}_l q^v_c(l) = \mathbb{P}\left( X \in (\psi^{1}_c(l), \psi^{2}_c(l)]\right),\\
\mathbb{E}_l q^{h}_c(l) = \mathbb{P}\left(X \leq \psi^{0}_c(l)\right),\
\mathbb{E}_l q^{hs}_c(l) = \mathbb{P}\left(X \leq  \psi^{1}_c(l)\right).
\end{gather*}

The dynamics of $Z(l)$ is Markovian and the transition probability from $Z(l)$ to some object $c\in \mc{K}$ is given as $P\left(Z(l),c\right)$. We define the following terms for the ease of presentation,
\begin{gather*}
\tilde{p}(c,l) = P(Z(l),c)\mathbb{E}_l (q^v_c(l)+q^h_c(l)+q^{hs}_c(l)),\\
\tilde{p}_1(c,l) = P(Z(l),c)\mathbb{E}_l (q^v_c(l)+q^{hs}_c(l)),\\
 \tilde{p}_2(c,l) = P(Z(l),c)\mathbb{E}_l q^m_c(l).
\end{gather*}

Formally, the transition of $\bm{\Psi}(l)$ is given as in equations \eqref{eq:psiAdapt} and \eqref{eq:psiSAdapt}, where $X$ is the inter arrival time distribution. {\color{blue} The updated values $\bm{\theta}^s(l+1), \bm{\theta}(l+1)$ are described shortly.}
\begin{equation}
\label{eq:psiAdapt}
\psi^{0}_c(l+1)= 
\begin{cases}
(\psi^{0}_c(l) - X)^{+} &\text{w.p.}~(1 - \tilde{p}(c,l))\\
\theta_{c_{typ}}(l+1)&\text{w.p.}~\tilde{p}(c,l). 
\end{cases}
\end{equation}

\begin{multline}
\label{eq:psiSAdapt}
(\psi^{1}_c(l+1),\psi^{2}_c(l+1))= \\
\begin{cases}
\left((\psi^{i}_c(l) - X)^{+}\right)_{i=1}^{2} & \text{w.p.}~ (1-\tilde{p}_1(c,l)-\tilde{p}_2(c,l))\\
(0,0) & \text{w.p.}~ \tilde{p}_1(c,l)\\
(\theta^s_{c_{typ}}(l+1), \theta_{c_{typ}}(l+1)) &\text{w.p.}~\tilde{p}_2(c,l). 
\end{cases}
\end{multline}
\color{black}

\paragraph{$\bm{\vartheta}$ Adaptation} 
We concisely represent the evolution of $\bm{\vartheta}(l)$ and $\bm{\theta}(l)$ as~\eqref{eq:ttlAdapt}, 
\begin{equation}
\centering
\begin{aligned}
\bm{\vartheta}(l+1) &= \mc{P}_{H}\left(\bm{\vartheta}(l)+\eta(l) {\widehat{w}(l)}\left(h^*_{t(l)} - Y(l)\right)\mathbf{e}^T_{t(l)}\right).\\
\bm{\theta}(l+1) &= \sum_t  L_t\vartheta_t(l+1)\mathbf{e}^T_t.
\end{aligned}
\label{eq:ttlAdapt}
\end{equation}
Here $\eta(l)= \frac{\eta_0}{l^{\alpha}}$ is a decaying step size for $\alpha\in (1/2,1)$, $\mc{P}_{H}$ denotes the projection over the set $H=\{\mathbf{x}\in \mathbb{R}^T: 0 \leq x_i \leq 1\}$, $\bm{L}$ is the upper bound of the {\ttl}
which is provided as an input, and 
\begin{equation*}
\widehat{w}(l)=
\begin{cases}
w(l),~\text{for byte hit rate,}\\
1,~\text{for object hit rate.}
\end{cases}
\end{equation*}

Recall, $Y(l)$ takes value $1$ if it is a cache hit at the $l$-th arrival or takes the value $0$ otherwise. We split $Y(l)$ in two components, $Y(l) = Y_{\mcs{K}}(l)+ \beta_h(l)$. The term $Y_{\mcs{K}}(l)=Y(l)\mathbbm{1}(c(l)\in \mc{K})$ denotes the contribution from \emph{recurring objects} and  $\beta_h(l)=Y(l)\mathbbm{1}(c(l)\notin \mc{K})$ denotes the contribution from the  \emph{rare objects}. 
The expected (w.r.t. $\mc{F}(l)$) cache hit from object $c\in \mc{K}$ is expressed by $g_c(\cdot)$ and it is given by the following measurable function,
\begin{align}\label{eq:avgMissFunc}
g_c(\psi_c(l), Z(l)) =\left(\mathbb{E}_l q^h_c(l) + \mathbb{E}_l q^{hs}_c(l)\right),~\forall c\in \mc{K}.
\end{align}
Given the history $\mc{H}(l)$, we obtain the conditional expectation of the update  of $\bm{\vartheta}(l+1)$ due to recurrent objects as (complete expression later),   
\begin{align}
\label{eq:avgMiss}
\mathbf{g}\left(\bm{\vartheta}(l), \bm{\vartheta}^s(l), \mc{S}(l)\right) 
= \mathbb{E}_l\left( \widehat{w}(l)\left(\mathbf{h}^*_{t(l)} -  Y_{\mcs{K}}(l)\right)\mathbf{e}_{t(l)}^T  \right).
\end{align}

Further, for all $l$, define the martingale difference $$\bm{\delta M}(l) = \widehat{w(l)}\left(\mathbf{h}^*_{t(l)} - Y_{\mcs{K}}(l)\right)\mathbf{e}^T_{t(l)} - \mathbf{g}\left(\bm{\vartheta}(l), \bm{\vartheta}^s(l), \mc{S}(l)\right).$$ 
Finally, in order to remove the projection operator we define the reflection due to this projection operator (i.e. $(\bm{x}- \mc{P}_H(\bm{x}))$) as $\bm{r}(l)$. We can rewrite the equation representing the  adaptation of $\bm{\vartheta}(l)$ in Equation~\eqref{eq:ttlAdapt} as;
\begin{equation}
\begin{aligned}
\bm{\vartheta}(\st{l+1}) &= \mc{P}_H\left(\vphantom{\mathbf{e}^T_{t(l)}}\bm{\vartheta}(l)+ \right. \eta(l)\mathbf{g}(\bm{\vartheta}(l), \bm{\vartheta}^s(l), \mc{S}(l)) \\
& \hspace{2cm}  \left. + \eta(l)\left(\bm{\delta M}(l) - \beta_h(l)\mathbf{e}^T_{t(l)}\right) \right) \\
&= \bm{\vartheta}(l)+ \eta(l)\mathbf{g}(\bm{\vartheta}(l), \bm{\vartheta}^s(l), \mc{S}(l)) \\ 
& \hspace{1cm}  + \eta(l)\left(\bm{\delta M}(l) - \widehat{w(l)}\beta_h(l)\mathbf{e}^T_{t(l)}\right) + \bm{r}(l).  
\end{aligned}
\label{eq:ttlAdapt2}
\end{equation}

\paragraph{$\bm{\vartheta}^s$ Adaptation} 
The evolution of $\bm{\vartheta}^s(l)$ and $\bm{\theta}^s(l)$ is represented as~\eqref{eq:ttldifAdapt}. 
\begin{equation}
\label{eq:ttldifAdapt}
\begin{aligned}
\bm{\vartheta}^s(\st{l+1})&= 
\mc{P}_{H}\left(\bm{\vartheta}^s(l)+\eta_s(l){{{w}(l)}}(s^*_{t(l)}-s(l))\mathbf{e}^T_{t(l)}\right)\\
\bm{\theta}^s(\st{l+1})&=\sum_t L_t\vartheta_t(\st{l+1})\up\left(\vartheta_t(\st{l+1}), \vartheta^s_t(\st{l+1});\epsilon\right)\mathbf{e}^T_t .
\end{aligned}
\end{equation}
Here $\eta_s(l)=\eta_0/l$ and $\epsilon$ is a parameter of the algorithm. The projection $\mc{P}_{H}$ and $\bm{L}$ are defined as in d-TTL. The $\up(\cdot,\cdot;\epsilon)$ is a \emph{threshold function}, defined earlier as, a twice differentiable non-decreasing function, $\up(x,y;\epsilon) : [0,1]^2 \to [0, 1]$, which takes value $1$ for $x\geq 1-\epsilon/2$ and takes value $y$ for $x\leq 1-3\epsilon/2$, and its partial derivative w.r.t. $x$ in between is bounded by $4/\epsilon$.

The random variable $s(l)$ is defined in algorithm~\ref{alg:f-TTL} and can be split into three components $s(l)= s_{\mcs{K}}(l) + s_{1}(l) + s_{2}(l)$. Here $s_{\mcs{K}}(l)=s(l)\mathbbm{1}(c(l)\in \mc{K})$ denotes the contribution of the \emph{recurring objects}. However, the contribution of the \emph{rare objects} has two parts, 1) stationary part, $s_1(l)= \theta^s_{t(l)}(l)\mathbbm{1}(c(l)\notin \mc{K})$, and 2)non-stationary `noise', $s_2(l) = (s(l)-\theta^s_{t(l)}(l))\mathbbm{1}(c(l)\notin \mc{K})$. 
The contribution of object $c$ to the expectation(w.r.t. $\mc{F}(l)$ ) normalized size is given as,
\begin{multline}
\label{eq:avgSizeFunc}
g^s_c(\bm{\vartheta}(l), \bm{\vartheta}^s(l),  \psi_c(l))= \theta^s_{t}(l)\mathbb{E}_l q^m_c(l)+ \theta_{t}(l)(1-\mathbb{E}_l q^m_c(l)) \\  
-\psi^{0}_c(l)\mathbb{E}_l q^{h}_c(l)-\psi^{1}_c(l)\mathbb{E}_l q^{hs}_c(l),~\forall c\in \mc{K}.
\end{multline}         

Finally, we have the conditional expectation of the update due to recurrent objects, w.r.t. $\mc{F}(l)$, as (full expression later),
\begin{equation}
\label{eq:avgSize}
\bm{g}^s \left(\st{\bm{\vartheta}(l), \bm{\vartheta}^s(l), \mc{S}(l)}\right)=\mathbb{E}_l\left( w(l)(s^*_{t(l)}-s_{\mcs{K}}(l) - s_1(l))e^T_{t(l)} \right).
\end{equation}

The adaptation of $\bm{\vartheta}^s$ for the f-TTL case (d-TTL is trivial), including similar definitions of the martingale difference, $\bm{\delta M}^s(l)$, and the reflection error, $\bm{r}^s(l)$, as
\begin{align}
\bm{\vartheta}^s(l+1)= & \bm{\vartheta}^s(l) + \eta_s(l)\bm{g}^s\left(\bm{\vartheta}(l), \bm{\vartheta}^s(l), \mc{S}(l)\right) \nonumber\\
&+ \eta_s(l)\left(w(l)s_2(l)\bm{e}^T_{t(l)} + \bm{\delta M}^s(l)\right) + \bm{r}^s(l).
\label{eq:ttlDifAdapt3}
\end{align}

\begin{eqholder*}[ht!]
\label{eqh:adaptation}
\begin{equation*}
\eqref{eq:avgMiss} = \dm{\sum}_t \left(  \dm{\sum_{\substack{c: c_{typ}= t\\ c\in \mc{K}}}} \widehat{w_c} p\left(Z(l),c\right) \Big( h^{*}_t   - g_c( \psi_c(l), Z(l))\Big) + \widehat{\bar{w}_t}\,p(Z(l),K+t)h^{*}_t \right) \mathbf{e}^T_t
\end{equation*}
\begin{equation*}
\eqref{eq:avgSize}= \dm{\sum}_t \left(\dm{\sum_{\substack{c: c_{typ}= t\\ c\in \mc{U}}}} w_c p\left(Z(l),c\right) \Big(s^*_t 
- g^s_c(\bm{\vartheta}(l), \bm{\vartheta}^s(l), \psi_c(l))\Big) +\bar{w}_t\,p(Z(l),K+t)\left(s^*_t - \theta_t^s(l)\right) \right)\mathbf{e}^T_t.
\end{equation*}
$\widehat{\cdot}$ differentiates between byte and object hit rates.

\hrulefill
\end{eqholder*}

%% file: static.tex
\subsection{Static Analysis}
Here we analyze the caching process with fixed parameters $\bm{\theta}$ and $\bm{\theta}^s$ which plays a crucial role in the proof of our main result. As discussed earlier fixing the latent parameters inadvertently fixes the above parameters.  Using `small' set based techniques~\cite{meyn2012markov} we show that the static system converges fast towards a stationary distribution. However, as the stationary distribution is hard to analyze, we use renewal theory~\cite{smith1958renewal} to obtain the properties of the system under stationary distribution.

\paragraph{Strong Mixing} The first step towards the convergence of the proposed dynamic algorithm is to show that the static system converges in an almost sure sense. We  show that the stochastic process $\{\mc{S}(l)\}$ converges almost surely and the convergence happens at a geometric rate. The system under static setting forms a discrete time Markov chain (DTMC), $$ \bm{\mc{S}} = \{\mc{S}(l)\equiv \left(\bm{\Psi}(l), Z(l)\right) : l \in \mathbb{N}\} $$ on state space ${\mathfrak{S} = [\mathbf{0},\bm{\theta}]\times [\mathbf{0},\bm{\theta}^s] \times [\mathbf{0},\bm{\theta}]\times [K+T]}$. 

Let the Markov kernel associated with this Harris chain be $\mc{P}(x,A)$, for all $A \subseteq \mc{B}(\mathfrak{S})$ (the Borel set on the space), $x\in\mathfrak{S}$. Further define the $n$-step kernel 
$$\mc{P}^n(x,A) \equiv \mathbb{P}(\mc{S}(n)\in A |\mc{S}(0)=x).$$ 
\begin{theorem}
\label{thm:HarrisRecur}
\s{Under Assumption~\ref{a:Arr} with $\|\mathbf{L}\|_{\infty}$-rarity condition}, the Harris chain $\bm{\mc{S}}$ over state space $\mathfrak{S}$ admits a unique invariant measure $\mu^*$. Furthermore, there exists constant $C>0$ and $\gamma\in(0,1)$ for each measurable and bounded function $f:\mathfrak{S} \rightarrow \mathbb{R}^d$, with finite $d$, the following holds
\begin{equation}
\label{eq:uniformErgodic}
\| \mathbb{E}f(S(n)) - \mathbb{E}_{S\sim \mu^*} f(S) \| \leq C\gamma^n,~\forall n\in \mathbb{N}.
\end{equation}  
\end{theorem}
\begin{proof}
Given a Harris chain over $\mathfrak{S}$, a set $A\in \mc{B}(\mathfrak{S})$ is called a \emph{small} set if there exists an integer $m>0$ and a non zero measure $\nu_m$ such that for all $x\in A$ and for all $B\in \mc{B}(\mathfrak{S})$,  $\mc{P}^m(x,B)\geq \nu_m(B)$.
By showing that under the arrival process of Assumption~\ref{a:Arr}, $\mathfrak{S}$ itself is a \emph{small} set with non zero measure $\nu_m$ we can guarantee that there exist a unique invariant measure $\mu^*$ and further the chain $\bm{\mc{S}}$ is uniformly ergodic, i.e. $$\| \mc{P}^n(S(0),\cdot) - \mu^* \|_{TV} \leq (\gamma')^{n/m}, ~\forall S(0)\in \mathfrak{S}.$$
Here $\gamma' = 1- \nu_m(\mathfrak{S})$ (See Chapter 16 in~\cite{meyn2012markov}). Finally from the properties of the total variation norm we obtain that for any measurable and bounded (from above and below) $f$ the above inequality~\eqref{eq:uniformErgodic} holds with $C = 2\sup \|f\| < \infty$ and $\gamma = (\gamma')^{1/m} \in (0,1)$. 

We show the content request process, i.e. the combination of the arrival process and the labeling process, possesses a structure that enables the wiping out of history. Our argument is based on two separate cases which are complementary to each other and results in wiping out of the effect of the initial states in two separate manner. Consider any object $c\in \mc{K}$. As the DTMC representing the process $Z$ is irreducible and aperiodic, there exists some finite $m_0$ such that  $\delta' = \dm{\min_{z \in [K+T]}} \mathbb{P}(Z(\st{m_0-1}) = c | Z(0)= z) > 0 $.

$\mathbf{C1:} \mathbb{P}(X \geq L_{\max}) \geq \delta_1>0$. 

In this case an inter arrival time of at least $L_{\max}$ effectively wipes out the history. Consider the system state $$\mc{S}_{rec}= \left(\bm{\Psi}=(0, \theta^s_{c_{typ}}, \theta_{c_{typ}})\bm{e}^{|\mcs{K}|}_c, Z = c\right).$$ The event, that the $(m_0-1)$-th inter arrival time is greater than $L_{\max}$ and the state of the labeling DTMC is $c$ at the $(m_0-1)$-th step, i.e. $E = \{X(m_0-1) > L_{max}\} \cap \{Z(m_0-1) = c\}$, which happens with probability at least $\delta'_1 \delta_1 > 0$, results in $\mc{S}(m_0-1) = \mc{S}_{rec}$. 
Therefore, $$\mc{P}^{(m_0-1)}\left( \mc{S}(m_0-1) = \mc{S}_{rec}\right) \geq \delta'_1 \delta_1.$$ 
Finally, letting  $\nu_{m_0}(\cdot)= \delta_1' \delta_1 \mc{P}(\mc{S}_{rec}, \cdot)$, we have $\mathfrak{S}$ to be a \emph{small} set with non zero measure $\nu_{m_0}$. So the theorem holds for $\gamma = (1-\delta'_1\delta_1)^{1/m_0}$.

$\mathbf{C2:} \mathbb{P}(X \geq L_{\max}) = 0$ and for TPM $P$, $\exists c,c'\in [K+T]$ s.t. $P(c,c) > 0$ and $P(c',c')>0$.

The first part of the condition ensures that inter arrival time is bounded away from zero with positive probability, which will be used, alongside the existence of self loop, in wiping out the history. Applying Paley-Zygmund inequality, under this condition we obtain, $\mathbb{P}(X> 1/2 \lambda)\geq  \tfrac{1}{4\lambda^2 L_{\max}^2}\equiv \delta_2$.  Let $\mathbb{P}( \sum^{n_0}_{i=1} X_i > L_{max})\geq \delta_2^{n_0} > 0$ for integer $n_0 \equiv 2 L_{max} \lambda$, where $X_i$s are independent copies of the inter arrival time.  

Consider two states $c, c'$ which have self loops, i.e. $P(c,c)>0$ and $P(c',c')>0$. 
Due to irreducibility and aperiodicity of  the process $Z$ there exists a finite integer $m_1$ such that $\delta'_2 = \dm{\min_{z \in [K+T], d\in \{c,c'\}}} \mathbb{P}(Z(m_1) = d | Z(0) = z) > 0 $.
Let $E_1$ be the event that; (i) the process $Z$ reaches state $c$, and (ii) the process $Z$ remains in the state $c$ such that all other objects are evicted from both levels. Also, by $E_2$ denote the event that; (i) the process $Z$ reaches from state $c$ to state $c'$, and (ii)  the process $Z$ remains in the state $c$ such that all other objects are evicted from both levels. Formally, we have the events
\begin{gather*}
E_1 =  \{Z(m_1) = c| Z(0)=z\} \cap  \left(\dm{\cap}_{i=1}^{n_0} \{Z(m_1+i) = c\}\right) \\ \cap \{\sum^{n_0}_{i=1} X(m_1+i) > L_{max}\}\\
E_2 =  \{Z(2 m_1 + n_0) = c'| Z(m_1 + n_0)= c\} \cap \\ \left(\vc{\cap}{2(m_1+n_0)}{i=2m_1+n_0} \{Z(i) = c\}\right) \cap \{\sum^{2(m_1+n_0)}_{i=2m_1+n_0} X(i) > L_{max}\}.
\end{gather*}
The event $E=E_1\cap E_2$ happens with probability at least $ (\delta'_2( \delta_2 P(c,c))^{n_0})^2 > 0$. We recall that the parameters $\bm{\theta}$ and $\bm{\theta}^s$ are fixed. Now we consider two separate conditions depending upon the pdf of inter arrival time $X$; letting $\delta''_2\equiv\mathbb{P}( X > \theta^s_{c'_t})$, 1) $ \delta''_2 = 1$, and 2) $\delta''_2< 1$. 

In subcase $1$ ($\delta''_2=1$), the content $c'$ is always evicted before it gets into the main cache $\mc{C}$. Therefore, following event $E$ the system ends in the state $\mc{S}(2(m_1+n_0))=\mc{S}_{rec}$, irrespective of the initial state. Where 
$$\mc{S}_{rec}= \begin{cases}\left(\bm{\Psi}=(0, \theta^s_{c_{typ}}, \theta_{c_{typ}})\bm{e}^{|\mcs{K}|}_c, Z = c\right)\text{ if } c\in \mc{K},\\ \left(\bm{\Psi}=(\bm{0}, \bm{0}, \bm{0}), Z=c\right) \text{ o/w.}\end{cases}.$$
Finally, letting  $\nu_{2(m_1+n_0)}(\cdot)= \delta'^2_2\delta^2_2 P(c,c)^{2n_0}\mc{P}(\mc{S}_{rec}, \cdot)$, we have $\mathfrak{S}$ to be a \emph{small} set with non zero measure $\nu_{2(m_1+n_0)}$. The theorem holds for $\gamma = (1-\delta'^2_2\delta^2_2 P(c,c)^{2n_0})^{\tfrac{1}{2(m_1+n_0)}}$.

In subcase $2$ ($\delta''_2<1$), consider additionally (to event $E$) the last inter arrival time is less than $\theta^s_{c'_t}$; i.e. $E'= E\cap\{X(2(m_1+n_0))\leq \theta^s_{c'_t}\}$. The event $E'$ happens with probability at least $(1-\delta''_2)(\delta'_2( \delta_2 P(c,c))^{n_0})^2$. Finally, the event $E'$ takes the system into the state $\mc{S}'(2(m_1+n_0))=\mc{S}'_{rec}$ or all initial conditions. In this case we have        
$$\mc{S}'_{rec}= \begin{cases}\left(\bm{\Psi}=(\theta_{c_{typ}}, 0, 0)\bm{e}^{|\mcs{K}|}_c, Z = c\right)\text{ if } c\in \mc{K},\\ \left(\bm{\Psi}=(\bm{0}, \bm{0}, \bm{0}), Z=c\right) \text{ o/w.}\end{cases}.$$
Here, letting  $\nu_{2(m_1+n_0)}(\cdot)= (1-\delta''_2)\delta'^2_2\delta^2_2 P(c,c)^{2n_0} \mc{P}(\mc{S}'_{rec}, \cdot)$, we have $\mathfrak{S}$ to be a \emph{small} set with non zero measure $\nu_{2(m_1+n_0)}$. The theorem holds for $\gamma = (1-(1-\delta''_2)\delta'^2_2\delta^2_2 P(c,c)^{2n_0})^{\tfrac{1}{2(m_1+n_0)}}$.
\color{black}
\end{proof}

\begin{remark}
Condition $C1$ is true for popular inter arrival distributions such as Poisson, Weibull, Pareto and Phase-type. The consideration of these two cases let us use the Paley-Zygmund type bound in case $C2$ without using a bound on second moment of the arrival process. Further, the presence of one self loop may be used to wipe out the history of all the other states. Therefore, two self loops in the Markov chain is sufficient in wiping out the entire history. In fact, it is necessary under the current assumptions on the inter arrival time distribution, as shown next.
\end{remark}

\begin{remark}[Non-mixing of a two-level Cache.]
\s{Now we present an example where, for some $\bm{\theta}$, $\bm{\theta}^s$ and inter arrival distribution $X$, ergodicity of the Harris chain does not hold if; 1) there is at most one self loop in irreducible and aperiodic DTMC representing process $Z$ and 2) $\mathbb{P}(X \geq L_{\max}) = 0$. Let there be only recurrent content of a single type and fix $L_{\max}=2$, $\theta = 1.5$ and $\theta^s =0.2$. Further, let the inter arrival time $X$ be distributed with absolutely continuous pdf and support $[0.3,0.4]$. Consider the labeling process, $Z$, given by the irreducible and aperiodic Markov chain; with three states $\{a,b,c\}$ and the transition probability given as $P(a,b)= P(a,a) = 0.5$, $P(b,a)=P(b,c)=0.5$ and $P(c,a)=1$. For the object $a$, corresponding to state $a$, the inter arrival time distribution $X_a$ has support of $[0.3,1.2]$. This implies if initially $a$ is in higher level cache it is never evicted from the higher level cache. On the contrary, if $a$ is initially in lower level cache, it never gets into the higher level cache.}  
\end{remark}

\paragraph{Stationary Properties} We now characterize the normalized size and hit rate  for the f-TTL caching algorithm under the static setting using results from renewal theory. This  encompasses the d-TTL algorithm as it is essentially identical to f-TTL for $\bm{\theta}^s=\bm{\theta}$.  The byte hit rate and object hit rate is differentiated by $\hat{w_c}$ in the analysis.
\begin{eqholder*}[t!]
\begin{equation}
\label{eq:hitrateAvg}
\bm{h}(\bm{\theta},\bm{\theta}^s)= \left\{\left(\dm{\sum_{\substack{c:c_{typ} = t\\ c \in \mc{K} }}} \widehat{w_c}\lambda_c \left(p_c^2\left( \theta_{t} \right) + \left(1-p_c\left(\theta_{t} \right)\right)p_c\left(\theta_t^s\right)\right)\right)\Bigg/\left( \widehat{\bar{w}_t}\alpha_t \lambda + \dm{\sum_{\substack{c: c_{typ} = t\\c \in \mc{K}}}} \widehat{w_c}\lambda_c\right)~,\forall t\in [T]\right\}
\end{equation}

\begin{equation}
\label{eq:sizerateAvg}
\bm{s}(\bm{\theta},\bm{\theta}^s) = \left\{\left( \dm{\sum_{\substack{c:c_{typ} = t \\ c \in \mc{K} }}}
w_c\lambda_c \left(p_c\left( \theta_{t} \right) \hat{s}_c\left(\theta_t\right) 
+ \left(1-p_c\left(\theta_{t} \right)\right)\hat{s}_c\left(\theta_t^s\right)\right)
+ \bar{w}_t\alpha_t\lambda\theta_t^s \right)\Bigg/
\left(\bar{w}_t\alpha_t\lambda + \dm{\sum_{\substack{c:c_{typ} = t \\ c \in \mc{K}} }}w_c\lambda_c\right)~,\forall t\in [T] \right\}
\end{equation}
Here $\hat{s}_c(\theta)= \int_{0}^{\theta} x dp_c(x) + \theta(1-p_c(\theta))$ and $\widehat{\cdot}$ differentiates between byte and object hit rates.

\rule{\textwidth}{0.5pt}
\end{eqholder*}

\begin{theorem}
\label{thm:Renewal}
\s{Under Assumption~\ref{a:Arr} with $\|\mathbf{L}\|_{\infty}$-rarity condition, the f-TTL algorithm with fixed parameters $\bm{\theta}\preceq \mathbf{L}$ and $\bm{\theta}^s\preceq \mathbf{L}$}, achieves the average hit rate ,  $\bm{h}(\bm{\theta},\bm{\theta}^s)$~\eqref{eq:hitrateAvg} a.s. and the average normalized size,  $\bm{s}(\bm{\theta},\bm{\theta}^s)$~\eqref{eq:sizerateAvg} a.s.
\end{theorem}
\begin{proof}
\textbf{Hit Rate.} In order to characterize the hit rate  under the arrival process in \ref{a:Arr}, we first show that the combined hit rate  and virtual hit rate  contribution of \emph{rare content} arrivals becomes zero almost surely.
\begin{claim}
	\label{clm:rarehit}
	Under fixed TTLs, $\bm{\theta}\preceq \mathbf{L}$ and $\bm{\theta}^s\preceq \mathbf{L}$, and Assumption~\ref{a:Arr} with $\|\mathbf{L}\|_{\infty}$-rarity condition, the hit rate of `rare content' asymptotically almost surely (a.a.s.) equals zero.
\end{claim}
\begin{proof}  
	We first note that on the $l$-th arrival we obtain a hit from a rare content of type $t$ only if it is $L_{\max}$-rare content of some type $t$, i.e. the arrival satisfies the two conditions: 1) the $l$-th arrival is of a rare content $c$ of type $t$ and 2) the previous arrival of the same rare content $c$ happened (strictly) greater than $L$ units of time earlier. Therefore, the hit event from rare content on $l$-th arrival is upper bounded by the indicator $\sum_{t\in[T]}\beta_t(l;L_{\max})$ ($\beta_t(l)$ for brevity) and the asymptotic hit rate of all the rare objects combined is upper bounded as  
	\begin{multline*}
		\dm{\lim_{n\to \infty}} \frac{1}{n}\sum_{l=1}^n \sum_{t\in [T]}\beta_t(l) \\
		=  \sum_{t\in [T]} \dm{\lim_{m\to \infty}}\dm{\lim_{n\to \infty}}\left(\frac{1}{n}\dm{\sum_{l=1}^{m-1}} \beta_t(l)+ \frac{1}{n}\sum^{n}_{l= m}\beta_t(l)\right) \\
		\leq \sum_{t\in [T]} \dm{\lim_{m\to \infty}}\dm{\lim_{n\to \infty}}\frac{1}{n-m}\sum^{n}_{l = m}\beta_t(l) = 0.~\text{w.p.} 1
	\end{multline*}
	The first term in the addition goes to zero by first taking the limit over $n$ for finite $m$, whereas, the second term goes to zero, for each type separately, as a consequence of the $\|\mathbf{L}\|_{\infty}$-rarity condition~\eqref{eq:rarity} and the adaptation $\bm{\theta}\preceq \mathbf{L}$ and $\bm{\theta}^s\preceq \mathbf{L}$.
	\color{black}
\end{proof}

Recall the arrival process notations presented earlier. To characterize the hit rates it suffices to consider only the \emph{recurring objects}, due to the above claim. For the $j$-th arrival (for $j\geq1$) of object $c$ of type $t$, the indicator to hit event in cache $\mc{C}$ is given as  $\mathbbm{1}_c(j)= \mathbbm{1}\left( X_c(j)\leq \theta_{t}\right)\mathbbm{1}\left( X_c(j-1)\leq \theta_{t}\right)$ and (real) hit in $\mc{C}_s$ is 
$\mathbbm{1}^s_c(j)= \mathbbm{1}\left( X_c(j)\leq \theta^s_{t}\right)\mathbbm{1}\left( X_c(j-1)>\theta_{t}\right)$. The hit from the first arrival from $c$ is (with slight abuse of notation) $\mathbbm{1}_c(0)=\mathbbm{1}\left( A_c(0) \leq \max\{\psi^0_c(0), \psi^1_c(0)\}\right)$.
We can represent the hit rate  under f-TTL at time $u > 0$  is given as, 
\begin{equation}
\label{eq:hitexp1}
h_{t}(u) = \frac{\dm{\sum_{\substack{c:c_{typ} = t\\c \in \mc{K}}}}\widehat{w_c}\left( \tfrac{\mathbbm{1}_c(0)}{u}  +  \frac{1}{u}\dm{\sum_{j=1}^{N_c(u)-1}}\left(\mathbbm{1}_c(j)+\mathbbm{1}^s_c(j) \right) \right)}
{\dm{\sum_{\substack{c:c \in \mc{U}, c_{typ} = t}}} \tfrac{\widehat{w_c} N_c(u)}{u}}.
\end{equation}

Due to the strong Markov property of the regeneration cycles given by the hitting times of state $c$, $X_c(j)$ are i.i.d. for all $j\geq 1$ and for all object $c\in \mc{K}$.  Therefore, it forms a renewal process~\cite{smith1958renewal}. For $c\in \mc{K}$, as $u\to \infty$, $N_c(u)/u \to \lambda_c$ from elementary renewal theorem. In a similar way, for each type $t$, as $u\to \infty$, $\dm{\sum_{c\notin \mc{K}:c_{typ}=t}} N_c(u)/u \to \alpha_t\lambda$. Finally, due to the renewal reward theorem the asymptotic hit rate  for type $t$ with f-TTL is almost surely same as given in the theorem (the expression is in a placeholder).      

\textbf{Normalized size.} Next we try to characterize the normalized size of the process, that is the size per arrival in the system.
The proof follows a similar idea but the only difference being the \emph{rare objects} have non negligible contribution towards normalized size for nonzero $\bm{\theta}^s$. For the $j$-th arrival of the object $t$ the time it is in the cache is given as 
$s_c(j)=\min\{\theta_t, X_c(j+1)\}\mathbbm{1}(X_c(j)\leq\theta_t)+\min\{\theta^s_t, X_c(j+1)\}\mathbbm{1}(X_c(j)>\theta_t)$.    We can represent the normalized size under f-TTL at time $u > 0$  is given as, 
\begin{equation}
\label{eq:hitexp1}
s_{t}(u) = \frac{\dm{\sum_{\substack{c:c_{typ} = t\\c \in \mc{U}}}}w_c\left( \tfrac{s_c(0)+ s_c(\st{N_c(u)-1})}{u}   +  \frac{1}{u}\dm{\sum_{j=1}^{N_c(u)-2}}s_c(j)  \right)}
{\dm{\sum_{\substack{c:c \in \mc{U}, c_{typ} = t}}} \tfrac{w_c N_c(u)}{u}}.
\end{equation}
However if $c$ is a \emph{rare content} of type $t$ we additionally have that the first term is almost surely $0$ and the second term is almost surely $\theta^s_t$. This is because the hit rate  and virtual hit rate  both are equal to zero, almost surely for a rare content. This simplifies the normalized size contribution of the rare object of type $t$ to $\bar{w}_t\theta^s_t$.  For the stationary objects using renewal reward theorem we can obtain the rest of the terms in the expression as given in the theorem statement (the expression is in a placeholder).

Note:  At each arrival of object $c$ we add to $s(l)$ the maximum time this object may stay in the cache while we subtract the time remaining in the timer. This subtraction compensates for the possible overshoot in the previous arrival. Therefore, up to time $u$ the sample-path wise addition of the term $s(l)$ for a object $c$ is identical to the total bytes-sec the object has been in the cache with a small overshoot for the last arrival.  Therefore, taking weighted average w.r.t. the arrival rates  for object $c$ we obtain the normalized size of the system.   Also in the limit $u\to \infty$ the contribution of the overshoot towards normalized size becomes zero. Therefore, the average over $l$ of the term $s(l)$ in the equation~\eqref{eq:ttldifAdapt} represents the normalized size almost surely. 
\end{proof}

The `$\mathbf{L}$-feasible' hit rate  region for  d-TTL caching and f-TTL caching under arrival \ref{a:Arr} is given in the following corollary which follows directly from the definitions. 
\begin{corollary}
\label{thm:Capacity}
\s{Under Assumption~\ref{a:Arr} with $\|\mathbf{L}\|_{\infty}$-rarity condition}, the set of `$\mathbf{L}$-feasible' hit rate  for d-TTL caching algorithm is $$\left\{\bm{h}(\bm{\theta},\bm{\theta}): \bm{\theta}\preccurlyeq \mathbf{L}\right\}.$$ 
Moreover, for f-TTL caching algorithm the set of `$\mathbf{L}$-feasible' hit rate , normalized size tuples is 
$$\left\{\left(\bm{h}(\bm{\theta},\bm{\theta}^s),\bm{s}(\bm{\theta},\bm{\theta}^s)\right): \bm{\theta}\preccurlyeq \mathbf{L},\bm{\theta}^s\preccurlyeq \bm{\theta}\right\}.$$
\end{corollary}

We end this section with the following proposition that gives some properties of the functions $\bm{h}(\bm{\theta},\bm{\theta^s})$ and $\bm{s}(\bm{\theta},\bm{\theta}^s)$.
\begin{proposition}\label{prop:funcproperty}
Under Assumption~\ref{a:Arr} for each $t$, the functions $h_t(\theta_t,\theta^s_t)$ and $s_t(\theta_t,\theta^s_t)$ are continuously differentiable for all $\theta_t\geq 0$, $\theta_t^s\geq 0$. Moreover, the function $h_t(\theta_t,\theta^s_t)$ is 1) strictly increasing (up to value $1$) in $\theta_t$ for $\theta_t\geq \theta^s_t \geq 0$ and 2) non decreasing in $\theta^s_t$.
\end{proposition}
\begin{proof}
The inter arrival time of any object $c\in \mc{K}$ is given as $$X_c = \sum_n \mathbb{P}(\inf\{l: Z(l)=c\wedge Z(0)=c\}=n) X^{(n)},$$ where $X^{(n)}$ is the sum of $n$ independent copies of the r.v. $X$ (inter arrival time). The first term represents the first passage time p.m.f. for the object $c$ under the Markovian labeling. As the p.d.f. of $X$ is absolutely continuous by Assumption~\ref{a:Arr}, for each $c\in \mc{K}$ the p.d.f. of $X_c$ is absolutely continuous. Further, the terms $p_c(\cdot)$ are c.d.f. of absolutely continuous random variables (a.c.r.v.) w.r.t Lebesgue measure. This means that both $h_t(\theta_t,\theta^s_t)$ and $s_t(\theta_t,\theta^s_t)$ are both continuously differentiable for all $\theta_t\geq 0$ and $\theta_t^s\geq 0$.

The p.d.f. of $X$, the inter arrival time, has simply connected support by assumption. Due to the continuity property of the convolution operator we can conclude that the p.d.f. of $X_c$ also has a simply connected support for each $c\in\mc{K}$. This means the c.d.f.-s $p_c(\cdot)$ are strictly increasing in $\theta_{c_{typ}}$ till it achieves the value $1$ and then it remains at $1$. We have the following,
\begin{align*}
\tfrac{\partial}{\partial \theta_t}\left(p_c^2\left( \theta_{t} \right) + \left(1-p_c\left(\theta_{t} \right)\right)p_c\left(\theta_t^s\right)\right)\\
= \left(2p_c\left( \theta_{t} \right) - p_c\left(\theta_t^s\right) \right)p'_c\left( \theta_{t} \right).\\
\tfrac{\partial}{\partial \theta^s_t}\left(p_c^2\left( \theta_{t} \right) + \left(1-p_c\left(\theta_{t} \right)\right)p_c\left(\theta_t^s\right)\right)\\
=\left(1-p_c\left(\theta_{t} \right)\right)p'_c\left(\theta_t^s\right).
\end{align*}
As $p'_c(\cdot)$ have simply connected support for all $c\in \mc{K}$,  given $h_t<1$ and $\theta_t\geq \theta^s_t \geq 0$ the function $h_t$ is strictly increasing in $\theta_t$. Also $h_t$ is trivially non decreasing in $\theta_t^s$.
\end{proof}

%% file: dynamic.tex
\subsection{Dynamic Analysis}
We are now ready to prove our main result, Theorem~\ref{thm:SAConv} which shows the d-TTL and f-TTL under adaptation~\eqref{eq:psiAdapt} and \eqref{eq:psiSAdapt} resp., achieves any `$\mathbf{L}$-feasible' hit rate where $\mathbf{L}$ is a parametric input. Our proof relies upon the tools developed in stochastic approximation theory. See \cite{kushner2003stochastic} and the references therein for a self-contained overview of stochastic approximation theory. Also, see \cite{borkar1997stochastic, bhatnagar2001two} for results on multi-timescale stochastic approximation.

\begin{proof}[\textbf{Proof of Theorem~\ref{thm:SAConv}}]
We consider the d-TTL and f-TTL of Section~\ref{sec:cachingalgorithms} and the arrival process as given in Assumption \ref{a:Arr} in Section~\ref{ssec:Arr}. The proof of stochastic approximation is based on the convergence of two timescale separated system using ODE methods. It is adapted from~\cite{borkar1997stochastic}, \cite{bhatnagar2009natural}, and \cite{kushner2003stochastic}.  

First we state the required conditions for the convergence of stochastic approximation. The conditions are standard and they ensure that,
\begin{enumerate}
\item The step sizes allow appropriate time scale separation,
\item The expected (w.r.t. $\mc{F}(l)$) updates are bounded and smooth,
\item Under fixed parameters the system exhibits strong ergodicity,
\item Noise effects are negligible over large enough time windows,
\item The faster timescale is approximated well by a projected ODE with certain `nice' properties.
\end{enumerate}
These conditions together lead to the approximation of the whole system by a `nice' projected ODE evolving in the slower timescale and its convergence depicts the convergence of the parameters $\bm{\vartheta}(l)$ and $\bm{\vartheta}^s(l)$ (in almost sure sense).\footnote{\color{blue}More detailed discussions on the conditions and the adaptation of the proof are provided as a remark after the proof.} {\color{blue} Further, the almost sure convergence in the parameters $\bm{\vartheta}(l)$ and $\bm{\vartheta}^s(l)$, implies almost sure convergence of the hit rate and the normalized size of the system, as they are continuous and bounded functions of $\bm{\vartheta}$ and $\bm{\vartheta}^s$.}

\begin{condition} [Adapted from~\cite{borkar1997stochastic},~\cite{bhatnagar2009natural}, and \cite{kushner2003stochastic}] Sufficient conditions for convergence:

\begin{itemize}
\item[$A0$.] The step sizes satisfy \\ 1) $\sum_l \eta(l)= \sum_l \eta_s(l) =\infty$, \\ 2) $\sum_l \eta(l)^2 <\infty$, $\sum_l \eta_s(l)^2 <\infty$ and \\ 3) $\dm{\lim_{l\to \infty}} \eta(l)=\dm{\lim_{l\to \infty}} \eta_s(l) = \dm{\lim_{l\to \infty}} \tfrac{\eta_s(l)}{\eta(l)}= 0$. 

\item[$A1$.] The update terms at each iteration satisfy \\ 1) $sup_l \mathbb{E}\|\widehat{w(l)}Y(l)\mathbf{e}^T_{t(l)}\|_1 < \infty$ and  \\ 2) $sup_l \mathbb{E}\|w(l)s(l)\mathbf{e}^T_{t(l)}\|_1 < \infty$.

\item[$A2$.] For any sequence $\{b(l)\}$ we define $m_{b}(n)=\min\{k:  \dm{\sum_{l=1}^{k+1}}b(l)> n\}$. 
For $b(\cdot)\in \{  \eta_s(\cdot), \eta(\cdot)\}$ the noise terms satisfy,\\ 1) $\dm{\lim_{n\to \infty}} \dm{\sum_{l= m_{b}(n)}^{m_{b}(n+1)-1}} b(l) \beta_h(l)=0$ a.s. and \\ 2)$\dm{\lim_{n\to \infty}} \dm{\sum_{l= m_{b}(n)}^{m_{b}(n+1)-1}} b(l) |s_2(l)| = 0$ a.s.

\item[$A3$.] The set $H$ is a hyper-rectangle. 

\item[$A4$.] There are non negative, measurable functions $\rho_1(\cdot)$ and $\rho_1^s(\cdot)$ of $\bm{\vartheta}$ and $\bm{\vartheta}^s$ such that $\| \mathbf{g}(\bm{\vartheta}, \bm{\vartheta}^s, \mc{S})\|_1\leq \rho_1(\bm{\vartheta}, \bm{\vartheta}^s)$ and $\| \mathbf{g}^s(\bm{\vartheta}, \bm{\vartheta}^s, \mc{S})\|_1\leq \rho_1^s(\bm{\vartheta},\bm{\vartheta}^s)$, uniformly over all $\mc{S}$. Moreover the functions $\rho_1(\cdot)$ and $\rho_1^s(\cdot)$ are bounded for bounded $(\bm{\vartheta},\bm{\vartheta})$-set.

\item[$A5$.] There are non negative, measurable and bounded functions  $\rho(\cdot)$ and $\rho^s(\cdot)$  such that  
$\dm{\lim_{\bm{x},\bm{y} \to \mathbf{0}}} \rho(\bm{x},\bm{y})=0$ and $\dm{\lim_{\bm{x},\bm{y} \to \mathbf{0}}} \rho^s(\bm{x},\bm{y}) = 0$. Moreover for all $\mc{S}$, $$\| \mathbf{g}(\bm{\vartheta}, \bm{\vartheta}^s, \mc{S})- \mathbf{g}(\bm{\vartheta}', \bm{\vartheta}'^s, \mc{S}) \|_1 \leq \rho(\bm{\vartheta}-\bm{\vartheta}', \bm{\vartheta}^s-\bm{\vartheta}'^s ).$$ 
Also similar inequality holds for $\bm{g}^s(\cdot)$ with $\rho^s(\cdot)$ uniformly for all $\mc{S}$.

\item[$A6$.] Let $\bar{\mathbf{g}}(\cdot)$ be a continuous function of $\bm{\vartheta}$ and $\bm{\vartheta}^s$. Fix any $\bm{\vartheta}$ and $\bm{\vartheta}^s$. Let $\bm{\xi}(l)=\mathbf{g}(\bm{\vartheta},\bm{\vartheta}^s, \mc{S}(l))- \bar{\mathbf{g}}(\bm{\vartheta},\bm{\vartheta}^s)$. For some constant $C>0$ and all $n$ the conditions, 1)$\mathbb{E}\bm{\xi}(n) = 0$ and 2) $ \sum_{i=n}^{\infty} |\mathbb{E}_n\bm{\xi}(l)|\leq C$, hold w.p. $1$.
   
\item[$A7$.] For each fixed $\bm{\vartheta}^s$ there exists a unique $\bm{\vartheta}(\bm{\vartheta}^s)$ which correspond to a unique globally asymptotically stable point of the projected ODE,  $\dt{\bm{\vartheta}} = \bar{\mathbf{g}}(\bm{\vartheta},\bm{\vartheta}^s) + \bm{r}^s(\bm{\vartheta},\bm{\vartheta}^s)$. Here $\bm{r}^s(\cdot)$ represents the error due to projection on $H$. 

\item[$A8$.] The functions  $\mathbf{g}^s(\bm{\vartheta}(\bm{\vartheta}^s), \bm{\vartheta}^s, \mc{S})$ follows inequalities similar to the ones in $A4$ and $A5$ for functions $\rho_2(\cdot)$ and $\rho_3(\cdot)$, respectively. 

\item[$A9$.] There exists a $\bar{\mathbf{g}}^s(\cdot)$ which is a continuous function of $\bm{\vartheta}^s$ such that statement identical to $A6$ holds  for the sequence,  $$\{\mathbf{g}(\bm{\vartheta}(\bm{\vartheta}^s), \bm{\vartheta}^s, \mc{S}(l)):l\in \mathbb{N}\}.$$

\item[$A10$.] There exist a twice continuously differentiable real valued function $f(\bm{\vartheta}^s)$ such that 
$\bar{\bm{g}}^s(\bm{\vartheta}^s)= -\nabla f(\bm{\vartheta}^s)$ for $\bm{h}^*$ which is $(1-2\epsilon)\mathbf{L}$-feasible.  
\end{itemize}
\label{a:stochApprox}
\end{condition}

We first show that all the Assumptions hold for our system. 

\begin{lemma}
\label{lemm:Assumption}
\s{Under TTL upper bound $\mathbf{L}$ and Assumption~\ref{a:Arr} with $\|L\|_{\infty}$-Rarity condition}, the conditions~\ref{a:stochApprox}.$A0$ to \ref{a:stochApprox}.$A10$ hold for the f-TTL and d-TTL caching. 
\end{lemma}
\begin{proof}
This is the key lemma that allows us to use the standard results in the multi-timescale stochastic approximation literature. The assumption that both number of recurring objects and the number of types of rare objects in the system are finite, along with the bound $\mathbf{L}$ play a crucial role in proving most of the boundedness assumptions. Furthermore, the `rarity' condition helps us in controlling the noise from non-stationary arrival. Finally, the most important part of the proof lies in the analysis of the two ODEs representing the evolution of $\bm{\vartheta}$ and $\bm{\vartheta}^s$.

$\mathbf{A0}$: Follows from definition of $\eta(l)$ and $\eta_s(l)$ for all $l\in \mathbb{N}$.

$\mathbf{A1}$: The assumption $A1$ is satisfied because both $Y(l)$ and $s(l)$ are well behaved. Specifically, 
\begin{align*}
\sup_l \mathbb{E}\|\widehat{w(l)}Y(l)\mathbf{e}^T_{t(l+1)}\|_1 &\leq w_{\max},\\
\sup_l \mathbb{E}\|w(l)s(l)\mathbf{e}^T_{t(l)}\|_1 &\leq  \sup_l w(l)(\theta_{t(l)} + \psi(l))\\
&\leq 2w_{\max}L_{\max}.
\end{align*}

$\mathbf{A2}$:  \color{black} Recall $\beta_t(l;R)$ is the indicator of the event that: (1) the $l$-th arrival is of a rare object $c$ of type $t$, \textit{and} (2) the previous arrival of the same rare object $c$ happened (strictly) less than $R$ units of time earlier, as given in the definition of rarity condition~\eqref{eq:rarity}. Let $R= L_{\max}$ and for brevity denote $\beta_t(l;L_{\max})$ as $\beta_t(l)$, for each type $t$ and arrival $l$. The event that at $l$-th arrival it is a hit from a rare object, $\beta_h(l)$ is upper bounded by $\dm{\sum_{t\in [T]}}\beta_t(l)$ with probability $1$. Further, $s_2(l)$ is non zero only if the $l$-th arrival results in a virtual hit or a hit from a rare object. This implies $s_2(l)$ is also upper bounded by $2L_{\max}\dm{\sum_{t\in[T]}}\beta_t(l)$ a.s.  Therefore, we have 
\begin{multline*}
\dm{\sum_{l= m_{\eta}(n)}^{m_{\eta}(n+1)-1}} \eta(l) \max\{|s_2(l)|, \beta_h(l)\}\\ 
\leq  \max\{1, 2L_{\max}\}\dm{\sum_{t\in [T]}}  \dm{\sum_{l= m_{\eta}(n)}^{m_{\eta}(n+1) -1}} \eta(l)\beta_t(l).
\end{multline*}
From the definitions, for $\alpha\in(1/2,1)$, we obtain $m_{\eta}(n)= (\tfrac{(1-\alpha)n}{\eta_0})^{1/(1-\alpha)} \pm \Theta(1)$. Therefore, there are $O(n^{\tfrac{\alpha}{1-\alpha}}) = O(m_{\eta}(n)^{\alpha})$ terms in the above summation and we have the following bound
$$\dm{\sum_{t\in [T]}} \dm{\sum_{l= m_{\eta}(n)}^{m_{\eta}(n+1) -1}} \eta(l)\beta_t(l) \leq 
\dm{\sum_{t\in [T]}}\tfrac{\eta_0}{m_{\eta}(n)^{\alpha}}\dm{\sum_{l= m_{\eta}(n)}^{\substack{m_{\eta}(n)+ \\ O(m_{\eta}(n)^{\alpha})}}} \beta_t(l).$$
As $n$ goes to $\infty$, $m\equiv m_{\eta}(n)\to \infty$ and $N^t_m\equiv m_{\eta}(n)^{\alpha} = \omega(\sqrt{m})$ for $\alpha\in (1/2,1]$ and $t\in [T]$. Therefore, due to $\|\mathbf{L}\|_{\infty}$-rarity condition~\eqref{eq:rarity} and the finiteness of $T$, the above term goes to $0$ almost surely as $n\to \infty$. This proves the assumption with step size $\eta(\cdot)$. The assumption with step size $\eta_s(\cdot)$ can be proved similarly.
\color{black}

$\mathbf{A3}$: Follows from definition of the set $H$.

$\mathbf{A4}$: From the definitions $\| \mathbf{g}(\bm{\vartheta}, \bm{\vartheta}^s, \mc{S})\|_1\leq 2 Tw_{\max}$ easily follows. Further $\| \mathbf{g}^s(\bm{\vartheta}, \bm{\vartheta}^s, \mc{S})\|_1\leq 2T w_{\max} L_{\max}$. 

$\mathbf{A5}$: Using equations~\eqref{eq:avgMissFunc} and ~\eqref{eq:avgSizeFunc} we bound the terms in assumption $A5$. 
Fix any state $\mc{S}$ and let $Z = c$ for this state. First note that given the state $\mc{S}$ the function $\mathbf{g}(\bm{\vartheta}, \bm{\vartheta}^s, \mc{S})$ is independent of $\bm{\vartheta}, \bm{\vartheta}^s$ so $A5$ holds for $\rho = 0$.

For a given $\mc{S}$, the function  $\mathbf{g}^s(\bm{\vartheta}, \bm{\vartheta}^s, \mc{S})$ is linear w.r.t.  $\bm{\theta}^s$ and   $\bm{\theta}$. However, the term $\bm{\theta^s}$ has nonlinear dependence on $\bm{\vartheta}^s$ and $\bm{\vartheta}$, with bounded slope $1$ and $1/\epsilon$, respectively. It is easy to verify that the following function will satisfy the conditions in $A5$ 
$$\rho^s(\bm{x},\bm{y}) = w_{\max}L_{\max}\left((1+1/\epsilon)\|\bm{x}\|_1+2\|\bm{y}\|_1\right).$$

$\mathbf{A6}$: To show the validity of Assumption $A6$ we use Theorem~\ref{thm:HarrisRecur}. 
Note that $\mathbf{g}(\bm{\vartheta}, \bm{\vartheta}^s, \mc{S})$ is a measurable and bounded function satisfying the inequality~\eqref{eq:uniformErgodic} for $C = 2Tw_{\max}$, w.r.t. $\|\cdot\|_1$ norm. 
Further, due to the existence of $\mu^*$ and the above boundedness we obtain the average as 
$\bar{\mathbf{g}}(\bm{\vartheta}, \bm{\vartheta}^s)= \mathbb{E}\mathbf{g}(\bm{\vartheta}, \bm{\vartheta}^s, \mc{S})$. From the ergodicity of the system we obtain $\|\mathbb{E}\bm{\xi}(l)\|_1 = 0$. Moreover due to uniform ergodicity we have 
$ \|\mathbb{E}_n\bm{\xi}(l)\|_1 \leq 2T\gamma^{(l-n)}$ for all $l\geq n$. The bound 
$\sum_{i=n}^{\infty} |\mathbb{E}_n\bm{\xi}(l)|\leq 2T/(1-\gamma)$ follows easily.

We now show that the function $\bar{\mathbf{g}}(\bm{\vartheta}, \bm{\vartheta}^s)$ is a continuous function of $\bm{\vartheta}$ and $\bm{\vartheta}^s$. Let $\bm{\theta}$ and $\bm{\theta}^s$ be the corresponding parameters. We claim that,
\begin{align*}
\bar{\mathbf{g}}(\bm{\vartheta}, \bm{\vartheta}^s) =
\dm{\sum_t}  \left(\dm{\sum_{\substack{c:c_{typ}=t}}} \widehat{w_c}\pi(c)\right)\left( h^{*}_t  - h_{t}(\theta_t, \theta_t^s) \right) \mathbf{e}^T_t.
\end{align*}
Here $\bm{h}(\bm{\theta}, \bm{\theta}^s)$ is as given in \eqref{eq:hitrateAvg}. To see the validity of the claim notice that the term $\mathbb{E}\,\widehat{w(l)}Y(l)$ almost surely converges to the average hit rate of the system which is obtained in Theorem \ref{thm:Renewal}.   From Proposition~\ref{prop:funcproperty}, for each $t$, $h_{t}(\theta_t, \theta_t^s)$ is continuous with respect to $\bm{\theta}$ and $\bm{\theta}^s$. But we know that $\bm{\theta}$ is identical to  $\bm{\vartheta}$ up to scale and $\bm{\theta}^s$ is a continuous function of both $\bm{\vartheta}$ and $\bm{\vartheta}^s$ from definition. Additionally we know that composition of continuous functions are continuous. Consequently $\bar{\mathbf{g}}(\bm{\vartheta}, \bm{\vartheta}^s)$ is continuous in $\bm{\vartheta}$ and $\bm{\vartheta}^s$.  

$\mathbf{A7}$:
For the rest of the proof of assumption $A7$, for notational convenience we represent the average hit rate as $\bar{\mathbf{g}}(\bm{\vartheta})$, by fixing and dropping the term $\bm{\vartheta^s}$ .

The function $\bar{\mathbf{g}}(\bm{\vartheta})$ is separable in types and it is the negative derivative of a real valued continuously differentiable function $f(\bm{\theta})= - \dm{\sum_t} \int g_t(\vartheta_t)d\theta_t$. This implies that the limit points of the projected ODE, $\dt{\bm{\theta}} = \bar{\mathbf{g}}(\bm{\vartheta}) + \mathbf{r}(\bm{\theta})$ ($\bm{\vartheta}^s$ suppressed) are the stationary points  $\dt{\bm{\theta}} = \mathbf{0}$, lying on the boundary or in the interior of $H$. We next show that the stationary point under the assumptions of `typical' hit rate is unique. 

For the fixed $\bm{\vartheta}^s$ and fixed $t$ we can represent $\theta_t^s$ as a continuous and strictly increasing function of $\vartheta_t$. The parameter $\theta_t$ is scaled version of $\vartheta_t$. Further, $\theta^s_t$ is increasing in $\vartheta_t$ for $\vartheta^s_t\neq 0$ and  is non decreasing in $\vartheta_t$ for $\vartheta^s_t=0$. Therefore, due to the Proposition~\ref{prop:funcproperty} we can conclude that each function $h_t(\theta_t, \theta_t^s)$ strictly increases (w.r.t. $\vartheta_t$) till the function reaches value $1$ and after that that it remains at $1$. Intuitively, for a fixed $\vartheta^s_t$ increasing $\vartheta_t$ increases the hit rate as the objects are being cached with higher {\ttl} values in cache $\mc{C}$ and cache $\mc{C}_s$.

Now for each type $t$ there can be two cases, 1) we achieve target hit rate $h_{t}^{*}$ or 2) we achieve a hit rate less than $h_{t}^{*}$. The achieved hit rate can not be higher than that of the target hit rate due to the monotonicity and continuity property of the functions  $h_t(\theta_t, \theta_t^s)$ for a fixed $\bm{\vartheta}^s$.  In case (1) we find a $\vartheta_t^{*}\in H$, such that $h_t(\theta_t^*, \theta^{s}_t) = h^*_t$. Finally, due to the assumption that $\mathbf{h}^*$ is a `typical' hit rate, i.e. $h^*_t < 1$ for all $t$, and the monotonicity property of $h_t(\cdot)$ w.r.t. $\vartheta_t$ we conclude that there is a unique $\vartheta_t^*$ that achieves this $\mathbf{h}^*$.  However, in case (2) the ODE gets pulled to the boundary of the set by the reflection term, i.e. $\vartheta_t = 1$ and $\theta_t=L_t$. Therefore, the parameter $\vartheta^*_t$ can be expressed as a function of $\vartheta^s_t$.

\textbf{The function $\bm{\vartheta}(\bm{\vartheta}^s)$.} We now present and analyze the properties of the function $\bm{\vartheta}(\bm{\vartheta}^s)$. For each $t$ and a target hit rate $\bm{h}^*$, the function is defined separately and implicitly as 
\begin{equation*}
\label{eq:vartheta}
\vartheta_t(\vartheta_t^s) =\min\left(\{\vartheta_t: h_t\left(L_t\vartheta_t, L_t\vartheta_t\up(\st{\vartheta_t,\vartheta_t^s; \epsilon})\right) = h^*_t\}\cup L_t\right).
\end{equation*}
Intuitively, changing the value of $\vartheta_t$ will change the hit rate from the desired value of $h^*_t$. Taking derivative of $h_t\left(L_t\vartheta_t, L_t\vartheta_t\up(\vartheta_t,\vartheta_t^s)\right)$ w.r.t. $\vartheta_t$, yields the bound 
$$\tfrac{\partial h_t}{\partial \vartheta_t} > \dm{\sum_{\substack{c:c\in \mc{K}\\c_{typ}=t}}} \widehat{w_c}\lambda_c p_c(L_t\vartheta_t) p'_c(L_t\vartheta_t)\Bigg/\dm{\sum_{\substack{c:c_{typ}=t}}} \widehat{w_c}\lambda_c.$$

Proposition~\ref{prop:funcproperty} states that $h_t(\cdot)$ is continuously differentiable in $\theta_t$ and $\theta_t^s$. By definition $\theta_t$ and $\theta_t^s$ are continuously differentiable in $\vartheta_t$ and $\vartheta_t^s$. However, the space of continuously differentiable functions is closed under composition. Therefore, $h_t\left(L_t\vartheta_t, L_t\vartheta_t\up(\vartheta_t,\vartheta_t^s)\right)$ is continuously differentiable function of $\vartheta_t$ and $\vartheta_t^s$. We also note that 1) the p.d.f. of the inter arrival time for each object $c$ has simply connected support and 3) for a `typical' target $\bm{h}^*$, all $t$ and all $\vartheta_t(\vartheta^s_t)$, we have $\dm{\sum_{\substack{c:c\in \mc{K}, c_{typ}=t}}} \widehat{w_c}\lambda_c p_c(L_t\vartheta_t)p'_c(L_t\vartheta_t)> 0$. Applying a version of global implicit function theorem, as stated in Corollary $1$ in~\cite{sandberg1981global}, we conclude that for each $t$ there exists a unique continuously differentiable function  $\tilde{\vartheta}_t(\vartheta_t^s)$  which gives $h_t\left(L_t\tilde{\vartheta}_t(\vartheta_t^s), L_t\tilde{\vartheta}_t(\vartheta_t^s)\up(\tilde{\vartheta}_t(\vartheta_t^s),\vartheta_t^s)\right)=h^*_t$.\footnote{In corollary $1$~\cite{sandberg1981global}, the condition (1) is true due to the existence of a $\vartheta_t$ for all $\vartheta^s_t$ for `typical' hit rate. The condition (3) holds as the functions $h_t(\cdot)$ are all measurable.} Finally, the pointwise maximum of a constant function and a continuous function is continuous and uniquely defined. Therefore, we conclude that for each $t$, $\vartheta_t(\vartheta_t^s)=\max\{\tilde{\vartheta}_t(\vartheta_t^s), 1\}$ is unique and continuous.   

\textbf{Special case of d-TTL.} Until this point it has not been necessary to differentiate the d-TTL cache from f-TTL cache. However, any hit rate that is $\bm{L}$-feasible will be achieved under the d-TTL scheme. This is not true for the f-TTL in general. Specifically, we note that  for a $\bm{L}$-feasible hit rate target the case (2) above will not happen under d-TTL caching for any type $t$. By definition if a hit rate is not achievable under d-TTL with $\theta_t = L_t$,  for at east one $t$, then it is not $\bm{L}$-feasible. 

\textbf{f-TTL Continuation:} The remaining part of this proof deals with assumptions on the f-TTL caching algorithm as for d-TTL algorithm $\bm{\vartheta}^s=\bm{1}$ and no further analysis is necessary.  
 
$\mathbf{A8}$: We have $\| \mathbf{g}^s(\bm{\vartheta}(\bm{\vartheta}^s), \bm{\vartheta}^s, \mc{S})\|_1\leq 2T w_{\max}L_{\max}$ directly from $A4$. The function $\bm{\vartheta}(\bm{\vartheta}^s)$ as noted earlier is component wise separable and  it is continuous in each component $\vartheta_t^s$. Therefore, using the bound in $A5$ we can prove similar bounds for $\mathbf{g}^s(\cdot)$. 

$\mathbf{A9}$: For the function $\bar{\mathbf{g}}^s(\bm{\vartheta}^s)=\mathbb{E}\mathbf{g}^s(\bm{\theta}(\bm{\vartheta}^s), \bm{\vartheta}^s, \mc{S})$ the statements identical to $A6$ hold with a different constant $TL_{\max}$. The continuity of $\bar{\mathbf{g}}^s(\bm{\vartheta}^s)$ follows from the fact that of $\|\mathbf{g}^s(\bm{\theta}(\bm{\vartheta}^s), \bm{\vartheta}^s, \mc{S})- \mathbf{g}^s(\bm{\theta}(\bm{\vartheta'}^s), \bm{\vartheta'}^s, \mc{S})\|_1\to 0$ as  
$\bm{\vartheta}^s\to \bm{\vartheta'}^s$, uniformly over $\mc{S}$.

Similar to the hit rate calculation we can compute the function $\bar{\mathbf{g}}^s(\bm{\vartheta}^s)$ using $\bm{s}(\bm{\theta},\bm{\theta}^s)$ from equation~\eqref{eq:sizerateAvg} in Theorem~\ref{thm:SAConv} as follows,
$$\bar{\mathbf{g}}^s(\bm{\vartheta}^s) = \dm{\sum_t}\left(\dm{\sum_{\substack{c:c_{typ}=t}}} w_c\pi(c)\right) \left(s^*_t - s_t(\theta_t(\vartheta^s_t), \theta_t^s(\vartheta^s_t)\right)\mathbf{e}^T_t.$$

$\mathbf{A10}$: We now analyze the projected ODE evolving in the slower timescale given by $\dt{\bm{\vartheta}^s} =\bar{\bm{g}}^s(\bm{\vartheta}^s)+\bm{r}^s(\bm{\vartheta}^s)$, where $\bm{r}^s(\bm{\vartheta}^s)$ is the reflection term. The first consequence of the separability of the function $\bar{\bm{g}}^s(\cdot)$ among types is the representation of the same function as a negative gradient of real valued continuously differentiable function 
$f^s(\bm{\vartheta}^s)= -\sum_t \int \bar{g}^s_t(\vartheta^s_t)d(\vartheta^s_t)$. Which readily makes the limit point of the ODE as the stationary points, i.e. the points where $\dt{\bm{\vartheta}^s}=\bm{0}$. 


\begin{claim}
Each stationary point of the slower ODE, given a $((1-2\epsilon)\bm{L})$-feasible tuple  $(\bm{h}^*,\bm{s}^*)$, corresponds to a TTL pair $\left(\hat{\bm{\vartheta}}, \hat{\bm{\vartheta}}^s\right)$ and an achieved hit rate, normalized size tuple 
$\left(\hat{\bm{h}},\hat{\bm{s}}\right)$, that, for each type $t$, satisfy one of the following three conditions:
\begin{enumerate}
\item Hit rate $\hat{h}_t = h^*_t$ and normalized size $\hat{s}_t = s^*_t$.
\item Hit rate $\hat{h}_t = h^*_t$ and normalized size $\hat{s}_t > s^*_t$, while $\hat{\vartheta}^s_t=0$.
\item Hit rate $\hat{h}_t = h^*_t$ and normalized size $\hat{s}_t < s^*_t$, while $\hat{\vartheta}^s_t=1$.
\end{enumerate}
\label{clm:stat}
\end{claim}
\begin{proof}
The separability of the ODE in each type $t$ enables us to analyze each coordinate of the ODE separately. First of all it is not hard to check that the points mentioned above gives $\bar{\bm{g}}^s(\bm{\vartheta}^s)=0$, i.e. they are stationary. The first case holds as the expectation of the update becomes zero. However, for the other two cases the projection term induces stationary points on the boundary. The harder direction is to rule out existence of other stationary points under $((1-2\epsilon)\bm{L})$-feasibility of the tuple.
  
We now recall the properties of the convergence of the projected ODE in the faster timescale given a fixed $\bm{\vartheta}^s$. Specifically, for a fixed $\bm{\vartheta}^s$ for each $t$ there can be two cases (1) for some unique $\bm{\theta}$ the hit rate $\bm{h}^*$ is achieved or (2) the hit rate is not achieved for types $t\in T'\subseteq T$ and $\theta_t=L_t$.

However, the scenario (2) is not possible for $(1-2\epsilon)\bm{L}$-feasible hit rates. For the sake of contradiction let us assume that there is a stationary point where $\theta_t=L_t$. This implies that $\theta_t^s=L_t$ and the cache is working in the d-TTL mode. The hit rate of object $t$ in f-TTL is always smaller than that of d-TTL with same $\theta_t$ value. This means the `typical' hit rate $\bm{h}^*$ can not be $(1-2\epsilon)\bm{L}$-feasible as it lies on the boundary of $\bm{L}$-feasible region. 

We are now left with the scenario that the hit rate is achieved, i.e. $\hat{h}_t = h_t^*$. We can further differentiate three cases  (1a) $\theta_t \leq (1-3\epsilon/2)L_t$, (1b) $ (1-3\epsilon/2)L_t < \theta_t < (1-\epsilon/2)L_t$ and (1c) $\theta_t \geq (1-\epsilon/2)L_t$. First of all we can rule out (1c) from the same logic as scenario (2). In case (1a) due to $\bm{L}$-feasibility we achieve the desired tuple as there exist some stationary point in our mentioned space. In the case (1b)  external force works on the ODE where we start moving f-TTL towards d-TTL. Nonetheless, the projected ODE for $\bm{\vartheta}^s$ tracks the average size of the system and reaches a stationary point only if the desired normalized size is achieved. Consequently, it achieves the targeted tuple with $\theta_t > L_t(1-2\epsilon)$. Therefore, none of the above three cases, (1a), (1b) and (1c), occurs. 
\end{proof}

Finally, to conclude the proof we have shown that for all $t$ the function $\bar{g}_t^s(\vartheta_t^s)$ are continuously differentiable in $\vartheta_t^s$. Recall that the function $\vartheta_t(\vartheta_t^s)$ is defined as
$\vartheta_t(\vartheta_t^s)=\max\{\tilde{\vartheta}_t(\vartheta_t^s), 1\}$ where $\tilde{\vartheta}_t(\vartheta_t^s)$ is a continuously differentiable function of $\vartheta_t^s$. For any $(1-2\epsilon)\bm{L}$-feasible hit rate target we know that $\vartheta_t(\vartheta_t^s)<L_t$ implying $\vartheta_t(\vartheta_t^s) \equiv \tilde{\vartheta}_t(\vartheta_t^s)$.\footnote{The assumption $A10$ fails to hold for the the hit rate and normalized size tuples at the `boundary' of $\bm{L}$-feasible region as $\vartheta_t(\vartheta_t^s)$ is no longer continuously differentiable. This implies that the convergence may happen to any `asymptotically stable' point in $H$. These points are really hard to characterize for the complex caching process. Refer~\cite{kushner2003stochastic} for details.} Therefore, for each $t$ the function $\bar{g}_t^s(\vartheta_t^s)$ is continuously differentiable (the space closed under composition) and the function $f^s(\bm{\vartheta}^s)$ is twice continuously differentiable. 
\end{proof}
  
We state the following theorem which is a direct adaptation from the works by~\cite{borkar1997stochastic},~\cite{bhatnagar2009natural}, and~\cite{kushner2003stochastic}.  
\begin{theorem}[Adapted from~\cite{borkar1997stochastic}, \cite{bhatnagar2009natural} and~\cite{kushner2003stochastic}]
\label{thm:Convergence}
Under the validity of $A0$ to $A10$ in Condition~\ref{a:stochApprox}, there is a null set $\mc{N}$ such that for any sample path $\omega\notin \mc{N}$, $(\bm{\vartheta}(l),\bm{\vartheta}^s(l))^{\omega}$ converges to a unique compact and connected subset of the stationary points given in the Claim~\ref{clm:stat}, as $l\to \infty$.  
\end{theorem}
    
This completes the proof by noting that the stationary points yield only the scenarios mentioned in the main theorem statement.     
\end{proof}

\begin{remark}[On Adaptation of Conditions and  Proof of Theorem~\ref{thm:Convergence}]
	\color{blue}
	The proof outline of two time scale stochastic algorithm is same as~\cite{borkar1997stochastic, bhatnagar2009natural}. However, the system described in~\cite{bhatnagar2009natural} does not consider correlated noise in the stochastic approximation dynamics. We follow the basic steps from Ch. 6 in ~\cite{kushner2003stochastic} to carry the results to our system by ensuring convergence of the two ODEs working in different time scales.

	The assumption A.0 is standard in the two time scale stochastic approximation literature (see e.g. \textit{equations (10,11) in~\cite{bhatnagar2009natural}}). In A.0 part (2) is sufficient to asymptotically wipe out the contribution from the martingale difference $\bm{\delta M}$, $\bm{\delta M}^s$. In A.0 last equality of part (3) separates the two timescale making the $\bm{\theta}$ parameter vary at a faster timescale. A.1 is adapted from \textit{$A.1.1$ in Ch. $6$ in~\cite{kushner2003stochastic}}. A.1. ensures uniform integrability of the iterates. A.2 is sufficient for \textit{$A.1.5$ in Ch. $6$ in~\cite{kushner2003stochastic}}---the Kushner-Clark condition---to hold for the noise sequences. A3 is characterizing the projection set similar to \textit{$A.3.1$ in Ch. $4$ in~\cite{kushner2003stochastic}}. A.4 and A.5 are boundedness and smoothness conditions on the mean ODE describing the slower and faster time scale dynamics. A.4 and A.5 are mentioned for a single time scale in \textit{$A.1.6$ and $A.1.7$ resp., in Ch. $6$ in~\cite{kushner2003stochastic}}. Also, the sufficiency of  similar conditions for both the time scales can be argued similarly as in~\cite{borkar1997stochastic}. A.6 is sufficient for \textit{$A.1.3$  in Ch. $6$ in~\cite{kushner2003stochastic}} (see, Example 1 under Ch. $6.2$ in~\cite{kushner2003stochastic} ). A.7 is one of the key assumptions for extending results to two time scale and it is adapted from assumption \textit{B2} in~\cite{bhatnagar2009natural}. A.8 and A.9 now works for the  mean ODE in the slow time scale, evolving with the average process from the faster time scale. Together they ensures assumption A4-A6 are valid for this mean ODE in the slow time scale.
	For arguing about the convergence of the ODEs in the two different time scale we use the \textit{Theorem 1.1 in in Ch. $6$ in~\cite{kushner2003stochastic}}. A.10 is necessary to argue that the convergence for the slower time scale happens to one of the possible many limit points (see, \textit{$A.2.7.$  in Ch. $6$ in~\cite{kushner2003stochastic}} and discussions therein). Finally, along the line of~\cite{borkar1997stochastic} we can argue how this ensures that the ODEs converge jointly as in Theorem 1.1. in~\cite{borkar1997stochastic}. 
	 
\end{remark}

\begin{remark}[On the Condition~\ref{a:stochApprox}]
The assumption of the target hit rate $\mathbf{h}^*$ to be `typical' as well as the condition that the inter arrival time p.d.f. is simply connected is introduced to prove convergence of the algorithm to a unique $\bm{\theta}^*$. If we relax these assumptions the convergence for d-TTL still happens and achieves the targeted hit rate. But it converges to a set of stationary points of the projected ODE for the faster timescale. However the two timescale analysis fails to hold as the uniqueness of the stationary point in the faster projected ODE is a standard assumption in the existing literature. On the contrary the absolute continuity of the p.d.f. of the inter arrival time is essential for the dynamics of the system to be continuous. Relaxing this assumption will deem the analysis infeasible even for d-TTL. The rarity condition can not be weakened significantly without violating Kushner-Clark condition on the noise sequence. We, also, note as the target approach the higher hit rate regions the f-TTL algorithm starts becoming less aggressive in meeting the normalized size. In this case f-TTL starts moving towards d-TTL algorithm by the use of threshold function $\up(\cdot,\cdot;\epsilon)$.
\end{remark}

%% file: poisson.tex
\subsection{Poisson Arrival, Independent Labeling}
In this section we prove Lemma~\ref{lemm:filtering} which states that full filtering yields the best normalized size if the hit rate is achievable in the full filtering mode. We also give a guideline about setting the parameter for the algorithm for a given hit rate target in Lemma~\ref{a:Zipf}.


\begin{proof}[\textbf{Proof of Lemma~\ref{lemm:filtering}}]
From the previous discussions, it suffices to prove the lemma for each type separately. So we consider the objects of type $t$ and drop the subscript $t$ (at times) for simplifying notations without loss of clarity. 
Given target hit rate $h^*$, let $(\hat{\theta}, \hat{\theta}^s)$ and $(\theta,0)$ be the {\ttl}s for the two f-TTL caching algorithms. For any TTL pair $(\theta,\theta^s)$ we have the hit rate of recurring object $c$ as 
$$h_c(\theta, \theta^s) = \left( \left(1- e^{-\lambda_c \theta}\right)^2 + e^{-\lambda_c \theta}\left(1-e^{-\lambda_c \theta^s}\right)\right).$$
Further, for Assumption~\ref{a:ArrPoisson} the normalized size of the object $c$ can be expressed as $s_c(\theta, \theta^s)=  \lambda_c^{-1}h_c(\theta, \theta^s)$. The combined hit rate and normalized size for type $t$ can be given as 
$$h(\theta, \theta^s)=  \tfrac{\dm{\sum_{c:c_{typ}=t}}w_c\pi_c h_c(\theta, \theta^s)}{\dm{\sum_{c:c_{typ}=t}}w_c\pi(c)},~s(\theta, \theta^s)= \tfrac{\dm{\sum_{c:c_{typ}=t}}w_c\pi_c s_c(\theta, \theta^s)}{\dm{\sum_{c:c_{typ}=t}}w_c\pi(c)}.$$

The difference of normalized size for a \emph{recurring object} $c$ among this two caches can be given as $$\Delta_c = h_c(\hat{\theta}, \hat{\theta}^s) - h_c(\theta, 0).$$  Due to monotonicity of hit rate we have $\theta > \hat{\theta}$ for `typical' target hit rate. Intuitively, as the total hit rate for the type remains fixed more filtering increase the  hit rate of the popular objects in f-TTL with $(\theta,0)$ compared to the other, whereas for the unpopular objects the order is reversed. Building on the intuition, for $\delta_1= 1-\tfrac{\hat{\theta}^s}{\hat{\theta}} \in [0,1)$, $\delta_2=\tfrac{\theta}{\hat{\theta}}-1>0$ and $x_c = e^{-\lambda_c \hat{\theta}}$ we can represent $\Delta_c$ as $q(x_c)$, where $q(\cdot)$ is 
$$q(x) = (1-x)^2 +x(1-x^{1-\delta_1})- (1-x^{1+\delta_2})^2.$$
Analyzing this polynomial we can conclude that the polynomial has exactly one root $x^*$ in the interval $(0,1)$. Therefore, it is easy to check that for $x_c \in  [x^*,1)$, $\Delta_c \geq 0$ and for  $x_c \in(0,x^*)$, $\Delta_c < 0 $. From definition of $x_c$, we deduce that there is a $c^*$ such that $\Delta_c < 0 $ for $c \leq c^*$ and  $\Delta_c \geq 0 $ otherwise. 

\color{black}
W.l.o.g. let $c$ be ordered (otherwise relabel) in the decreasing order of popularity ($c=1$ most popular). Therefore, first due to the equality of the hit rate for the two scenario we have 
\begin{equation}
\label{eq:hitrate}
\dm{\sum}_{c=c^*+1}^{|\mc{K}_t|}\hat{w}_c\pi_c\Delta_c = \dm{\sum}_{c=1}^{c^*}\hat{w}_c\pi_c|\Delta_c|,
\end{equation}
where $\hat{w}_c = 1$ if object hit rate is considered else, in case of byte hit rate, $\hat{w}_c = w_c$.
Further, for type $t$ we have the difference in the average normalized size over all objects as, 
\begin{subequations}
\begin{align}
&\left(\dm{\sum_{c:c_{typ}=t}}w_c\lambda_c\right)\left(s(\hat{\theta}, \hat{\theta}^s) -  s(\theta,0)\right) \nonumber\\
&= \dm{\sum}_{c=1}^{|\mc{K}_t|}w_c\Delta_c + \bar{w}_t\alpha_t\hat{\theta}^s, \label{eq:sub1}\\
&= \dm{\sum}_{c=c^*+1}^{|\mc{K}_t|}w_c\Delta_c - \dm{\sum}_{c=1}^{c^*}w_c|\Delta_c|+\bar{w}_t\alpha_t\hat{\theta}^s,\label{eq:sub2}\\
&\geq \tfrac{w_{c^*+1}}{\hat{w}_{c^*+1}}\dm{\sum}_{c=c^*+1}^{|\mc{K}_t|}\hat{w}_c\Delta_c - \tfrac{w_{c^*}}{\hat{w}_{c^*}}\dm{\sum}_{c=1}^{c^*}\hat{w}_c\pi_c|\Delta_c|+ \bar{w}_t\alpha_t\hat{\theta}^s,\label{eq:sub3}\\
&\geq \tfrac{w_{c^*+1}}{\hat{w}_{c^*+1}}\left(\dm{\sum}_{c=c^*+1}^{|\mc{K}_t|}\hat{w}_c\Delta_c - 
\dm{\sum}_{c=1}^{c^*}\hat{w}_c\pi_c|\Delta_c|\right)+ \bar{w}_t\alpha_t\hat{\theta}^s,\label{eq:sub4}\\
&\geq \tfrac{w_{c^*+1}}{\hat{w}_{c^*+1}}\left(\dm{\sum}_{c=c^*+1}^{|\mc{K}_t|}w_c\Delta_c - \frac{1}{\pi_{c^*}}\dm{\sum}_{c=1}^{c^*}w_c\pi_c|\Delta_c|\right)+ \bar{w}_t\alpha_t\hat{\theta}^s,\label{eq:sub5}\\
&= \tfrac{w_{c^*+1}}{\hat{w}_{c^*+1}}\dm{\sum}_{c=c^*+1}^{|\mc{K}_t|}w_c\left(1- \tfrac{\pi_c}{\pi_{c^*}}\right)\Delta_c + \bar{w}_t\alpha_t\theta_1^s \label{eq:sub6}\\
&\geq \bar{w}_t\alpha_t\hat{\theta}^s. \label{eq:sub7}
\end{align} 
\end{subequations}
The equalities~\eqref{eq:sub1} and~\eqref{eq:sub2} are trivial, as well as 
the inequalities~\eqref{eq:sub3} and~\eqref{eq:sub4} in case of byte hit rate. 
Whereas, for object hitrate, the inequalities~\eqref{eq:sub3} and~\eqref{eq:sub4} hold as, by assumption, the size of a recurrent object is non-decreasing w.r.t. probability $\pi_c$, i.e. $w_c\geq w_{c'}$ for $c<c'$. The inequality~\eqref{eq:sub5} holds as $\frac{\pi_c}{\pi_{c^*}}\geq 1 $  for $c \leq c^*$ and the equality~\eqref{eq:sub6} is due to the relation~\eqref{eq:hitrate}. Finally, the last inequality~\eqref{eq:sub7} follows from, $ \frac{\pi_c}{\pi_{c^*}}\leq 1$ and $\Delta_c>0$ for $c> c^*$. This completes the proof. 
\color{black}
\end{proof}

The above lemma strengthens the statement in Theorem~\ref{thm:Convergence} under the assumption of Poisson traffic with independent labeling, as stated in Corollary~\ref{corr:Poisson},  which we prove next. 

\begin{proof}[\textbf{Proof of Corollary~\ref{corr:Poisson}}]
  We know from Theorem~\ref{thm:SAConv} that for any type $t,$ the
  f-TTL algorithm converges to one of the three scenarios. In scenario
  $1$ and $3,$ the statement of the corollary immediately holds. We
  now show by contradiction that convergence to scenario $2$ never
  happens. Suppose the algorithm converges to scenario $2$. The
  algorithm achieves average hit rate $h^*_t$ and average normalized size
  $\tilde{s}_t > s^*_t$, by the supposition. However, as the tuple
  $(\bm{h}^*, \bm{s}^*)$ is $(1-2\epsilon)\bm{L}$-feasible only the
  two following cases take place.

  In the first case there exists a tuple $(\theta_t, \theta_t^s)$,
  with $\theta_t^s>0$, such that for type $t$ we achieve the tuple
  $(h^*_t, s^*_t)$ (a.a.s.). Due to Lemma~\ref{lemm:filtering}, this
  implies the normalized size under scenario $2$, $\tilde{s}_t \leq
  s^*_t$. This leads to a contradiction.

  In the second case (complement of the first) there exists a tuple
  $(\tilde{\theta}_t, 0)$ that achieves $(h^*_t,s^*_t)$. However, as
  $h^*_t<1$ ($\bm{h}^*$ is `typical'), under full filtering
  ($\theta_s^t=0$) \s{the hit rate monotonically increases with the TTL value. Therefore,} there is a unique value of 
  $\theta_t$ for which  $h^*_t$ is achieved (a.a.s.).
  Therefore, convergence in scenario $2$ happens to pair
  $(\tilde{\theta}_t, 0)$ (a.a.s.) and the average normalized size
  $\tilde{s}_t = s^*_t$. However this is again a contradiction.
\end{proof}

\subsection{Poisson Arrival and Zipfian Labeling}
Finally, we present a guideline for parameter setup for Poisson arrival and Zipfian distribution---a popular model in caching literature. Consider the following Zipfian popularity distribution where the object label distribution is as follows.

\assumptiongroup
\begin{assumptiongrp}
\label{a:Zipf}
Poisson Arrival and Zipfian Labeling:
\begin{itemize}[leftmargin=*]
\item  The inter arrival times are i.i.d. and follows exponential distribution with rate $\lambda=n\lambda_0$, for constant $\lambda_0>0$.

\item  Let $\pi(c)$ be the probability of a recurring object $c$. Then for each type $t\in [T]$, the marginal probability of recurring objects is $q_t = \dm{\sum_{c: c_{typ} = t, c\in \mc{K}}} \pi(c)$. Whereas, on each
turn, some rare object of type t is chosen with probability $\alpha_t$ and following the condition in~\eqref{eq:rarity}.

\item Arrange the recurrent objects of type $t$ in decreasing order of the popularity $\mc{K}_t = \left(c: c_{typ} = t, c\in \mc{K} \right)$. For each type $t$, the popularity of recurrent objects follows a Zipfian distribution with parameter $\beta_t$. Specifically, let $c_k = \mc{K}_t(k)$ be the $k$-th popular recurrent object of type $t$, then $\pi(c_k) = q_t Z_t^{-1} k^{-\beta_t}$, for $k = 1,\dots, K_t$.

\item All the objects have unit size, i.e. $w_c=1$ for all $c$.
\end{itemize}
\end{assumptiongrp}

We next discuss the choice of appropriate truncation parameter
$\mathbf{L}$ for Poisson arrival and Zipfian distribution. 
\begin{lemma}
\label{lemm:tuning}
\s{
Suppose, for some constant $\gamma>0$, the parameter $\mathbf{L}$ satisfy, $$\mathbf{L}=\Big(L_t: t \in [T], L_t =
\tilde{\Omega}\Big(n^{-(1-\frac{\beta_t \gamma}{\beta_t-1})}\Big)\Big).$$
Then, under Assumption~\ref{a:Zipf} with $\|\mathbf{L}\|_{\infty}$-rarity condition,  
 for both d-TTL and f-TTL algorithms an object hit rate  $\bm{h}^* = (1-
\Omega(n^{-\gamma}))\bm{1}^T$, is $\mathbf{L}$-feasible. 
}
\end{lemma}

\begin{proof}[\textbf{Proof of Lemma~\ref{lemm:tuning}}]
Let $c$ be a type $t$ object and $\lambda_c = \lambda q_t Z_t^{-1}c^{-\beta_t}$. For d-TTL algorithm the hit rate for type $t$ objects is given as,
\begin{equation*}
h_t(\theta_t) = \frac{q_t}{\alpha_t + q_t} \left(\dm{\sum}_{c = 1}^{|\mc{K}_t|} Z_t^{-1}c^{-\beta_t}(1- e^{-\lambda_c\theta_t}) \right).
\end{equation*}
Further define $r_t^*= 1- \tfrac{h_t^*}{1+\alpha_t/q_t}$ and $\delta r_t = \tfrac{r_t^*}{\log(n)}$. For $c_{typ}^* = (\delta r_t Z_t (\beta_t -1 ))^{-\frac{1}{\beta_t-1}}$, we have $\dm{\sum_{c = c_{typ}^*+1}^{|\mc{K}_t|}}\lambda_c\leq \delta r_t$. 
\begin{equation*}
\dm{\sum}_{c=1}^{|\mc{K}_t|} Z_t^{-1}c^{-\beta_t} e^{- \lambda_c L_t} \leq  e^{- \lambda_{c_{typ}^*} L_t}  + \delta r_t.
\end{equation*}
Choosing $L_t = \frac{1}{\lambda_{c_{typ}^*}} \ln \left( \frac{1}{0.99 r_t^*-\delta r_t}\right)$, $ e^{- \lambda_{c_{typ}^*} L_t} =  (0.99r^* - \delta r_t)$  and the hit rate for {\ttl} $L_t$, is strictly greater than  $h_t^*$. 
Furthermore, as $r_t^*= \mathcal{O}(n^{-\gamma})$ and $\lambda = n\lambda_0$, we have $$L_t=\Omega\left(n^{-(1-\frac{\beta_t \gamma}{\beta_t-1})}(\log n)^{1+\frac{\beta_t}{\beta_t-1}}\right).$$

For f-TTL we arrive at the order-wise same result for $L_t$ by observing the hit rate expression for the f-TTL is only constant times away from that of d-TTL.
\end{proof}

\begin{remark}
Lemma~\ref{lemm:tuning} implies that under Poisson arrival with arrival rate
$\lambda = \Omega(n)$ and  Zipfian popularity distribution with
$\min_t \beta_t > 1$, we can achieve arbitrarily high hit rates using parameter
$\bm{L}$ of the order $O(1)$ for large enough $n$ and appropriate rarity condition.
The above lemma considers only unit sized object, but can be extended to more general object sizes 
that are bounded.
\end{remark}

%% file: bhr.tex
\section{Byte hit rate performance of d-TTL and f-TTL} \label{sec:bhr}

In this section, we evaluate the byte hit rate performance and convergence of d-TTL and f-TTL using the experimental setup described in Section~\ref{sec:exp-setup}. 

\subsection{Hit rate performance of d-TTL and f-TTL}
\begin{figure} [!h]
\centering
\includegraphics[width=0.9\linewidth]{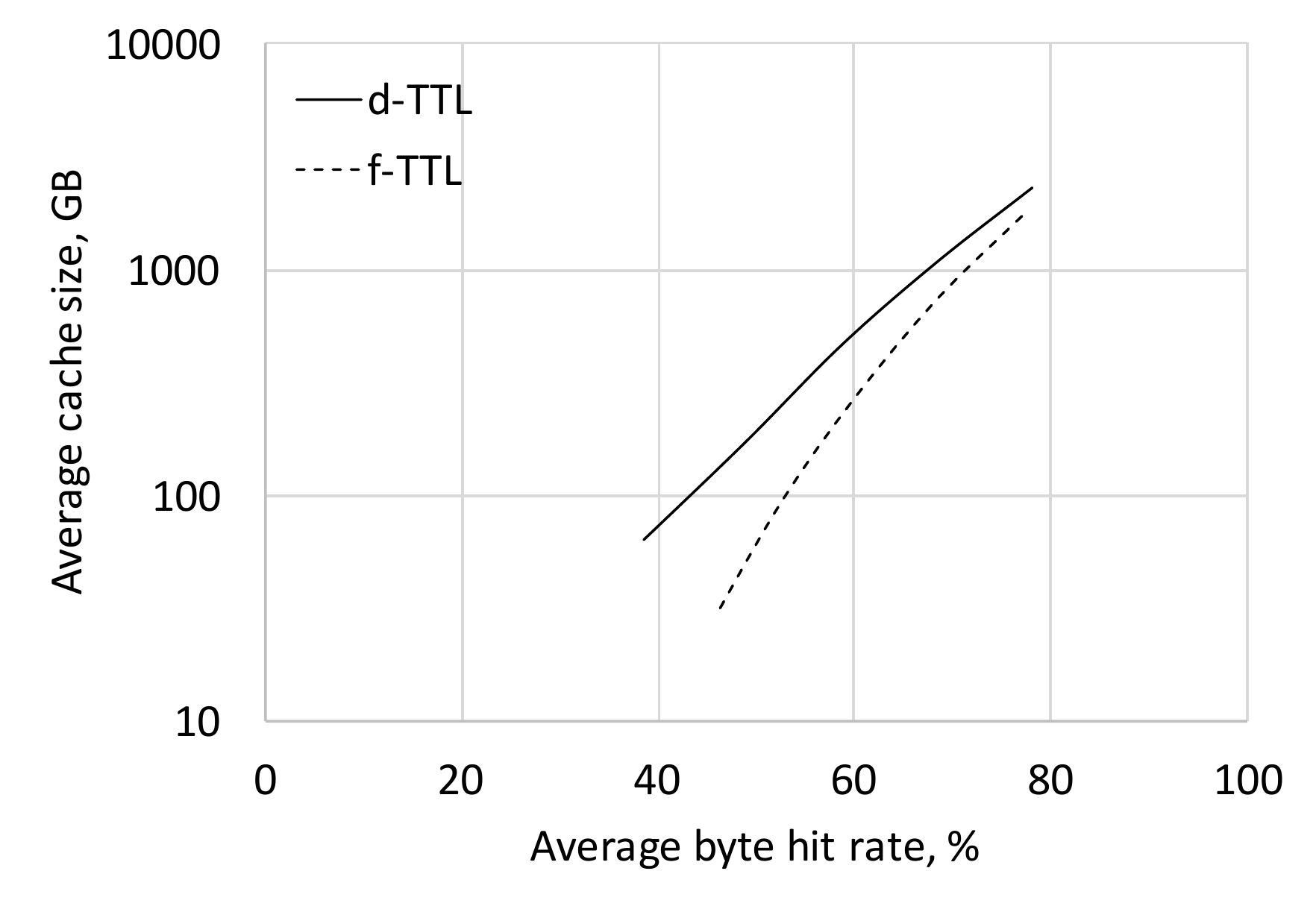}
\caption{Hit rate curve for byte hit rates.}
\label{fig:hrc-bhr-vos}
\end{figure}

To obtain the HRC for d-TTL, we fix the target byte hit rate at 80\%, 70\%, 60\%, 50\% and 40\% and measure the hit rate and cache size achieved by the algorithm. Similarly, for f-TTL, we fix the target hit rates at 80\%, 70\%, 60\%, 50\% and 40\%. \s{Further, we set the target normalized size of f-TTL to 50\% of the normalized size of d-TTL}. The HRCs for byte hit rate is shown in Figure \ref{fig:hrc-bhr-vos}.

From ~\ref{fig:hrc-bhr-vos}, we see that, for a given hit rate, f-TTL requires lesser cache space than d-TTL. On average, f-TTL requires a cache that is  39\% smaller than d-TTL to achieve the same byte hit rate. Further note that achieving a specific byte hit rate value requires more cache size than achieving the same value for object hit rate. For instance, the d-TTL algorithm requires a cache size of 469GB to achieve a 60\% byte hit rate, whereas a 60\% object hit rate is achievable with a smaller cache size of 6GB (Figure~\ref{fig:hrc-ohr-vos}). This discrepancy is due to the fact that popular objects in production traces tend to be small (10's of KB) when compared to unpopular objects that tend to be larger (100's to 1000's of MB).

\subsection{Convergence of d-TTL and f-TTL for byte hit rates}

In this section we measure the byte hit rate convergence over time, averaged over the entire time window and averaged over 2 hour windows for both d-TTL and f-TTL. We set the target hit rate to 60\% and a target normalized size that is 50\% of the normalized size of d-TTL.

\begin{figure} [!h]
\centering
\includegraphics[width=0.9\linewidth]{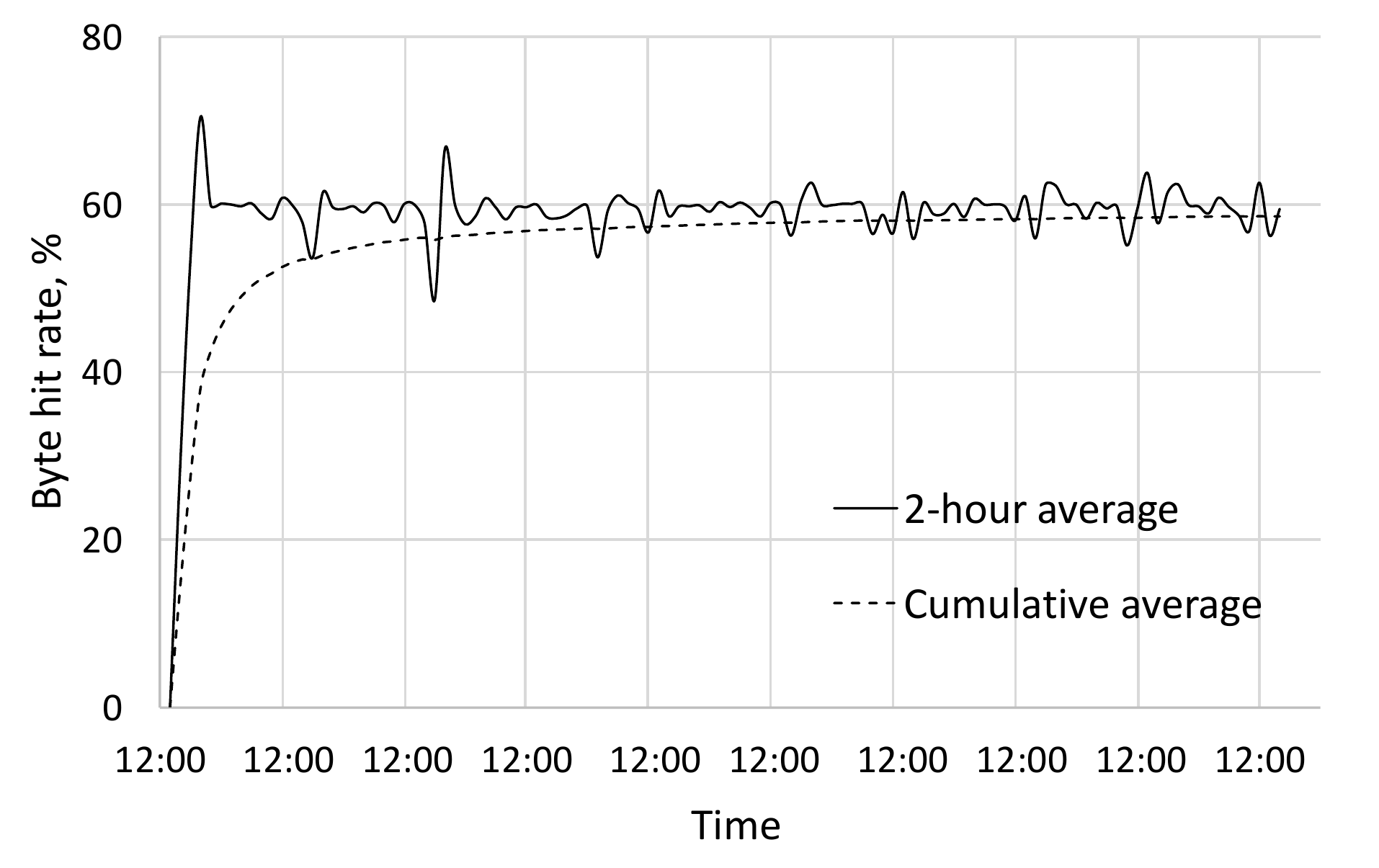}
\caption{Byte hit rate convergence over time for d-TTL;  target byte hit rate=60\%.}
\label{fig:hc-bhr-vos-d-ttl}
\end{figure}

\begin{figure} [!h]
\centering
\includegraphics[width=0.9\linewidth]{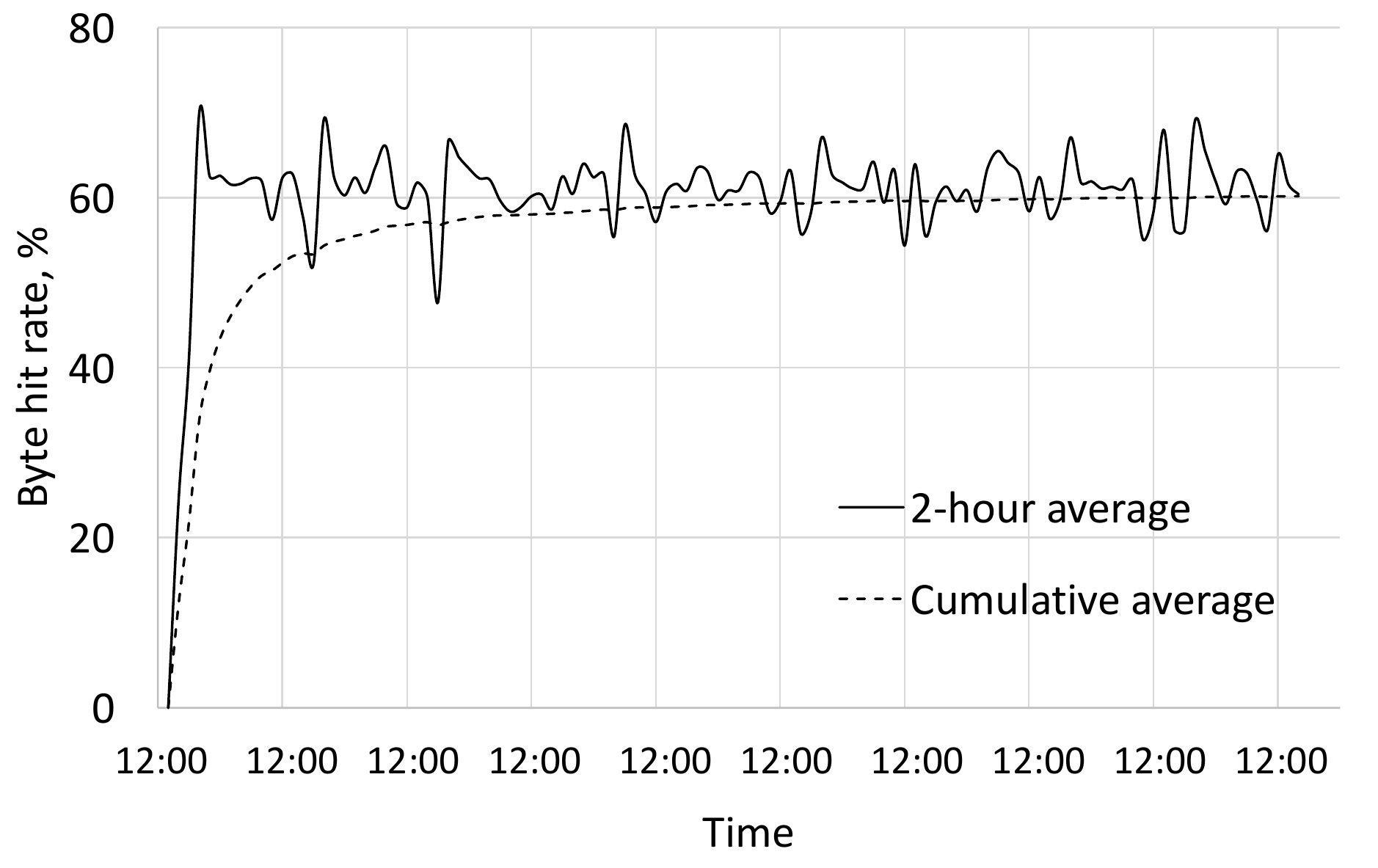}
\caption{Byte hit rate convergence over time for f-TTL;  target byte hit rate=60\%.}
\label{fig:hc-bhr-vos-f-ttl}
\end{figure}

From Figures~\ref{fig:hc-bhr-vos-d-ttl} and \ref{fig:hc-bhr-vos-f-ttl}, we see that d-TTL has a cumulative error of less than 2.3\% on average while achieving the target byte hit rate and f-TTL has a corresponding error of less than  0.3\%.
Moreover, we see that both d-TTL and f-TTL tend to converge to the target hit rate, which illustrates that both d-TTL and f-TTL are able to adapt well to the dynamics of the input traffic. We also observe that the average byte hit rates for both d-TTL and f-TTL have higher variability compared to object hit rates (Figures~\ref{fig:hc-ohr-vos-d-ttl} and~\ref{fig:hc-ohr-vos-f-ttl} \footnote{Refer to Section 6 in the main paper.}, due to the the fact that unpopular content in our traces have larger sizes, and the occurrence of non-stationary traffic can cause high variability in the dynamics of the algorithm.
  
In general, we also see that d-TTL has lower variability for both object hit rate and byte hit rate compared to f-TTL due to the fact that d-TTL does not have any bound on the normalized size while achieving the target hit rate, while f-TTL is constantly filtering out non-stationary objects to meet the target normalized size while also achieving the target hit rate.

%% file: sensitivity.tex



\section{ Effect of  target normalized size on  f-TTL} \label{sec:sen-f-ttl-size}

In Section~\ref{sec:empirical}, f-TTL is implemented by setting a normalized size target that is 50\% of the normalized size of d-TTL. This helps f-TTL achieve the same target hit rate as d-TTL but at half the expected cache size. In this section, we evaluate the performance of f-TTL when we change the normalized size target. Specifically, we measure the average hit rate and cache size achieved by f-TTL when we set the target object hit rates to 60\% and 80\% and the target normalized size to 45\%, 50\% (as in Section~\ref{sec:empirical} and 55\% of the normalized size of d-TTL. 

\begin{table}[!ht]
\small
\caption{Impact of normalized size target on the performance of f-TTL.}
\label{tab:sen-f-ttl-size}
\centering
\begin{tabular}{|p{0.55in}|p{0.55in}|p{0.55in}|p{0.55in}|p{0.55in}|}
\hline
Target  OHR (\%) & Normalized size target (\%) & OHR achieved (\%) & Cache size achieved (GB)   \\
\hline
\multirow{3}{*}{60} 	& 45 & 59.36 & 2.94 \\
				& 50 & 59.36 & 2.96 \\
				& 55 & 59.36 & 2.98 \\
\hline
\multirow{3}{*}{80} 	& 45 & 78.55 & 54.81 \\
				& 50 & 78.55 & 55.08 \\
				& 55 & 78.55 & 55.32 \\
\hline
\end{tabular}
\end{table}

From Table~\ref{tab:sen-f-ttl-size}, we see that with a target hit rate of 60\%, f-TTL is able to achieve the target hit rate with a small error of 0.64\% in all three target normalized size scenarios. Similarly, f-TTL is also able to achieve the target hit rate of 80\% in all three target normalized size scenarios with a slightly larger error of 1.45\%. Both these scenarios show that f-TTL is able to the target hit rate at different target normalized sizes with high accuracy.

We also measure the average cache size achieved by f-TTL in all three scenarios. In the case of the 60\% target hit rate, we see that f-TTL achieves a hit rate of 59.36\% with the smallest average cache size when the target normalized size of f-TTL is 45\% of that of d-TTL and the largest average cache size when the target normalized size of f-TTL is 55\% of that of d-TTL. This shows that f-TTL more aggressively filters out non-stationary content to achieve the target hit rate at smaller normalized size targets. The opposite happens at higher target normalized sizes. Similar behavior is observed when setting the target hit rate to 80\%. 

As discussed in Section~\ref{sec:T2P}, when a target normalized size is unachievable, the target can instead be used to control the aggressiveness of f-TTL in filtering out non-stationary content.
